\documentclass[letterpaper,11pt]{article}
\pdfoutput=1 

\usepackage{jheppub}

\usepackage[utf8]{inputenc}
\usepackage{subfig}
\usepackage{units}

\DeclareRobustCommand{\Sec}[1]{Sec.~\ref{#1}}

\DeclareRobustCommand{\App}[1]{App.~\ref{#1}}

\DeclareRobustCommand{\Fig}[1]{Fig.~\ref{#1}}

\DeclareRobustCommand{\Eq}[1]{Eq.~(\ref{#1})}
\DeclareRobustCommand{\Eqs}[2]{Eqs.~(\ref{#1}) and (\ref{#2})}
\DeclareRobustCommand{\Eqss}[3]{Eqs.~(\ref{#1}), (\ref{#2}), and (\ref{#3})}
\DeclareRobustCommand{\Ref}[1]{Ref.~\cite{#1}}
\DeclareRobustCommand{\Refs}[1]{Refs.~\cite{#1}}

\newcommand{\df}{\text{d}}
\newcommand{\zcut}{{z_\text{cut}}}
\newcommand{\thetacut}{{\theta_\text{cut}}}
\newcommand{\SD}{\text{SD}}
\newcommand{\GFF}{\mathcal{F}}

\newcommand{\as}{\alpha_s}

\newcommand{\LambdaNP}{\Lambda_\text{NP}}

\newcommand{\be}{\begin{equation}}
\newcommand{\ee}{\end{equation}}
\allowdisplaybreaks

\preprint{MIT--CTP 4987}

\title{Casimir Meets Poisson: Improved Quark/Gluon Discrimination with Counting Observables}

\author[a]{Christopher Frye,}
\author[b]{Andrew J.~Larkoski,}
\author[c]{Jesse Thaler,}
\author[c]{and Kevin Zhou}

\affiliation[a]{Department of Physics, Harvard University, Cambridge, MA 02138}
\affiliation[b]{Physics Department, Reed College, Portland, OR 97202}
\affiliation[c]{Center for Theoretical Physics, Massachusetts Institute of Technology, Cambridge, MA 02139}

\emailAdd{frye@physics.harvard.edu}
\emailAdd{larkoski@reed.edu}
\emailAdd{jthaler@mit.edu}
\emailAdd{knzhou@mit.edu}

\abstract{
Charged track multiplicity is among the most powerful observables for discriminating quark- from gluon-initiated jets.
Despite its utility, it is not infrared and collinear (IRC) safe, so perturbative calculations are limited to studying the energy evolution of multiplicity moments.  
While IRC-safe observables, like jet mass, are perturbatively calculable, their distributions often exhibit Casimir scaling, such that their quark/gluon discrimination power is limited by the ratio of quark to gluon color factors.
In this paper, we introduce new IRC-safe counting observables whose discrimination performance exceeds that of jet mass and approaches that of track multiplicity.
The key observation is that track multiplicity is approximately Poisson distributed, with more suppressed tails than the Sudakov peak structure from jet mass.
By using an iterated version of the soft drop jet grooming algorithm, we can define a ``soft drop multiplicity'' which is Poisson distributed at leading-logarithmic accuracy.
In addition, we calculate the next-to-leading-logarithmic corrections to this Poisson structure.
If we allow the soft drop groomer to proceed to the end of the jet branching history, we can define a collinear-unsafe (but still infrared-safe) counting observable.
Exploiting the universality of the collinear limit, we define generalized fragmentation functions to study the perturbative energy evolution of collinear-unsafe multiplicity.
}

\begin{document} 
\maketitle


\section{Introduction}
\label{sec:intro}

The fantastic jet reconstruction performance of ATLAS and CMS \cite{Aaboud:2017jcu,Khachatryan:2016kdb}---along with increasingly sophisticated tools to predict jet properties from first principles \cite{Dasgupta:2013ihk,Frye:2016aiz,Larkoski:2013paa,Dasgupta:2015lxh,Dasgupta:2014yra,Larkoski:2015kga,Procura:2014cba,Waalewijn:2012sv}---has led to significant advances in the field of jet substructure \cite{Adams:2015hiv,Altheimer:2013yza,Altheimer:2012mn,Abdesselam:2010pt}. A key goal in jet substructure is to robustly discriminate quark-initiated jets from gluon-initiated jets \cite{Gallicchio:2011xq,Gallicchio:2012ez,Larkoski:2013eya,Larkoski:2014pca,Bhattacherjee:2015psa,Badger:2016bpw,Komiske:2016rsd,Davighi:2017hok,Gras:2017jty}, with many applications to new physics searches at the Large Hadron Collider (LHC) (see e.g.~\cite{CMS:2013kfa,FerreiradeLima:2016gcz,Bhattacherjee:2016bpy}). In the eikonal limit, quarks and gluons differ only by their respective color charges, $C_F = 4/3$ versus $C_A = 3$, such that gluon jets emit more soft gluon radiation than quark jets. At this order, the difference between quark and gluon radiation patterns is controlled entirely by the Casimir ratio $C_A/C_F = 9/4$, which drives (and limits) the expected separation power between quark and gluon jets.

One of the most powerful quark/gluon discriminants is hadron multiplicity, or its charged-particle-only variant, track multiplicity $n_{\rm tr}$ \cite{Gallicchio:2011xq,Gallicchio:2012ez,Aad:2014gea,Larkoski:2014pca,ATLAS:2016wzt,Aad:2016oit}.
This is an effective discriminant because the average track multiplicity within quark and gluon jets scales approximately as (see e.g.~\cite{Bolzoni:2012ii,Bolzoni:2013rsa})
\be
\label{eq:multiplicity_scaling}
\frac{\langle n_{\rm tr} \rangle_g}{\langle n_{\rm tr} \rangle_q} \simeq \frac{C_A}{C_F}.
\ee
Since multiplicity is not infrared and collinear (IRC) safe, though, it is difficult to predict its discrimination performance from first principles.\footnote{It is possible to calculate the evolution with energy of the multiplicity moments; see, e.g., \Ref{Ellis:1991qj} for a review.}  On the other hand, IRC-safe observables like jet mass and jet width are analytically tractable \cite{Banfi:2004yd,Ellis:2010rwa,Larkoski:2014uqa}, but they exhibit worse quark/gluon performance than multiplicity. The reason is that these discriminants are dominated by a single emission at leading-logarithmic (LL) accuracy, giving rise to Casimir scaling of the quark/gluon discrimination power,
\be
\label{eq:naive_casimir_scaling}
\text{(gluon mistag rate)} \simeq (\text{quark efficiency})^{C_A/C_F},
\ee
and therefore relatively weak separation between quark and gluon jets. This Casimir scaling behavior holds for any observable with a Sudakov form factor at LL accuracy, including a wide range of IRC-safe additive observables \cite{Larkoski:2013eya}. While one can try to interpolate between the IRC-unsafe and IRC-safe regimes using generalized angularities \cite{Larkoski:2014pca}, track multiplicity remains one of the best performing---yet analytically puzzling---quark/gluon discriminants.

In this paper, we introduce a new class of ``counting observables'' that are IRC safe, yet yield comparable quark/gluon performance to track multiplicity. Unlike additive observables, which are only sensitive to a single emission at LL order, these counting observables are directly sensitive to multiple emissions at LL, allowing them to exceed the performance estimate in \Eq{eq:naive_casimir_scaling}. Crucially, the quark/gluon performance of counting observables still depends on the color factors $C_A$ and $C_F$, but instead of being described by Sudakov form factors, these observables are described by Poisson distributions; this allows their discrimination power to improve as more emissions are included. These counting observables not only clarify the underlying reason why track multiplicity performs so well, but they also demonstrate the new kinds of analytic structures possible from IRC-safe but non-additive observables.\footnote{An alternative counting method was proposed in \Ref{Bhattacherjee:2015psa}, which considers associated subjets outside of the jet boundary.  Additionally, there has been interest in understanding the scaling of the cross section at high jet multiplicity \cite{Gerwick:2012hq,Bern:2013gka,Badger:2013yda,Aad:2014qxa}. Here, we focus on counting subjets within the jet of interest.
}

The counting observables we study are based on an iterated variant of soft drop declustering \cite{Larkoski:2014wba}.  As a grooming procedure, soft drop starts at the trunk of an angular-ordered clustering tree \cite{Dokshitzer:1997in,Wobisch:1998wt} and sequentially removes soft branches with small momentum fraction $z_{ij}$ until a hard branching is found.  At a step in the clustering tree where branches $i$ and $j$ split, the splitting is retained in the groomed jet if the momentum fraction satisfies
\begin{equation}
\label{eq:soft_drop_condition}
z_{ij} > \zcut \left(\frac{\theta_{ij}}{R_0}\right)^\beta\,,
\end{equation}
where $\theta_{ij}$ is an appropriately defined relative angle between branches $i$ and $j$, and $R_0$ is the jet radius. For appropriate choices of the soft drop parameters $\zcut$ and $\beta$, observables defined on the groomed jet are automatically infrared (but not necessarily collinear) safe. 
While the original soft drop procedure terminates once it finds a hard $1 \to 2$ splitting, the iterated variant we employ in this paper continues, following the hardest branch (the ``trunk'') through multiple levels until an angular cutoff scale $\thetacut$ is reached.

The simplest counting observable we can define using iterated soft drop (ISD) is just the total number of emissions from the trunk of the clustering tree that ISD records.  
In particular, this includes all emissions $n \in [1, n_{\rm max}]$ that satisfy the soft drop condition and lie outside the $\thetacut$ cone.  
We call this observable ``soft drop multiplicity'',
\be
\label{eq:nSDdefinition}
n_\SD(z_{\rm cut}, \beta,\thetacut) = \sum_{n} 1,
\ee
which depends on the choice of ISD parameters.  
It is complementary to the ``soft drop level'' observable $L_\SD(\beta)$ introduced in \Ref{Aad:2015rpa}, which also iteratively applies the soft drop condition, but changes the $z_{\rm cut}$ scale.  
As long as $\zcut > 0$, soft drop multiplicity is infrared safe.  

With $\thetacut > 0$ or $\beta < 0$, $n_\SD$ is collinear safe as well, so we can use analytic resummation tools to predict its discrimination power. 
We do this to resum large logarithms of $\zcut$ and $\thetacut$, which are of soft and collinear origin, respectively, and which lead to a double-logarithmic observable.
The analysis at LL order is straightforward, yielding a Poisson distribution whose average value is set by the phase space ``area'' of counted emissions. 
This leads to quark/gluon discrimination power which approaches that of track multiplicity, particularly in the case of $\beta = -1$.  
Moving from LL to next-to-leading-logarithmic (NLL) order, one finds a slight decrease in discrimination power, due in part to the jet-flavor mixing that appears at this accuracy.  
We implement the NLL calculation through a set of evolution equations that have a similar form to parton evolution.  

With $\thetacut = 0$ and $\beta \geq 0$, the soft drop multiplicity $n_\SD$ is no longer collinear safe, so we cannot predict its absolute discrimination power. 
That said, for the special case of $\beta = 0$ (which was initially introduced as the modified mass drop tagger \cite{Dasgupta:2013ihk,Dasgupta:2013via}), 
we can use renormalization group (RG) techniques to predict the \emph{evolution} of its discrimination power. 
When $\beta = 0$, soft drop multiplicity has purely collinear divergences, which can be absorbed into a generalized fragmentation function (GFF) that depends on the RG scale $\mu$ \cite{Elder:2017bkd}. 
After extracting this GFF at low scales (either from LHC data\footnote{Just as for parton distribution functions and ordinary fragmentation functions, extracting GFFs involves matching to fixed-order calculations, as described in \Ref{Elder:2017bkd}.  
These fixed-order calculations involve a mixture of quark and gluon final-state partons, so multiple event samples with different quark/gluon fractions are required to disentangle the contributions from quark and gluon GFFs.} 
or parton shower simulations), one can use a perturbative DGLAP-like evolution equation to predict the discrimination power 
achievable at higher scales. Intriguingly, in the limit of pure Yang-Mills, one can show that at lowest order, the soft drop multiplicity asymptotes to a true Poisson distribution at large values of $\mu$, such that it behaves like an idealized counting observable (albeit in a theory with only gluons).

The remainder of this paper is organized as follows. 
In \Sec{sec:var_def}, we define the ISD procedure, introduce soft drop multiplicity, and take a first look at its distribution using parton shower generators. 
In \Sec{sec:ll_scale}, we perform an LL analysis, focusing on the contrast between soft drop multiplicity's Poisson behavior and the more familiar Sudakov-peak behavior of additive observables. 
We extend our analytic calculations to NLL order in \Sec{sec:irc_safe} and compare our analytic distributions to those obtained from various parton showers.  
We consider the collinear-unsafe case of $\thetacut = 0$ and $\beta = 0$ in \Sec{sec:irc_unsafe}, deriving the corresponding RG evolution equations and presenting numerical results based on parton shower inputs. 
We present our conclusions in \Sec{sec:conclusion}.

In an appendix, we demonstrate that our analytical tools can also be used to study more general ISD observables, in particular the weighted multiplicity 
$\sum_n (z_n)^\kappa$ which weights each counted emission according to its momentum fraction $z_n$.
Soft drop multiplicity is a special case $(\kappa = 0)$ of this more general observable, and the one most useful for quark/gluon discrimination.


\section{Counting Observables from Soft Drop Declustering}
\label{sec:var_def}

\subsection{Iterated Soft Drop}
\label{subsec:ISD}

Our counting observables are defined using an iterated variant of the soft drop declustering algorithm. We briefly review soft drop here for convenience and to establish conventions.

The soft drop grooming procedure can be applied to any jet found using a standard jet algorithm of characteristic 
radius $R_0$.  After reclustering the jet using the Cambridge/Aachen (C/A)
algorithm \cite{Dokshitzer:1997in,Wobisch:1998wt}, soft drop involves sequentially undoing the cluster history to remove wide-angle soft radiation and identify hard 2-prong substructure.  For each C/A branching into subjets $i$ and $j$, there are quantities $z_{ij}$ and 
$\theta_{ij}$, which are defined differently for different collider environments:
\begin{align}
  \text{$e^+e^-$ collisions:}&
  &  z_{ij} &= {\min(E_i,E_j) \over E_i+E_j}\,, 
  &  \theta_{ij} &= \text{angle between $i,j$}\,, \\
  \text{$pp$ collisions:}&
  &  z_{ij} &= {\min(p_{Ti}, p_{Tj}) \over p_{Ti}+p_{Tj}}\,,
  &  \theta_{ij} &= \Delta R_{ij}\,,
\end{align}
where $\Delta R$ represents distance in the rapidity-azimuth plane.   The soft drop grooming algorithm can be summarized as follows:
\begin{enumerate}
  
  \item Traverse the C/A clustering tree, beginning at the trunk and sequentially examining each branching. 
    
  \item Upon arriving at a branching into subjets $i$ and $j$,
    check whether the soft drop condition is satisfied:
    \begin{equation}
      z_{ij} > \zcut \, \left({\theta_{ij} \over R_0}\right)^\beta\,,
    \end{equation}
 where $\zcut$ and $\beta$ are fixed parameters of the algorithm.
 If so, the algorithm terminates; stop grooming and return the jet as is.

  \item If the branching fails this condition, remove the softer of the two subjets ($i$ or $j$) from the groomed jet and 
    return to Step 2 on the next branching in the remaining clustering tree.
\end{enumerate}

\begin{figure}[t]
\centering
\includegraphics[width=.8\textwidth]{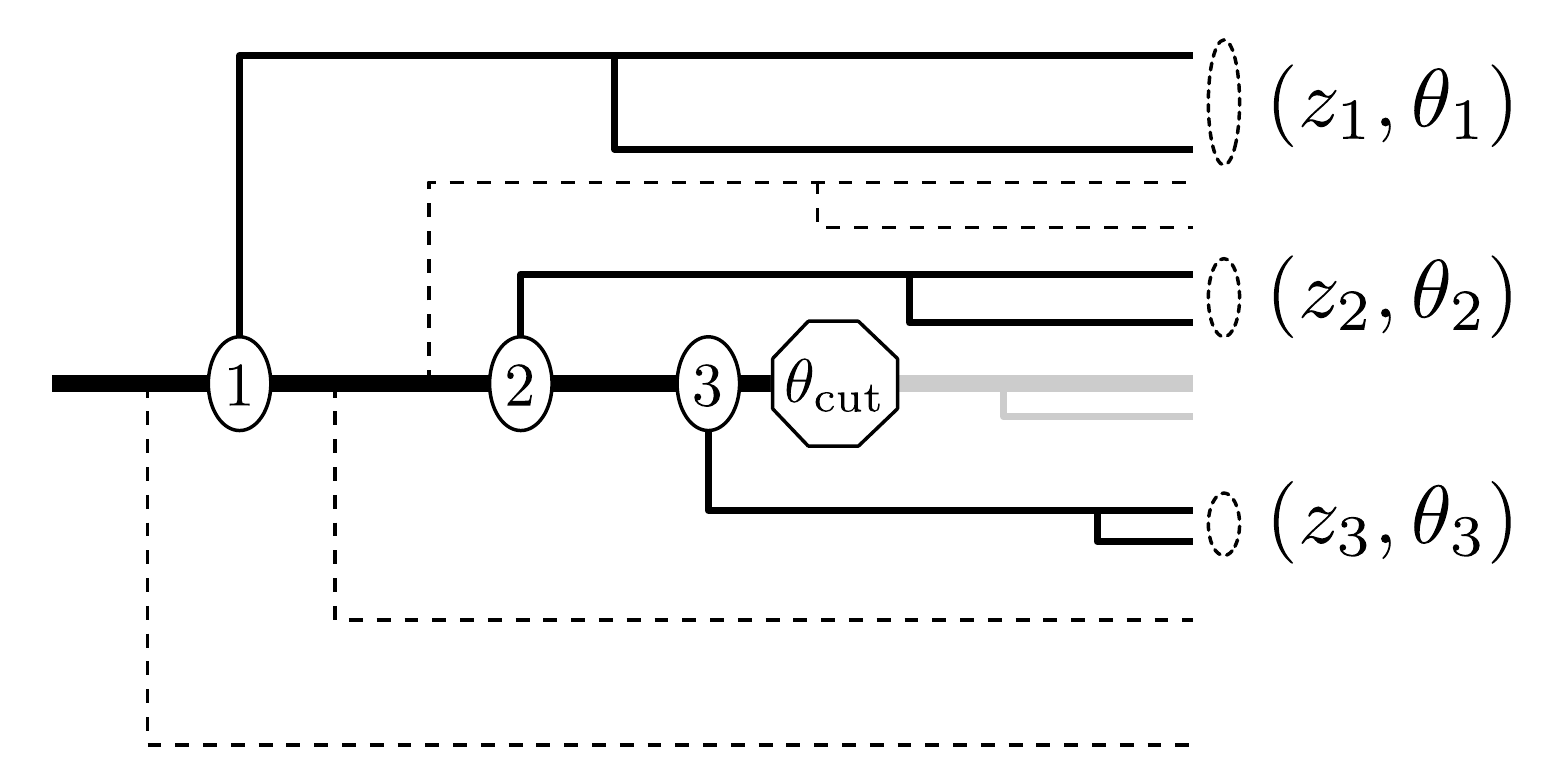}
\caption{Illustration of the ISD procedure. A C/A tree is declustered from the trunk (thick line), defined by the hardest $p_T$ branches.  If a node fails the soft drop condition, it is removed from consideration (dashed lines). If a node passes the soft drop condition after $n$ iterations, this defines the value of $(z_n, \theta_n)$.  The declustering stops at an angular scale of $\theta_{\rm cut}$, and subsequent nodes are not considered further (gray lines).}
\label{fig:sd_tree_diagram}
\end{figure}

Our analysis is based on ISD where the soft drop algorithm is iterated.  In this case, the procedure does not
terminate when a hard branching is found, but is instead iteratively applied to the harder of the two subjets.
This continues until an angular cutoff is reached, so in addition to $\zcut$ and $\beta$, ISD 
depends on an additional parameter $\thetacut$.  While ISD could be used as a grooming procedure in its own right, the primary purpose of ISD in this paper
is to determine which set of $(z_{ij},\theta_{ij})$ branchings contribute to the 
observables we define below.
For this purpose, the ISD algorithm proceeds as follows:
\begin{enumerate}
  
  \item[$1'$.] Set the counter $n$ equal to 1.  Traverse the C/A clustering tree, beginning at the trunk and sequentially examining each branching. 
    
  \item[$2'$.] Upon arriving at a branching into subjets 
    $i$ and $j$, check whether the branching angle satisfies
    \begin{equation}
      \label{eq:theta_cut}
      \theta_{ij} > \thetacut\,.
    \end{equation}
    If not, the algorithm terminates.

  \item[$3'$.] If $\theta_{ij} > \thetacut$, then check whether 
    the soft drop condition is satisfied:
    \begin{equation}
    \label{eq:SD_condition}
      z_{ij} > \zcut \, \left({\theta_{ij} \over R_0}\right)^\beta\,.
    \end{equation}
    If not, return to Step $2'$ on the harder of subjets $i$ and $j$. 

  \item[$4'$.] If the soft drop condition is satisfied, define 
    \begin{equation}
      z_n \equiv z_{ij}\,, \qquad \theta_n \equiv \theta_{ij}\,.
    \end{equation}
    Then increment $n \to n+1$ and return to Step $2'$ on the harder of subjets $i$ and $j$.
    
\end{enumerate}
Because we recurse to the harder subjet at each junction, we think of each $(z_n,\theta_n)$ splitting 
as an emission from the ``hard core'' of the jet and refer to the above procedure as traversing the ``trunk'' 
of the clustering tree. A schematic of this procedure is shown in \Fig{fig:sd_tree_diagram}.

To emphasize, we are not using ISD as an alternative grooming technique to soft drop. In fact, we have found no need to refer to the ISD-groomed jet explicitly in our analysis. Instead, we employ ISD simply as a method to obtain an IRC-safe set of $(z_n,\theta_n)$ values to define our counting observables. Of course, the specific values of $(z_n,\theta_n)$ depend on the precise choice of ISD procedure. In this paper, we focus on the soft drop multiplicity, which counts emissions from the trunk of the clustering tree, and have defined ISD accordingly. In \Sec{subsec:variations}, we consider variants of soft drop multiplicity, with corresponding variants to the ISD procedure.

To demonstrate the qualitative behavior of observables defined below in this section, we present results from parton shower simulations. 
We separately generate $pp \to Z+q$ and $pp \to Z+g$ events at center-of-mass energy 13 TeV using \textsc{MadGraph} 2.4.0 and let the $Z$ decay to neutrinos for simplicity. 
We then shower the events through \textsc{Vincia} 2.0.01 \cite{Giele:2007di,Giele:2011cb}, a plug-in to \textsc{Pythia} 8.215 \cite{Sjostrand:2007gs}, with default tuning parameters.\footnote{In 
\Sec{sec:irc_safe}, we show results from four different parton shower generators. 
Here, we use \textsc{Vincia} as a representative example since it makes predictions which are intermediate relative to the other generators.}
Jet are identified using the anti-$k_t$ algorithm \cite{Cacciari:2008gp} with radius $R_0 = 0.6$ in \textsc{FastJet} 3.1.3 \cite{Cacciari:2011ma}. 
We use a sample of events in which the hardest jet with $|\eta| < 2.5$ has $p_T$ between 450 and 550 GeV. 
We recluster and measure our observables on the hardest jet from each event using \textsc{FastJet}.
Because ISD is sufficiently different from ordinary soft drop, we do not use the \texttt{RecursiveTools} \textsc{fjcontrib} \cite{fjcontrib}, but rather directly traverse the C/A tree in our analysis.  
We plan to make our code available publicly in a future release of \textsc{fjcontrib}.

\subsection{Soft Drop Multiplicity}
\label{subsec:mult}

The $(z_n, \theta_n)$ values from ISD allow us to define a variety of interesting jet observables.
Here, we focus on soft drop multiplicity $n_\SD$, which is simply the total count of the recorded $(z_n, \theta_n)$ pairs.  This observable, defined already in \Eq{eq:nSDdefinition}, depends implicitly on the ISD parameters $\zcut$, $\beta$, and $\thetacut$. Among all of the observables we tested, $n_\SD$ appears to perform the best for quark/gluon discrimination.  We discuss more general observables in \Sec{subsec:variations} and \App{sec:weighted_nSD}. 

As defined above, ISD only follows the harder branch (i.e.~the trunk) at each junction of the clustering tree. 
Therefore, $n_\SD$ effectively counts emissions from the hard core of the jet, down to the angular resolution scale $\thetacut$. 
When $\zcut = \thetacut = 0$, $n_\SD$ is simply the depth of the trunk of the C/A tree. 

When $\zcut > 0$, the soft drop multiplicity is infrared safe, as all soft emissions at finite 
angles fail the soft drop condition in \Eq{eq:SD_condition}. When $\thetacut > 0$, soft drop 
multiplicity is also collinear safe, since an exactly collinear splitting along the trunk does not satisfy 
\Eq{eq:theta_cut}. Alternatively, $\beta < 0$ also gives collinear-safe distributions, since an exactly collinear 
splitting along the trunk does not satisfy \Eq{eq:SD_condition}. The borderline case of $\thetacut = 0$
and $\beta = 0$ is collinear unsafe, but it can be handled using RG methods, as shown in \Sec{sec:irc_unsafe}.

\begin{figure}[t]
\centering
\includegraphics[width=.5\textwidth]{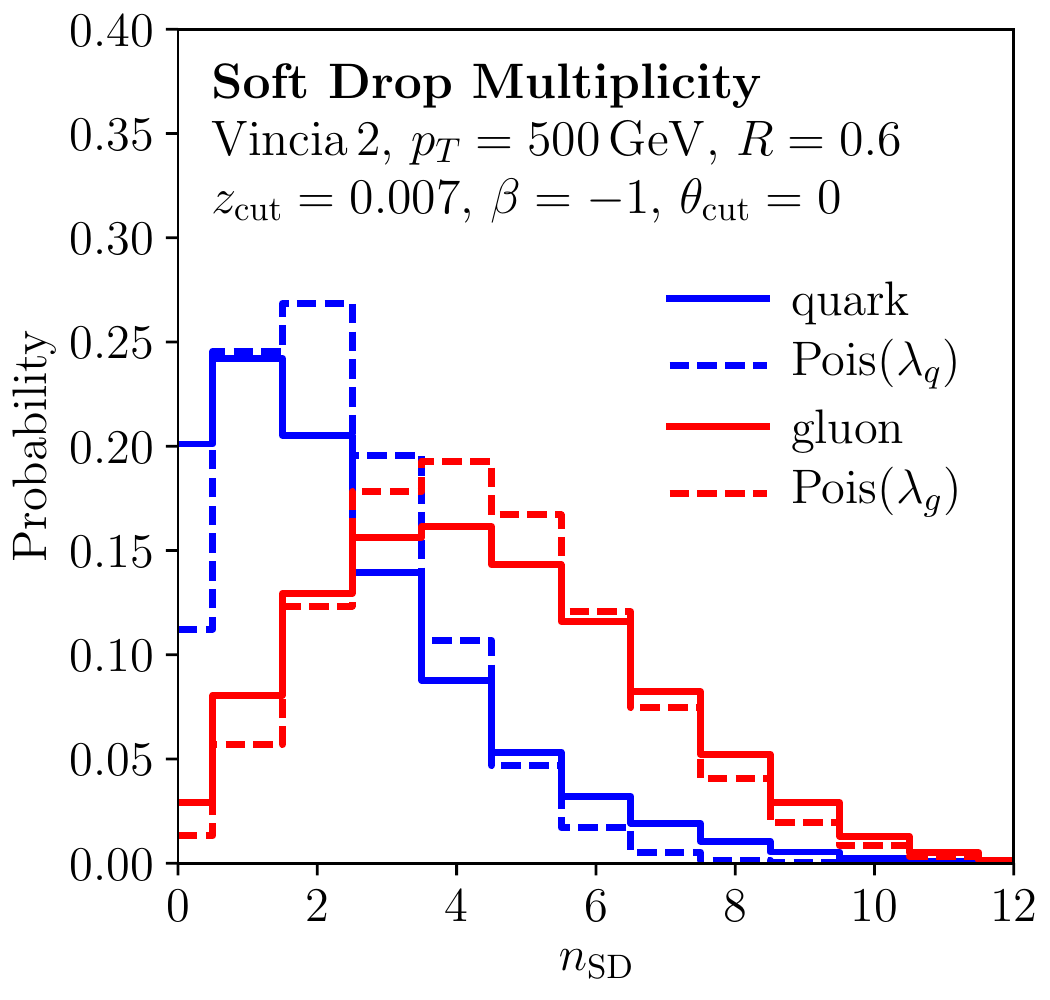}
\caption{Distribution of the soft drop multiplicity $n_\SD$ from \textsc{Vincia} 2.0.01.   Using the IRC-safe benchmark parameters in \Eq{eq:benchmark}, we find good quark/gluon discrimination power.  For comparison, we show Poisson distributions with the same means as dashed curves ($\lambda_q = 2.2$, $\lambda_g = 4.3$).}
\label{fig:sd_mul_dist}
\end{figure}

In \Fig{fig:sd_mul_dist}, we show the soft drop multiplicity distributions for quark and gluon jets as extracted from \textsc{Vincia}. Results are given using the benchmark parameters 
\begin{equation}\label{eq:benchmark}
\zcut = 0.007\,, \qquad \beta = -1\,, \qquad \thetacut = 0\,.
\end{equation}
This benchmark is chosen to maximize quark/gluon discrimination power while retaining perturbative calculability, as discussed in \Sec{sec:ll_scale}. 
The distributions are approximately Poisson and yield good quark/gluon discrimination power.

\subsection{Multiplicity Variants}
\label{subsec:variations}

While the focus of this paper is on soft drop multiplicity $n_\SD$, many other observables could be defined 
using the $(z_n, \theta_n)$ values recorded by ISD.  For example, the techniques developed in this
paper can be directly applied to the weighted soft drop multiplicity,
\begin{equation}
\label{eq:nSD_again}
n_\SD^{(\kappa)} = \sum_{n} z_n^\kappa\,.
\end{equation}
Note that soft drop multiplicity is a special case ($\kappa = 0$) of this more general observable, with the same criteria for IRC safety. 
We study the weighted multiplicity in detail in \App{sec:weighted_nSD}, but find its quark/gluon discrimination
power to be inferior to the discrete $\kappa = 0$ case. In fact, LL reasoning leads one to expect the soft drop multiplicity
$n_\SD$ to have the best discrimination power of any observable defined on the $(z_n, \theta_n)$ values; 
see the end of \Sec{sec:mean_discrim} for a short discussion.

Nevertheless, several other promising variants of soft drop multiplicity might prove useful:
\begin{itemize}
  
\item The weighted soft drop multiplicity in \Eq{eq:nSD_again} only refers to the momentum fractions $z_n$ in the
    sum over emissions. One could also consider an angle-weighted variant
    \be
    \sum_{n} z_n^{\kappa} \, \theta_n^\alpha\,,
    \ee
    or indeed any function of $z_n$ and $\theta_n$. The potential advantage of including $\theta_n$ information 
    is that even for $\thetacut = 0$, such observables would be collinear safe for $\alpha > 0$.
    
  \item Instead of counting emissions only from the trunk of the C/A tree, we could extend the sum to include
    all branchings down to the angular resolution $\thetacut$. This multiplicity variant
    would require a modification of the ISD algorithm: 
    in step $4'$, the recursion would be applied to both subjets $i$ and $j$, not just the harder one.
    This is a step in similarity towards full hadron multiplicity, reducing to it exactly
    when $\zcut = \thetacut = 0$. This variant of soft drop multiplicity is more difficult to study
    analytically, however, due to the nonlinear structure of the recursion. Moreover, it is not clear that this variant
    would provide a performance advantage over $n_\SD$. While gluons emitted from the hard core of a quark (gluon) jet give 
    rise to factors of $C_F$ ($C_A$), subsequent emissions from those gluons give rise to factors of $C_A$ regardless
    of the jet flavor; this might wash out quark/gluon discrimination power. 
  \item 
    The original soft drop algorithm uses a C/A tree to mimic the angular-ordered structure of the parton 
    shower. One could also study variants based on reclustering with the generalized-$k_t$ algorithm with 
    exponent $p$ \cite{Cacciari:2008gp,Cacciari:2011ma}.  The C/A algorithm used above corresponds to $p = 0$, while the $k_t$ algorithm uses $p = 1$. 
    For this variant, it would make sense to replace the angular cut $\thetacut$ with a cut 
    $d_{\text{cut}}$ on the generalized distance measure $d_{ij}$. 
\end{itemize}
    This last $k_t$ variant is of particular interest, given the discussion below in \Sec{sec:mean_discrim}.
    Nonperturbative physics typically dominates when $k_t \simeq \Lambda_{\rm QCD}$, so it makes sense to use a clustering algorithm
    where the clustering scale is ``parallel'' to the nonperturbative scale.  This variant of $n_\SD$ 
    would then allow the nonperturbative phase space to be clearly separated from the perturbative region and 
    avoided. This would open up as much perturbative phase space for measured emissions as possible. 
    We note that it is possible to mimic some of the LL structure of the $k_t$ variant by using ISD with $\beta = -1$, though there would be differences going to NLL order.

We defer an analysis of these variants to future work, anticipating that many of the analytic tools from this paper can be translated to these generalized contexts.  Experimentally, one might want to measure a track-based version of $n_\SD$, trading collinear safety for improved robustness to pileup, which could be studied with the help of track functions \cite{Chang:2013rca,Chang:2013iba,Elder:2017bkd}.  


\section{Leading-Logarithmic Analysis}
\label{sec:ll_scale}

At LL order, the only difference between quarks and gluons is encoded in the color factors $C_F$ and $C_A$, so Casimir scaling is a generic feature of many quark/gluon discriminants. Here, we review the case of additive observables (and close variants), where Casimir scaling of the Sudakov form factor yields a universal discrimination power at LL that depends only on $C_A/C_F$. We then show that the soft drop multiplicity is Poisson distributed, with its mean and variance satisfying Casimir scaling.

In general, any observable that is sensitive to multiple emissions at LL is ``Poisson-like'' distributed, in the sense that its variance $\sigma^2$ and mean $\mu$ both scale with the number $n$ of emissions counted, i.e.~$\sigma^2 = \mathcal{O}(\mu)$. In the limit of many emissions, all such observables converge to a normal distribution with decreasing relative width $w_{\rm rel} \sim \sigma/\mu \sim 1/\sqrt{n}$. Then as more emissions are counted, the discrimination power is not a universal function of $C_A/C_F$, but instead improves as $\mu$ increases and the quark/gluon distributions separate. 

In this section, we illustrate this behavior for soft drop multiplicity with distributions extracted from \textsc{Vincia}, using the setup described in \Sec{subsec:ISD}. 
We extract ROC (receiver operating characteristic) curves of the quark efficiency versus the gluon mistag rate, and explain their qualitative behavior.
In \App{sec:weighted_nSD}, we consider weighted soft drop multiplicity, with behavior that interpolates between that of Poisson- and Sudakov-distributed observables. 

\subsection{Review of Additive Observables}

A generic jet observable is defined on the momenta $p_i$ and quantum numbers $q_i$ of particles within a jet. 
An additive IRC-safe observable $f$ is one that reduces to the form  
\begin{equation}
\label{eq:additive_obs_def}
f\left(\{p_i, q_i\}\right) = \sum_{i \in \text{jet}} f(p_i)
\end{equation}
in the soft/collinear limit, so that the observable depends on a simple sum over the jet constituents, independent of $q_i$.\footnote{One could consider additive but IRC-unsafe observables which do depend on $q_i$.}  The function $f(p_i)$ can depend on global properties of the jet (e.g.~its $p_T$), but not on its substructure. Collinear safety implies that $f(p_i)$ is linear in the particle energies $E_i$. Examples of additive observables include the jet mass \cite{Clavelli:1979md,Catani:1991bd,Catani:1992ua}, the radial moments \cite{Gallicchio:2010dq}, and the angularities \cite{Berger:2003iw,Almeida:2008yp,Ellis:2010rwa}, among many others.

\begin{figure}
\subfloat[]{
\label{fig:emission_space}
 \includegraphics[scale=0.5]{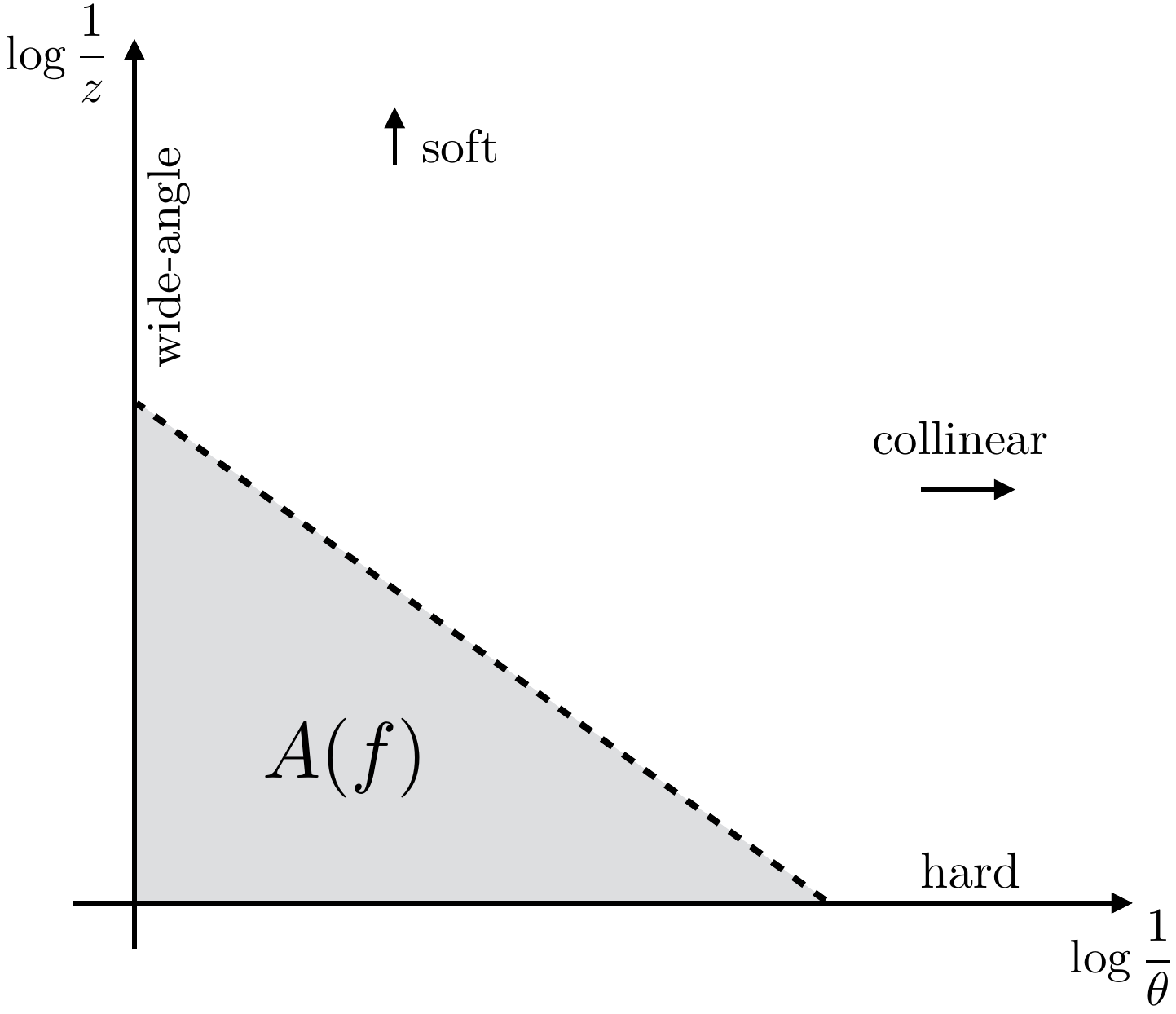}
}
\subfloat[]{
\label{fig:sd_mult_region}
 \includegraphics[scale=0.5]{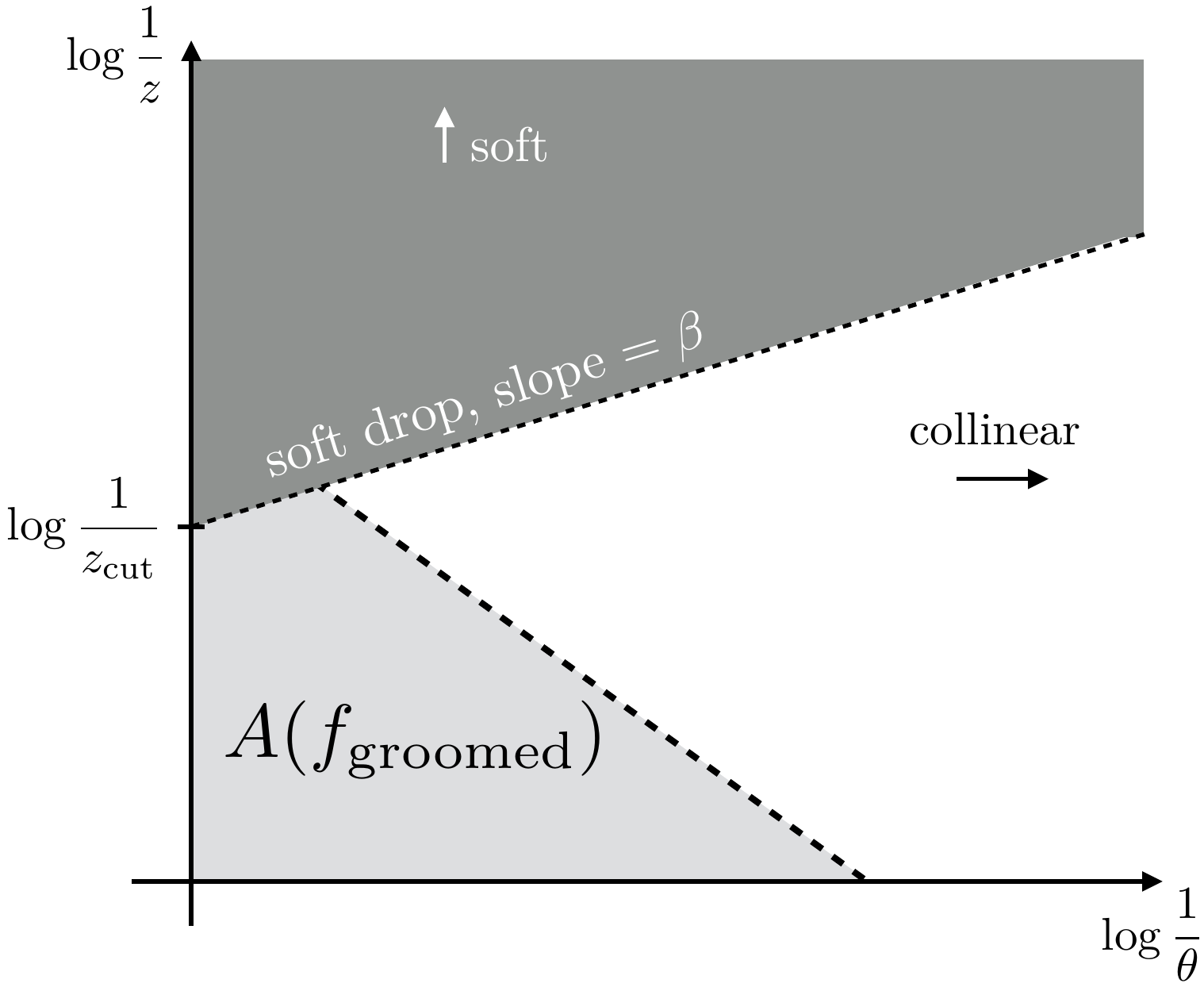}
}
\caption{Illustration of emission phase space, which is a close analog of the Lund diagram \cite{Andersson:1988gp}, where gluon emissions are uniformly distributed in the $(\log1/\theta, \log1/z)$ plane.  (a) The measurement of an additive observable $f$ imposes a Sudakov veto on the phase space area $A(f)$.  (b)  For groomed observables, the measurement of a quasi-additive observable $f_{\rm groomed}$ also imposes a Sudakov veto.
}
\label{fig:spaces}
\end{figure}

We now review the Casimir scaling of additive observables at LL order, as discussed in \Ref{Larkoski:2013eya}.\footnote{Casimir 
scaling of additive observables at LL is identical to the statement of Casimir scaling of the cusp anomalous dimension in QCD, which has a long history in QCD \cite{Gatheral:1983cz,Becher:2009qa}.  
Casimir scaling is known to hold through three loops \cite{Moch:2004pa} in the cusp anomalous dimension, but is not expected to hold exactly \cite{Frenkel:1984pz}.
At NLL and beyond, Casimir scaling is broken by the appearance of the non-cusp anomalous dimension.}  
For simplicity of the discussion below, we let $\as$ be a fixed coupling so that the expressions are more compact, but it is straightforward to include a running coupling at LL order.  At this order, we need only consider gluon emissions from the jet core that are both soft and collinear, described by the most singular terms in the splitting function. Parametrizing emissions by their angle $\theta$ and energy (or $p_T$) fraction $z$, real emissions are uniformly distributed in the $(\log1/\theta, \log1/z)$ plane.  The density in this emission phase space is
\begin{equation}
\label{eq:rhodensity}
\rho_i = \frac{2\as C_i}{\pi},
\end{equation}
where $C_i$ is the appropriate color factor, equal to $C_F = 4/3$ for quarks and $C_A = 3$ for gluons. The structure of emission phase space is shown in \Fig{fig:emission_space}.  Virtual emissions are encoded in the boundaries of the emission phase space, where $\log(1/\theta), \log(1/z) \to \infty$, such that the total emission probability at each $\as$ order is zero to maintain the normalization of the probability distribution.

Applying the strongly-ordered limit and the fact that $f(p_i)$ is linear in $E_i$, only a single dominant emission 
contributes to the observable at lowest order:
\begin{equation}
\sum_{i \in \text{jet}} f(p_i) \quad \mathop{\Longrightarrow}^\text{LL} \quad \max_{i \in \text{jet}} f(p_i).
\end{equation}
Therefore, the probability that the observable $f$ is less than some value $f_{\text{max}}$ is equal to the 
probability that there are no emissions in the region where $f(p_i) > f_{\text{max}}$. This implies a cumulative 
distribution function 
\begin{equation}
\int_0^{f_{\text{max}}} \text{d} f \, p(f) \equiv \Sigma_i(f_{\text{max}}) =  e^{-\rho_i A(f_{\text{max}})}\,,
\end{equation}
where $A(f_{\text{max}})$ is the forbidden area of emission phase space, shown in \Fig{fig:emission_space}:
\begin{equation}
\label{eq:areaveto}
A(f_{\text{max}}) = \int_{f(z,\theta) > f_{\text{max}}} \frac{\df \theta}{\theta} \frac{\df z}{z}.
\end{equation}
Note that the cumulative distributions for quarks and gluons are related by
\begin{equation}
\label{eq:sigma_casimir_scale}
\Sigma_g(f_{\text{max}}) = \Bigl[ \Sigma_q(f_{\text{max}})\Bigr]^{C_A/C_F},
\end{equation}
where $C_A/C_F = 9/4$. That is, the Sudakov form factors for $f$ are related by Casimir scaling. As a result, the 
ROC curve for quark/gluon discrimination, which simply plots $\Sigma_q(f)$ versus $\Sigma_g(f)$, takes the universal 
form of \Eq{eq:naive_casimir_scaling}.

From this logic, it is clear that the above analysis also extends to certain non-additive observables.  For example, jet observables defined on groomed jets are not additive, since the grooming procedure removes emissions that would otherwise contribute to the sum in \Eq{eq:additive_obs_def}.  But groomed observables of the quasi-additive form
\be
\label{eq:def_f_groomed}
f_{\rm groomed}\left(\{p_i, q_i\}\right) = \sum_{i \in \text{groomed jet}} f(p_i)
\ee
still exhibit Casimir scaling, since the measured value of $f_{\rm groomed}$ forbid emissions in the region $A(f_{\rm groomed})$ shown in \Fig{fig:sd_mult_region}.  More generally, Casimir scaling arises whenever the value of the measurement actively forbids emissions from some region of phase space.  This vetoed phase space region builds up a Sudakov form factor which in turn controls the discrimination power achievable at LL.

\begin{figure}[t]
\centering
\includegraphics[width=.49\textwidth]{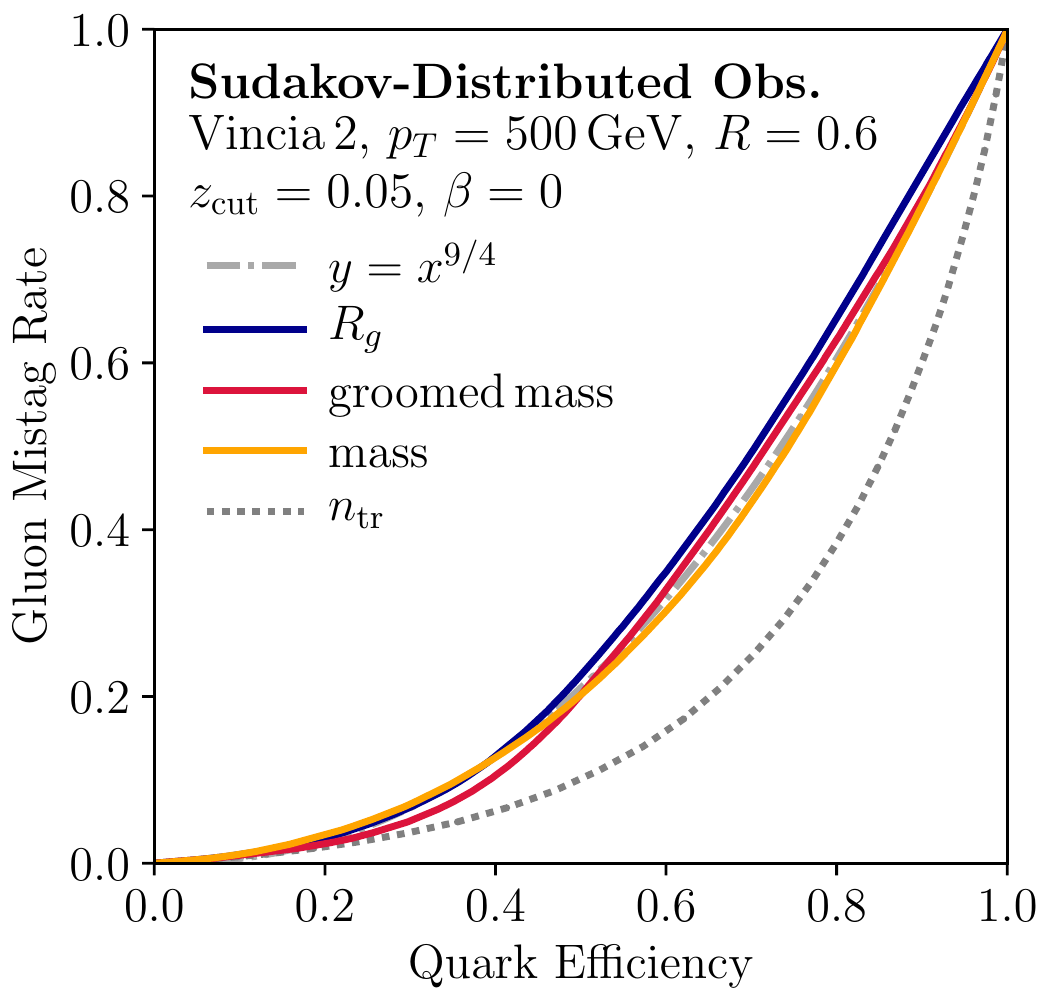}
\caption{\label{fig:casimir_examples} Comparison of the quark/gluon ROC curves for various Sudakov-distributed observables to the $y = x^{9/4}$ prediction from Casimir scaling.  Shown are the groomed jet radius, groomed jet mass, and ordinary jet mass.  As a useful benchmark, we also show the performance of track multiplicity $n_{\rm tr}$, which is known to be a very strong discriminant.}
\end{figure}

Beyond LL order, different Sudakov-distributed observables will exhibit different discrimination power due to higher-order or nonperturbative effects, but \Eq{eq:sigma_casimir_scale} is still a representative benchmark. In \Fig{fig:casimir_examples}, we show ROC curves for jet mass $m$, the soft-dropped jet mass $m_\SD$, and the groomed jet radius $R_g$, which all roughly follow the prediction from Casimir scaling. We also show track multiplicity $n_{\rm tr}$, which exhibits substantially better performance and provides a useful discrimination target.

\subsection{Soft Drop Multiplicity}
\label{sec:sd_mult_ll}

\begin{figure}
\centering
\includegraphics[scale=0.5]{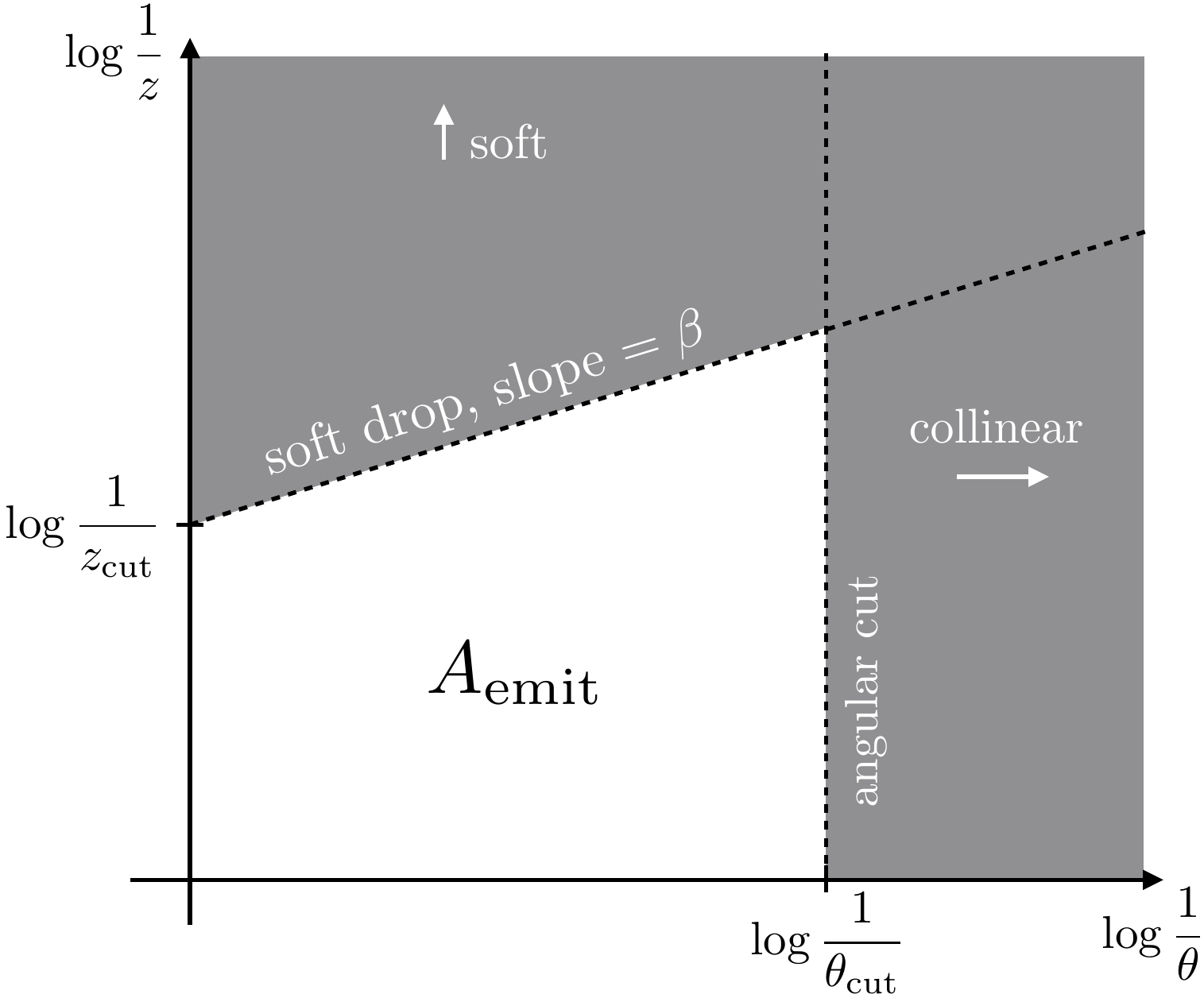}
\caption{Same as \Fig{fig:sd_mult_region}, but now highlighting the allowed emission region $A_\text{emit}$ that is counted by soft drop multiplicity.
}
\label{fig:spaces_with_Aemit}
\end{figure}

Soft drop multiplicity is not an additive observable, nor does the measured value of $n_\SD$ actively forbid emissions in any region of phase space.
As a result, $n_\SD$ does not exhibit Sudakov behavior and it instead satisfies a fundamentally different scaling relation.
Physically, this is because all emissions that pass the soft drop condition are weighted equally, so $n_\SD$ depends on multiple emissions even at leading accuracy. 
These emissions occur in the region of phase space passing the soft drop and angular cuts, shown in \Fig{fig:spaces_with_Aemit}.

Restricting to the IRC safe case with $\thetacut > 0$, the measured region has finite area in the emission plane,
\begin{equation}\label{eq:poissonmean}
A_\text{emit} = \log \frac{R_0}{\thetacut} \left( \log \frac{1}{2\zcut} + \frac{\beta}{2} \log \frac{R_0}{\thetacut} \right),
\end{equation}
and soft drop multiplicity simply counts the number of real emissions in this area. 
This expression actually holds for all $\beta\in(-\infty,\infty)$ as long as the angular cut $\thetacut$ imposes a non-trivial constraint on emissions. 
Since real emissions occur independently with uniform probability, they are described by a Poisson process, and the soft drop multiplicity is Poisson distributed at LL order:\footnote{Note that at this order, we do not account for color correlations, so the emissions are effectively Abelian.}
\begin{equation}\label{eq:sd_mult_poisson}
\mathcal{P}_i(n_\SD) = \text{Pois}(\lambda_i)[n_\SD], \qquad  \lambda_i = \rho_i A_{\text{emit}}.
\end{equation}
For reference, the Poisson distribution with mean $\lambda$ is
\begin{equation}
\text{Pois}(\lambda)[n] = \frac{\lambda^n e^{-\lambda}}{n!}.
\end{equation}
Since the variance of a Poisson distribution is also equal to $\lambda$, the means and variances of $n_\SD$ both satisfy Casimir scaling
\begin{equation}
\label{eq:mean_var_scaling}
\frac{\langle n_\SD \rangle_g}{\langle n_\SD \rangle_q} \simeq \frac{C_A}{C_F}, \qquad \frac{\text{Var}(n_\SD)_g}{\text{Var}(n_\SD)_q} \simeq \frac{C_A}{C_F},
\end{equation}
mirroring the behavior of track multiplicity in \Eq{eq:multiplicity_scaling}, but for an IRC-safe observable.

To be clear, in defining our resummation accuracy, we count large logarithms of $\zcut$ and $\thetacut$ in the mean/variance of the $n_\SD$ distribution. 
That is, we define LL and NLL exactly as for more familiar additive observables, 
with LL including all terms of the form $\as^n \log^{n+1}$ that appear in the exponent of the $n_\SD$ distribution, 
and NLL including those terms of the form $\as^n \log^n$.
With this definition, \Eq{eq:poissonmean} then shows that $n_\SD$ is indeed a double-logarithmic observable.  
In this section, we study this observable's general properties with fixed coupling, i.e.~in the double-logarithmic approximation, for purposes of illustration.  
In \Sec{sec:irc_safe}, LL and NLL results are computed using the appropriate running coupling.

\begin{figure}[t]
\subfloat[]{
\label{fig:ratio_comparison}
\includegraphics[width=.49\textwidth]{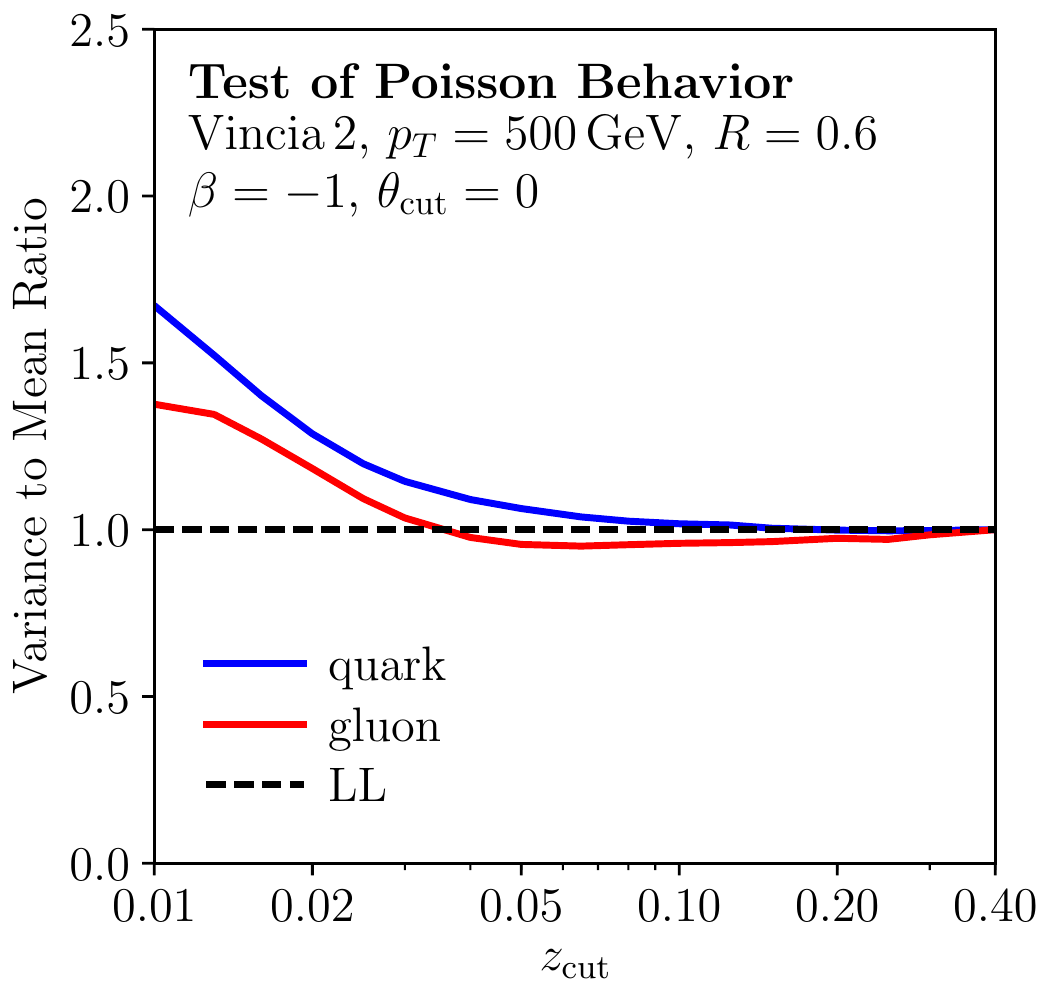} 
}
\subfloat[]{
\label{fig:ratio_comparison2}
\includegraphics[width=.49\textwidth]{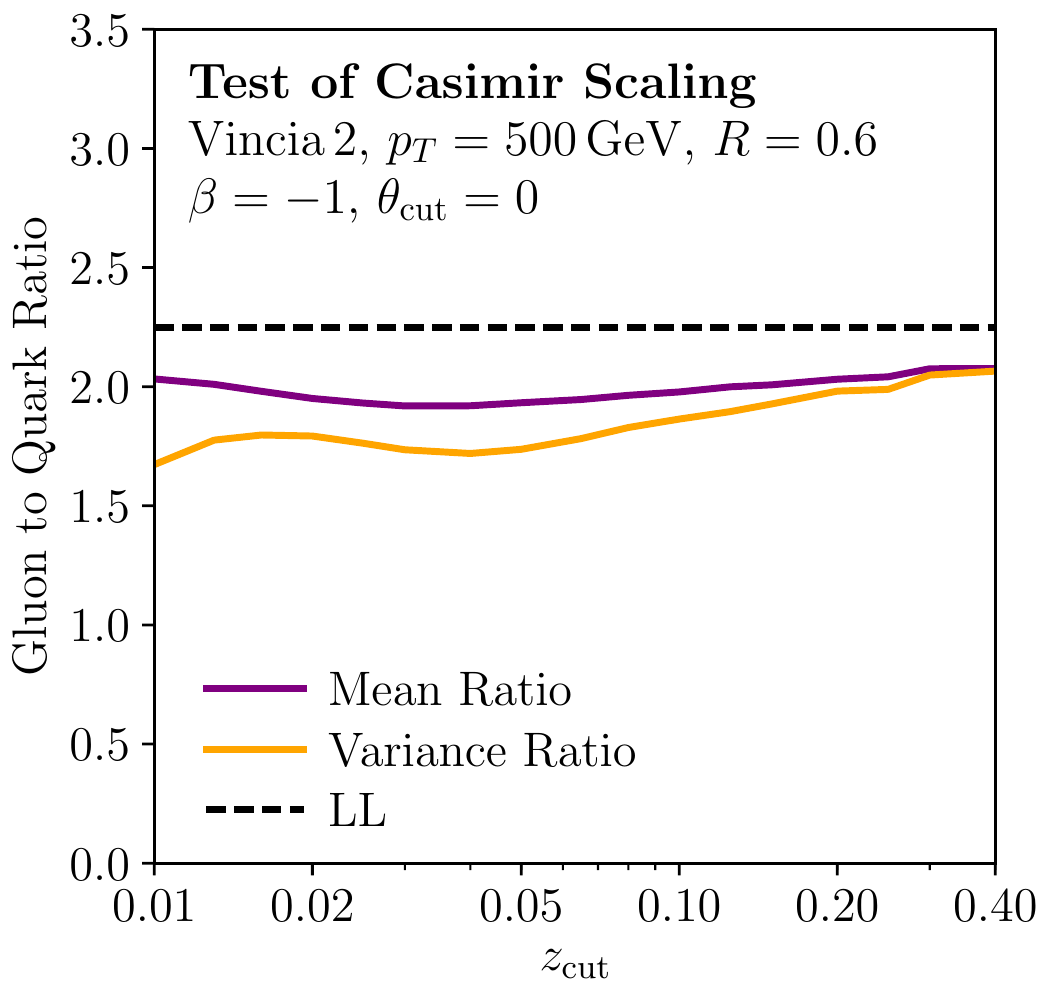} 
}
\caption{\label{fig:ll_scaling} (a) Variance to mean ratio of the soft drop multiplicity as a function of $z_{\mathrm{cut}}$.  The parameters $\beta$ and $\theta_{\mathrm{cut}}$ are set to the benchmark values in \Eq{eq:benchmark}, and the LL prediction of equal mean and variance is shown as a dashed line.  (b) Gluon to quark mean ratios and variance ratios, with the prediction of Casimir scaling shown as a dashed line.  In both cases, we see qualitative agreement between \textsc{Vincia} and the LL predictions down to $z_{\mathrm{cut}} = 0.02$.
}
\end{figure}

The above analysis provides several concrete predictions. Our most salient result is that, since the soft drop multiplicity is Poisson distributed at LL, we expect the ratio of the variance to the mean to be close to 1, as shown in \Fig{fig:ratio_comparison}. We also predict that the mean and variance satisfy the Casimir scaling relations in \Eq{eq:mean_var_scaling}, as shown in \Fig{fig:ratio_comparison2}. Though not shown here, we also checked the prediction that for $\beta = 0$, the mean soft drop multiplicity scales as
\begin{equation}
\lambda_i \propto \log \frac{1}{z_\mathrm{cut}} \log \frac{1}{\theta_{\mathrm{cut}}}.
\end{equation}
In general, we find good agreement for these predictions at large values of $z_{\mathrm{cut}}$, even out to $z_{\mathrm{cut}} \simeq 0.4$ where $\log z_{\mathrm{cut}}$ is not so large. 
For lower cut values, nonperturbative and higher-order effects cause these LL results to break down.
In \Sec{sec:mean_discrim}, we demonstrate how to choose parameters so that nonperturbative effects can be avoided,
and in \Sec{sec:NLL}, we compute the NLL corrections to the perturbative predictions discussed here.

\begin{figure}[t]
\centering
\includegraphics[width=0.5\linewidth]{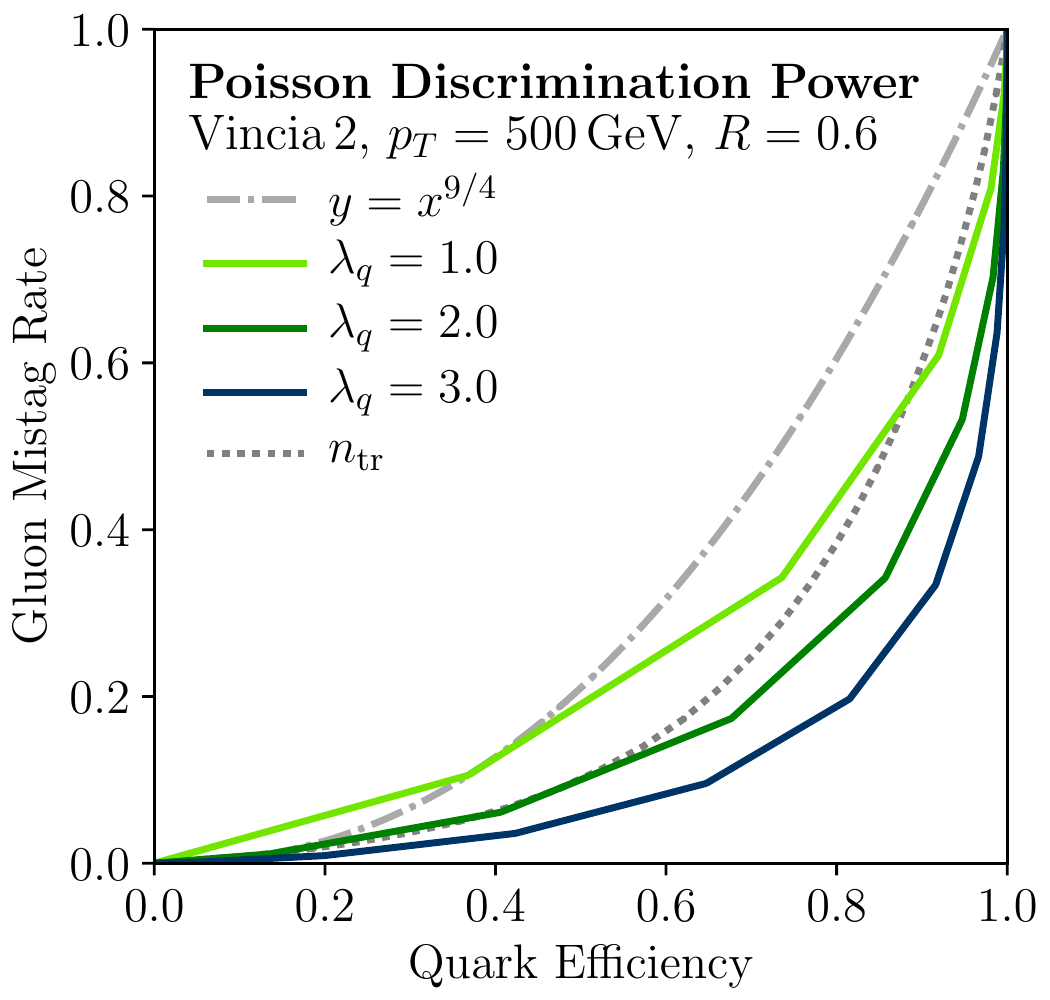}
\caption{\label{fig:poisson_casimir}
Expected quark/gluon discrimination power for Poisson-distributed observables. The mean observable value for quarks is $\lambda_q$, and we assume the mean for gluons is given by Casimir scaling $\lambda_g = (C_A/C_F) \lambda_q$. For reference, we show the $y = x^{9/4}$ curve for additive observables with Casimir scaling, as well as track multiplicity $n_{\rm tr}$ extracted from \textsc{Vincia}.  For mean quark values $\lambda_q \gtrsim 2$, a Poisson-like observable satisfying Casimir scaling would be competitive with track multiplicity.  The ROC curves are piecewise linear since the observable takes on discrete integer values.
}
\end{figure}

\subsection{Optimal Discrimination Power}
\label{sec:mean_discrim}

As a direct result of the properties exhibited in \Sec{sec:sd_mult_ll}, the discrimination power of soft drop multiplicity improves as the means $\lambda_i = \rho_i A_{\text{emit}}$ increase.  
This is because the mean of each distribution is proportional to the Casimir $C_i$, while the standard deviation is equal to the square root of the mean.
The overlap of the distributions is characterized by the relative width
\begin{equation}
\label{eq:relative_width_first}
w_{\text{rel}} \equiv \frac{\sqrt{\text{Var}(n_\SD)_i}}{\langle n_\SD \rangle_i} = \frac{1}{\sqrt{\lambda_i}}.
\end{equation}
Indeed, in the many-emission limit where the distributions are approximately Gaussian, have equal mean and variance, and satisfy Casimir scaling, 
the discrimination power is solely determined by the relative width.  As the cuts $z_{\mathrm{cut}}$ and $\theta_{\mathrm{cut}}$ are lowered, the means increase, causing the relative widths to narrow, reducing the overlap between the quark and gluon distributions, and improving the discrimination power. 

For reference, the discrimination power of Poisson distributions with different means is shown in \Fig{fig:poisson_casimir}, from which we see that track multiplicity has comparable discrimination power to a 
$\lambda_q \simeq 2$ observable.

To maximize the quark/gluon discrimination power, one should maximize the mean of the soft drop multiplicity distributions, which corresponds to taking $\zcut$ and $\thetacut$ as small as possible, for a given exponent $\beta$.  The validity of this analysis, however, is restricted to perturbation theory, so we must ensure that the values of the chosen parameters do not allow for distributions that are dominated by nonperturbative emissions. We can determine the parameters that enforce perturbative emissions by restricting the minimum relative $k_t$ appropriately.  

\begin{figure}[t]
\centering
\subfloat[]{
\label{fig:best_ps_a}
\includegraphics[width=.49\textwidth]{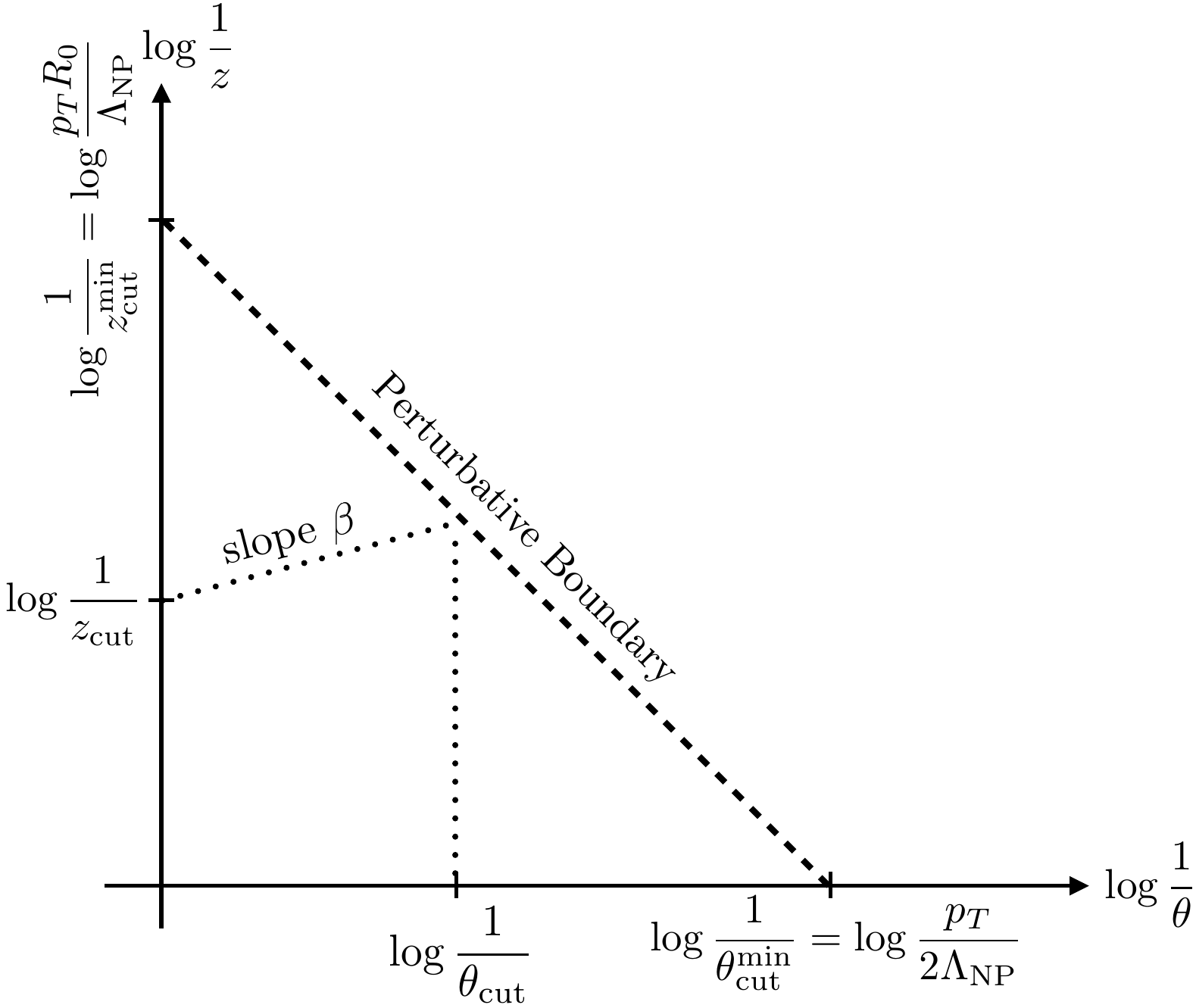}
}
\subfloat[]{
\label{fig:best_ps_b}
\includegraphics[width=.49\textwidth]{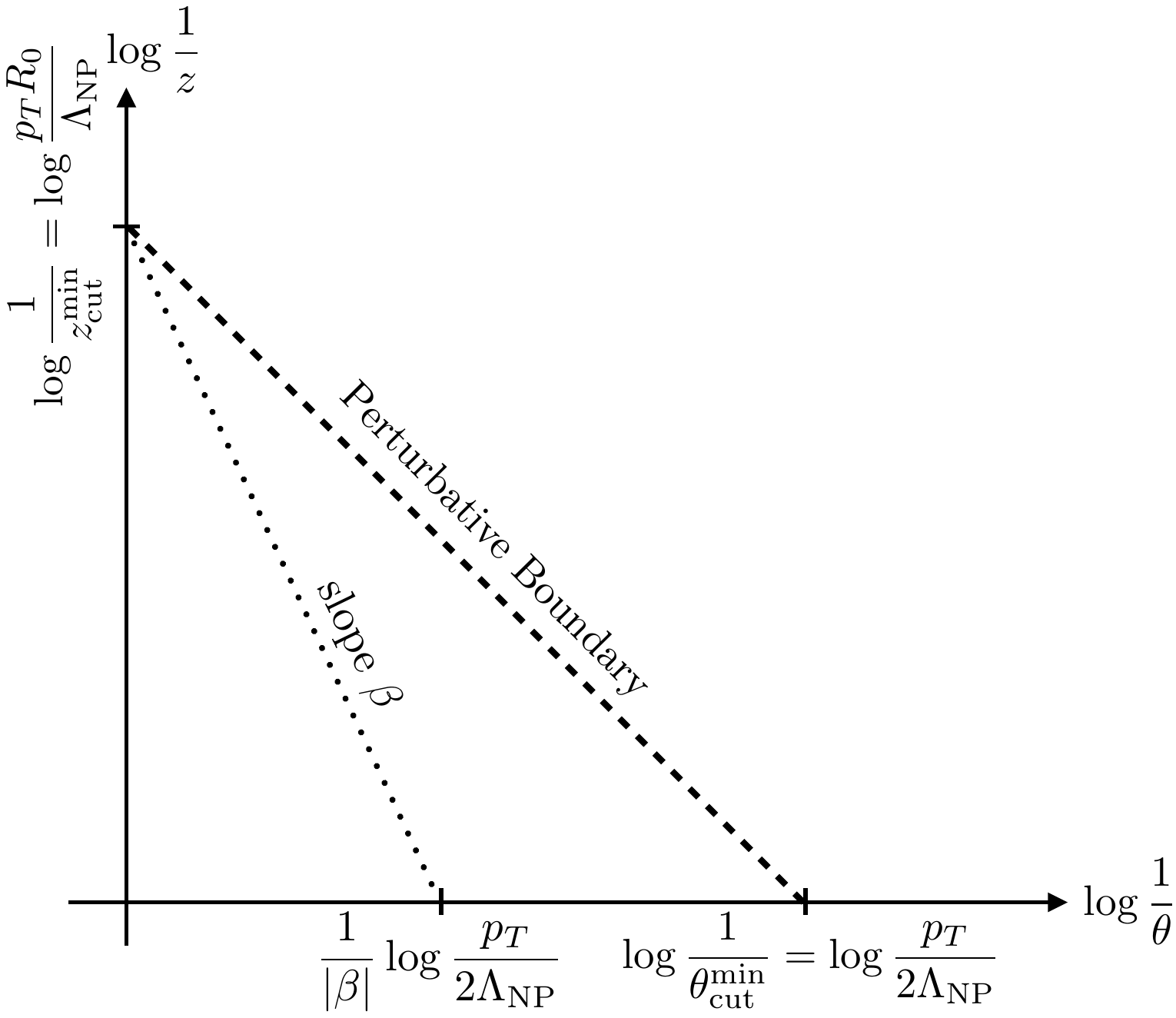}
}
\caption{Illustration of the optimal phase space configuration consistent with a perturbative analysis. The dashed line with slope $-1$ separates perturbative and nonperturbative emissions.  (a)  For $\beta > -1$, the value of $\theta_{\rm cut}$ has to be chosen to avoid allowed emissions above the nonperturbative boundary. (b) For $\beta < -1$, $\theta_{\rm cut}$ can be set to zero, with $z_{\rm cut}$ pushed to the nonperturbative boundary. To maximize the allowed perturbative phase space, one should take $\beta = -1$ and $z_{\mathrm{cut}}$ set to the  optimal value in \Eq{eq:optparams3}.
}
\label{fig:best_ps} 
\end{figure}

To enforce that an emission is perturbative, we require that the relative $k_t$ of the emission is larger than a perturbative cutoff scale $\LambdaNP$, i.e.
\begin{equation}\label{eq:np_bound1}
z\, \theta \gtrsim \frac{\LambdaNP}{p_T}\,,
\end{equation}
where $z$ and $\theta$ are the energy fraction and splitting angle of the emission, and $p_T$ is the transverse momentum of the jet. 
Below, we take $\LambdaNP = 2$ GeV unless otherwise noted.
For an emission that just passes soft drop, and therefore contributes to the soft drop multiplicity, we have
\begin{equation}
\label{eq:np_bound2}
z\gtrsim \zcut \frac{\theta^\beta}{R_0^\beta}\,.
\end{equation}
There are two regimes to consider. For $\beta > -1$ as in \Fig{fig:best_ps_a}, we can find the intersection of \Eqs{eq:np_bound1}{eq:np_bound2}. Setting $\theta\to\thetacut$, we find a restriction on $\thetacut$ to be perturbative:
\begin{equation}
\label{eq:thetacut_perturbative}
\thetacut\gtrsim \left(
\frac{\LambdaNP}{\zcut p_T R_0}
\right)^{\frac{1}{1+\beta}}R_0\,.
\end{equation}
To determine the optimal choice of $\zcut$ while enforcing perturbativity, we set $\thetacut$ to saturate this inequality and insert it into the double-log expression for the average soft drop multiplicity, \Eq{eq:poissonmean}. Maximizing this quantity, we find the optimal ISD parameters to be
\begin{align}\label{eq:optparams}
 \quad \left.\zcut\right|_\text{optimal} &=\frac{1}{2} \left(\frac{2\LambdaNP}{p_T R_0}\right)^{\frac{1}{2+\beta}}\,,\\
\quad \left.\thetacut\right|_\text{optimal} &=\left(
\frac{2\LambdaNP}{p_T R_0}
\right)^{\frac{1}{2+\beta}}R_0\,.
\label{eq:optparams2}
\end{align}
The factors of two arise because the energy fraction of the softer emission is (by definition) less than $1/2$.  
Inserting these results into the expression for the average soft drop multiplicity, we find the largest perturbative value for the mean soft drop multiplicity to be
\begin{equation}
\label{eq:zcutoptimal}
\quad \langle n_\text{SD}\rangle^{\beta > -1}_\text{optimal}\simeq \frac{\as}{\pi}\frac{C_i}{2+\beta}\log^2\left(
\frac{2\LambdaNP}{p_T R_0}
\right)\,.
\end{equation}

For $\beta < -1$, one can see from the $(\log1/\theta,\log1/z)$ phase space in \Fig{fig:best_ps_b} that an angular cutoff is not needed to avoid the 
nonperturbative region, so we can set $\thetacut=0$.  In this case, $\zcut$ saturates the bound \Eq{eq:np_bound1} for $\theta \to R_0$, yielding
\begin{equation}
\label{eq:optparams3}
\left.\zcut\right|_\text{optimal} = \frac{\LambdaNP}{p_T R_0},
\end{equation}
and the average soft drop multiplicity is
\begin{equation}
\langle n_\text{SD}\rangle^{\beta<-1}_\text{optimal}\simeq \frac{\as}{\pi}\frac{C_i}{|\beta|}\log^2\left(
\frac{2\LambdaNP}{p_T R_0}
\right)\,.
\end{equation}
Combining these regions for all $\beta\in(-\infty,\infty)$, the maximum attainable mean soft drop multiplicity with perturbative parameters is  
\begin{equation}\label{eq:opt_sdmult}
\langle n_\text{SD}\rangle_\text{optimal}\simeq \frac{\as C_i}{\pi}\min\left[\frac{1}{|\beta|},\frac{1}{|2+\beta|}\right]\log^2\left(
\frac{2\LambdaNP}{p_T R_0}
\right)\,.
\end{equation}
In particular, the mean is maximized for $\beta = -1$, giving the optimal perturbative discrimination power in this double-log approximation.  
This result can be understood directly from \Fig{fig:best_ps}, which shows that soft drop multiplicity with $\beta = -1$ can capture all of the perturbative emissions in phase space.

\begin{figure}[t]
\centering
\subfloat[]{\label{fig:sd_zcut_sweep:a} \includegraphics[width=0.49\textwidth]{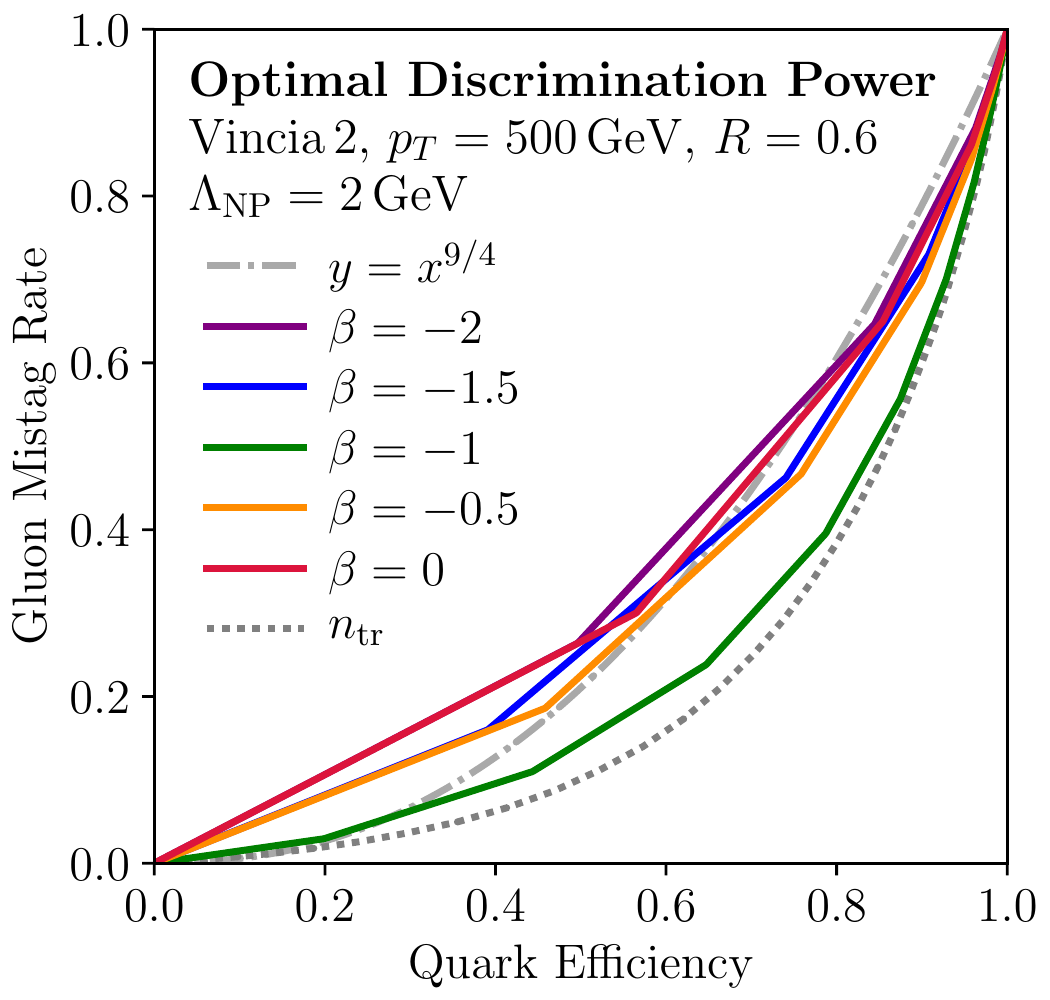}}
\subfloat[]{\label{fig:sd_zcut_sweep:b} \includegraphics[width=0.49\textwidth]{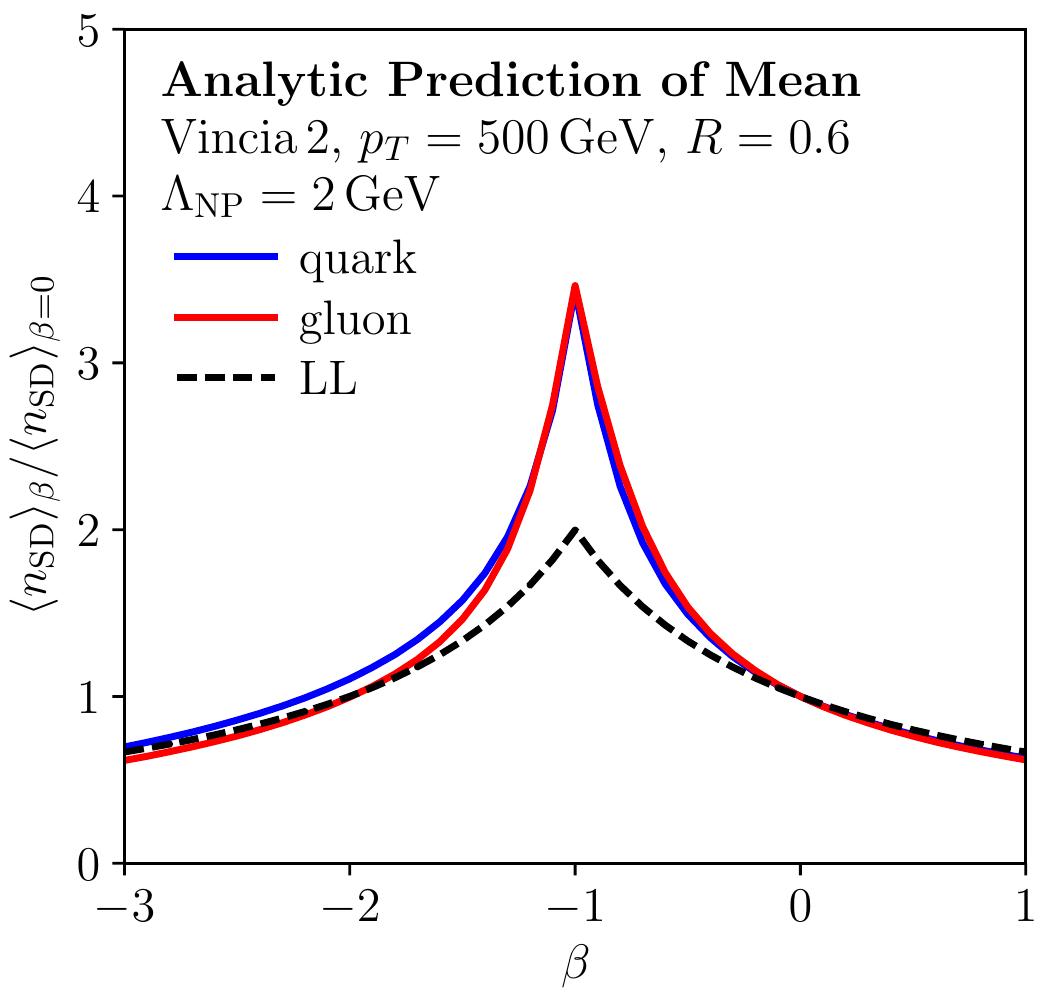}}
\caption{(a) Discrimination power of soft drop multiplicity as a function of $\beta$, with the optimal (perturbative) values of $\zcut$ and $\thetacut$ computed from \Eqss{eq:optparams}{eq:optparams2}{eq:optparams3} using $\Lambda_\text{NP} = 2$ GeV.
(b) Ratio of mean $n_\SD$ as a function of $\beta$ to mean $n_\SD$ at $\beta=0$.   The \textsc{Vincia} results 
for quarks and gluons agree with the double log prediction from \Eq{eq:opt_sdmult}, except near  $\beta = -1$ where nonperturbative effects become important.}
\label{fig:sd_zcut_sweep} 
\end{figure}

We can directly test this double-log prediction in parton shower generators.  
In \Fig{fig:sd_zcut_sweep:a}, we show the quark/gluon ROC curve for soft drop multiplicity with the optimal perturbative soft drop parameters, sweeping through $\beta$.  The best discrimination power found in \textsc{Vincia} is indeed observed near $\beta = -1$.
For a more quantitative test, \Eq{eq:opt_sdmult} predicts that the ratio of the optimal soft drop multiplicity for a given value of $\beta$ to the optimal soft drop multiplicity at $\beta = 0$ is 
\begin{equation}
\frac{\langle n_\text{SD}\rangle_\text{optimal}}{\langle n_\text{SD}\rangle^{\beta = 0}_\text{optimal}} = \min\left[\frac{2}{|\beta|},\frac{2}{|2+\beta|}\right]\,.
\end{equation}
In \Fig{fig:sd_zcut_sweep:b}, we compare this ratio to distributions extracted from \textsc{Vincia} and find good agreement away from $\beta = -1$.  
Note that when $\beta=-1$, the counted and nonperturbative regions share a boundary, while in all other cases the two regions only meet at a single point. 
This explains why nonperturbative sensitivity should be amplified when $\beta$ nears $-1$. This extra sensitivity could of course be mitigated by using a more conservative value of $\Lambda_\text{NP}$,
but there is a tradeoff between reducing nonperturbative effects and increasing discrimination power.

\begin{figure}[t]
\centering
\subfloat[]{\label{fig:sd_lambda_sweep:a} \includegraphics[width=0.49\textwidth]{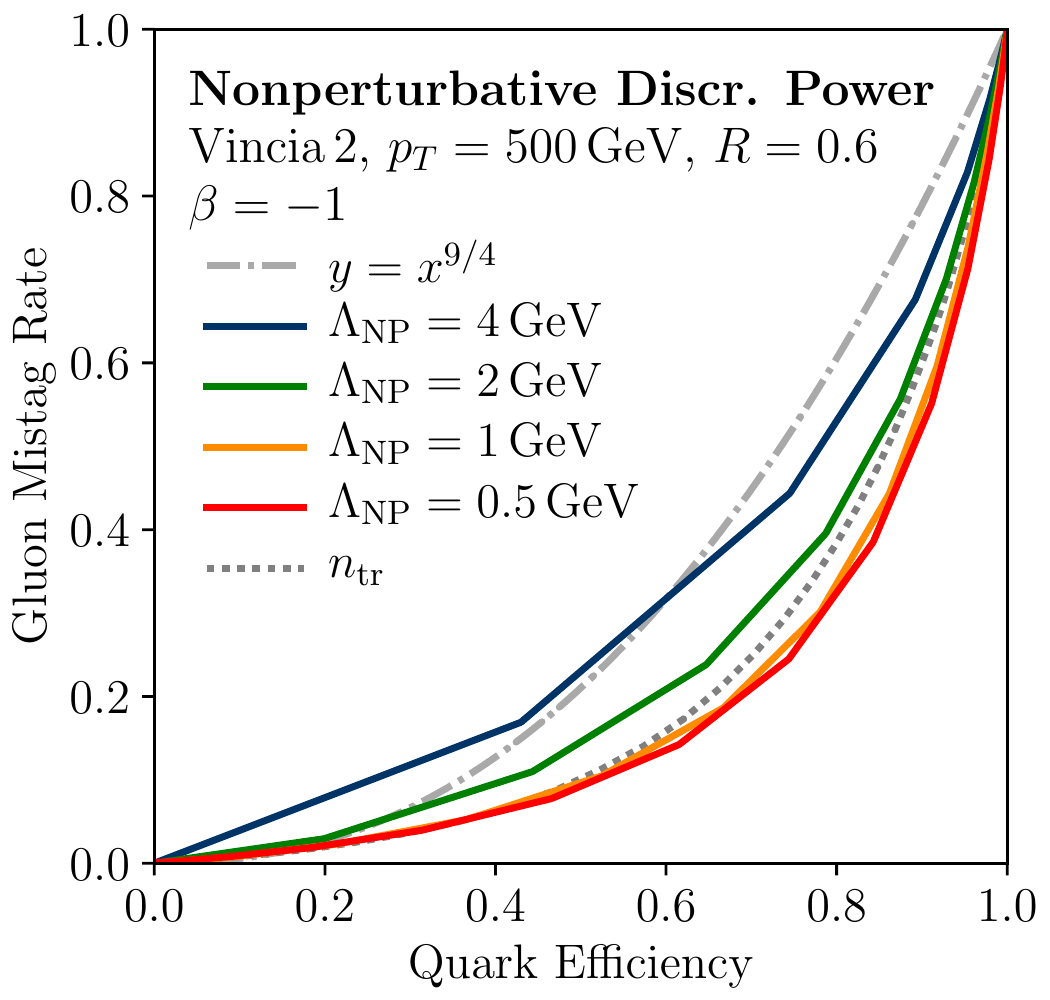}}
\subfloat[]{\label{fig:sd_lambda_sweep:b} \includegraphics[width=0.49\textwidth]{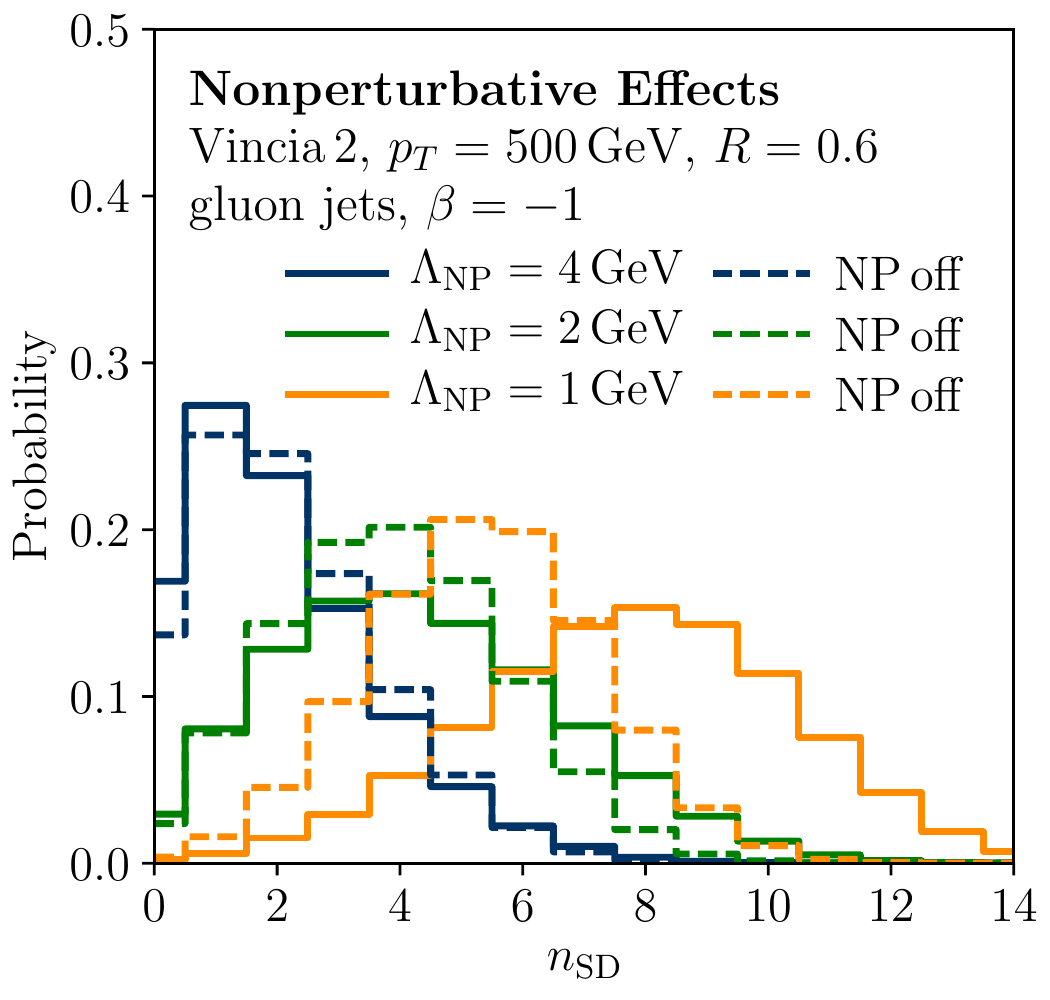}}
\caption{(a) Discrimination power of soft drop multiplicity as a function of $\LambdaNP$ with $\beta=-1$, $\thetacut=0$, and $\zcut$ computed from \Eq{eq:optparams3}.
(b) Impact of hadronization and underlying event in \textsc{Vincia} on gluon distributions.
}
\label{fig:sd_lambda_sweep} 
\end{figure}

In \Fig{fig:sd_lambda_sweep:a}, we show the effect that decreasing $\LambdaNP$ (and thus decreasing $\zcut$ and $\thetacut$) has
on the discrimination power, holding $\beta=-1$ fixed. Note that $n_\SD$ rivals $n_\text{tr}$ for $\LambdaNP = 1$ GeV, but that there is no gain in performance when $\LambdaNP$ is taken smaller.
In \Fig{fig:sd_lambda_sweep:b} we show the shift in gluon $n_\SD$ distributions from switched off hadronization and underlying event in \textsc{Vincia}. 
We take this as an indicator of nonperturbative sensitivity in the distributions. One can see that perturbative control is lost for $\LambdaNP < 2$ GeV.   For $p_T = 500$ GeV,  $\LambdaNP = 2$ GeV gives the benchmark parameters in \Eq{eq:benchmark}. 

Our perturbative analysis here was restricted to LL order and fixed coupling, and the inclusion of higher-order effects will affect the discrimination power of soft drop multiplicity. In particular, at NLL order, quark and gluon jet flavors can mix, so we expect that higher-order effects in general decrease the discrimination power from the LL prediction. We perform NLL calculations and compare our results to parton showers in \Sec{sec:irc_safe}. Beyond these higher-order effects, we have restricted the analysis to perturbative parameters. Allowing nonperturbative emissions to contribute to the soft drop multiplicity should improve the discrimination power, however, at the expense of loss of predictivity. We discuss in \Sec{sec:irc_unsafe} how to restore some of this predictive power in the nonperturbative regime with GFFs. 

One might wonder if the discrimination power could be further improved by weighting the emissions, e.g.\ by their energy, as in the weighted soft drop multiplicity of \Eq{eq:nSD_again}. 
At LL order, however, the soft drop multiplicity is provably the most powerful discriminant that can be defined on the $(z_n,\theta_n)$ values.\footnote{We 
thank Ben Nachman for discussions on this point.  Specifically, he demonstrated that the quark/gluon likelihood ratio is a monotonic function of $n_\SD$, with no other non-trivial $(z_n,\theta_n)$ dependence, 
thus providing further confirmation that $n_\SD$ is the optimal discriminant one can construct.} 
To see this, note that the normalized distribution of emissions in the $(\log 1/\theta, \log 1/z)$ plane is identical for quark and gluon jets at LL order, even including running coupling effects. 
Therefore, once the value of $n_\SD$ is known for a given jet, no additional discriminatory information can be gleaned from the $(z_n,\theta_n)$ values.  
Nevertheless, weighted soft drop multiplicity
provides an example of a more general observable that can be effectively studied with our analytic tools; we demonstrate this in \App{sec:weighted_nSD}.\footnote{One 
might be attracted to weighted soft drop multiplicity because it reduces sensitivity to soft emissions.
Presumably, the value of $\LambdaNP$ could be reduced somewhat without introducing significant nonperturbative effects. 
One cannot increase perturbative discrimination power in this way, however, since any gain in discrimination power from reducing $\LambdaNP$ must necessarily come with comparable nonperturbative sensitivity.} 


\section{Calculations for IRC-Safe Soft Drop Multiplicity}
\label{sec:irc_safe}

We now demonstrate that the LL predictions of the previous section can be reproduced by a set of perturbative evolution equations.  These equations describes how soft drop multiplicity evolves with decreasing $\thetacut$, similar to traditional parton evolution \cite{Altarelli:1977zs}.  This approach also admits a generalization to NLL, which we use to make precise predictions for comparison to parton showers.

When talking about the resummation of large logarithms at LL and NLL accuracy, we are specifically referring to factors of $\log \zcut$ and $\log \thetacut$, not to any logarithms associated with the $n_\SD$ observable (which is an integer).
As we already saw in \Eq{eq:poissonmean}, these logarithms control the size of the emission phase space, which in turn control the expected mean value of  $n_\SD$, so their resummation is essential for predicting the distribution of $n_\SD$.

\subsection{Leading-Logarithmic Evolution Equations}
\label{subsec:LLevol}

We begin by analyzing the soft drop multiplicity to LL accuracy.
This case is simple enough to keep the structure of the $\thetacut$ evolution transparent; the generalization to NLL just requires keeping track of more details.
To achieve LL accuracy, we need only consider soft-collinear gluons emitted from the hard core of a jet; flavor-changing effects are not present at this order.
Furthermore, the trunk of the clustering tree retains all but an $\mathcal O(\zcut)$ fraction of the original jet's energy, so for $\zcut \ll 1$, energy losses are negligible at this order as well.

Let $p_n^i(\thetacut)$ denote the probability that, given a jet of flavor $i$ and ISD parameter $\thetacut$, its soft drop multiplicity $n_\SD(\thetacut)$ is measured to be $n$.  Here, we leave the dependence on $\zcut$ and $\beta$ implicit, since they do not participate directly in the evolution equations.
Since $n_\SD$ is a discrete counting observable, $p_n^i(\thetacut)$ is finite and should satisfy the normalization condition 
$\sum_{n=0}^\infty p_n^i(\thetacut) = 1$ for each flavor $i$.

We can compute the distribution for $p_n^i(\thetacut)$ by solving a set of evolution equations.   Consider decreasing the resolution angle from $\thetacut$ to $\thetacut-\delta\thetacut$. 
The value of $n_\SD$ will increase by one if there is an emission in the interval
$[\thetacut-\delta\thetacut, \thetacut]$ that passes soft drop; otherwise $n_\SD$ will remain unchanged. That is, 
\begin{align}
  p_n^i(\thetacut-\delta\thetacut) &= p_{n-1}^i(\thetacut) \, \frac{\delta\thetacut}{\thetacut}\int_0^{1/2} \df z \, {\as(z\,\thetacut\,p_T) \over \pi} \, P_{i \to i}(z) \, \Theta_\SD(z,\thetacut) \nonumber\\
  &\hspace{5mm} + p_{n}^i(\thetacut) \left( 1 - \frac{\delta\thetacut}{\thetacut}\int_0^{1/2} \df z \, {\as(z\,\thetacut\,p_T) \over \pi} \, P_{i \to i}(z) \, \Theta_\SD(z,\thetacut) \right) .
\label{eq:LLevoleqn}
\end{align}
Here, $P_{i \to i}(z)$ is the splitting function for the hard parton $i$ to emit a collinear gluon of energy fraction $z$ (and remain as flavor $i$),
and $\Theta_\SD(z,\theta)$ imposes the soft drop condition,
\begin{equation}
  \Theta_\SD(z,\theta) \equiv \Theta\!\left(z - \zcut \, {\theta^\beta \over R_0^\beta}\right).
\end{equation}
At LL, $\as(z\,\thetacut\,p_T)$ runs with the 1-loop $\beta$ function.

Using \Eq{eq:LLevoleqn}, we can derive the linear first-order differential equation in $\thetacut$,
\begin{equation}
  {\df p_n^i \over \df \thetacut} = {p_n^i(\thetacut) - p_{n-1}^i(\thetacut) \over \thetacut} \int_0^{1/2} \df z \, {\as(z\,\thetacut\,p_T) \over \pi} 
  \, P_{i \to i}(z) \, \Theta_\SD(z,\thetacut)\,.
\label{eq:LLdiffeq}
\end{equation}
Because no emissions are recorded outside the jet radius $R_0$, there is a boundary condition $p_n^i(R_0) = \delta_{n,0}$.  With this boundary condition, the solution to \Eq{eq:LLdiffeq} is
\begin{align}
  p_0^i(\thetacut) &= e^{-I_{i \to i}(\thetacut,R_0)}\,, \label{eq:LLsoln1}
\\
  p_{n \geq 1}^i(\thetacut) &= 
  \int_\thetacut^{R_0} {\df \theta \over \theta} ~ e^{-I_{i \to i}(\thetacut,\theta)} 
  \left(\int_0^{1/2} \df z \, {\as(z\,\theta\,p_T) \over \pi} \, P_{i \to i}(z) \, \Theta_\SD(z,\theta) \right) \, p_{n-1}^i(\theta)\,,
\label{eq:LLsoln2}
\end{align} 
where
\begin{equation}
  I_{i \to i}(\theta_1, \theta_2) = \int_{\theta_1}^{\theta_2} {\df \theta \over \theta} \int_0^{1/2} \df z \,{\as(z\,\theta\,p_T) \over \pi} ~ P_{i \to i}(z) \, \Theta_\SD(z,\theta)\,.
\end{equation}
The expression in \Eq{eq:LLsoln1} corresponds to the case of no emissions between $R_0$ and $\thetacut$.  The expression in \Eq{eq:LLsoln2} computes the probability that ISD records $n-1$ emissions in the interval $[\theta,R_0]$, one final emission at $\theta$, then zero emissions in the interval $[\thetacut,\theta]$,
with an integral over the angle $\theta$ where the final counted emission occurs.

We can interpret \Eq{eq:LLsoln2} as a recursion relation in $n$ with \Eq{eq:LLsoln1} as the initial condition.  The first step in the recursion ($n=1$) gives
\begin{align}
  p_1^i(\thetacut) &= \int_\thetacut^{R_0} {\df \theta \over \theta} \, e^{-I_{i \to i}(\thetacut,\theta)} \, 
  \left( \int_0^{1/2} \df z  \, {\as(z\,\theta\,p_T) \over \pi}~P_{i \to i}(z) \, 
  \Theta_\SD(z,\theta) \right) \, e^{-I_{i \to i}(\theta,R_0)} \nonumber\\[5pt]
  &= e^{-I_{i \to i}(\thetacut,R_0)} I_{i \to i}(\thetacut,R_0)\,.
\label{eq:LLfindPoisson}
\end{align}
A similar simplification occurs for each value of $n$, and we recognize the Poisson distribution
we found in \Eq{eq:sd_mult_poisson}:
\begin{equation}
  p_n^i(\thetacut) = {1 \over n!} \, \Big[I_{i \to i}(\thetacut,R_0)\Big]^n\,e^{-I_{i \to i}(\thetacut,R_0)} \,.
\label{eq:Poisson}
\end{equation}
At LL, the soft drop multiplicity $n_\SD$ is thus Poisson distributed with 
mean $\lambda_i = I_{i \to i}(\thetacut,R_0)$.  With fixed coupling, the mean value agrees exactly with $\lambda_i = \rho_i A_{\text{emit}}$ found before (see \Eqs{eq:rhodensity}{eq:poissonmean}):
\begin{equation}
\left.I_{i \to i}(\thetacut,R_0)\right|_{\text{fixed }\as} = \frac{2 \as C_i}{\pi} \log\frac{R_0}{\thetacut}\left(
\log \frac{1}{2\zcut} + \frac{\beta}{2} \log \frac{R_0}{\thetacut}
\right).
\end{equation}

\subsection{Next-to-Leading-Logarithmic Corrections}
\label{sec:NLL}

The next-to-leading logarithms take the form $\as^n\,\log^n\zcut$ and $\as^n\,\log^n\thetacut$ in the logarithm of $p_n^i(\thetacut)$. 
To resum these, we must consider emitted partons that are not necessarily soft and that can be either quarks or gluons. 
This requires us to take energy losses and flavor changes into account at this accuracy. 
It is convenient to compute $p_n^i(\thetacut)$ by expressing it as
\begin{equation}
  p_n^i(\thetacut) = \sum_{j = q,g} \int_{1/2^n}^1 \df Z \, p_n^{i \to j(Z)}(\thetacut)\,.
\end{equation}
Here, $\df Z\,p_n^{i \to j(Z)}(\thetacut)$ is the differential probability that, given a jet of flavor $i$,
ISD counts $n$ emissions from its hard core that result in a flavor change from $i$ to $j$, and a remaining energy fraction
in the interval $[Z,Z+\df Z]$.\footnote{
The probability for the hard core to be left with energy fraction between 
$Z/(1-z)$ and $(Z+\df Z)/(1-z)$ is then $p_n^{i \to j[Z/(1-z)]}(\thetacut) \, \df Z/(1-z)$.
This is used in \Eq{eq:NLLevol}.
}  
These more differential distributions evolve with $\thetacut$ as
\begin{align}
  & p_n^{i \to j(Z)}(\thetacut-\delta\thetacut) \nonumber\\
  & \hspace{5mm} = 
  p_n^{i \to j(Z)}(\thetacut) \, \left(1 - {\delta\thetacut \over \thetacut} \int_0^{1/2} \df z \, {\as(z\,\thetacut\, Z \, p_T) \over \pi} \, P_{j \to \text{any}}(z) \, \Theta_\SD(z,\thetacut) \right) \nonumber\\
  & \hspace{5mm} + 
  \sum_{k} \, {\delta\thetacut \over \thetacut} \, \int_0^{1/2} \df z \, {\as\big(z\,\theta\,{Z \over 1-z} \,p_T\big) \over \pi} \, P_{k \to j}(z) \, \Theta_\SD(z,\theta)  \, 
  p_{n-1}^{i \to k[Z/(1-z)]}(\thetacut) ~ {1 \over 1-z} \, ,
\label{eq:NLLevol}
\end{align}
where
\begin{equation}
  P_{i \to \text{any}}(z) = \sum_{j} P_{i \to j}(z)\,.
\end{equation}
The middle line of \Eq{eq:NLLevol} is the probability that $n$ emissions are counted at resolution $\thetacut$,
and that only virtual or soft-dropped emissions (neither of which have an impact on energy fractions, up to $\zcut$ corrections) 
occur in the interval $[\thetacut-\delta\thetacut, \thetacut]$.
The second line is the probability that $n-1$ emissions are counted at resolution $\thetacut$ and result in a flavor conversion $i \to k$,
and that an additional counted emission causing further conversion $k \to j$ occurs in $[\thetacut-\delta\thetacut,\thetacut]$.

We now justify that these evolution equations do indeed resum large logarithms to NLL, with one caveat.
As is necessary for NLL resummation, these evolution equations contain NLO information about the jet's substructure.
To achieve NLL accuracy, we need to properly include the following double-emissions structures:  collinear plus collinear (C+C), soft plus collinear (S+C), soft plus soft (S+S), and hard plus soft-collinear (H+SC).
Since ISD is an angular-ordered algorithm, collinear emissions factorize in the cross section, so our evolution equations correctly include C+C and S+C double emissions.
The S+S case is included as well by letting $\as$ run with the 2-loop $\beta$ function in the CMW scheme \cite{Catani:1990rr}.
The one caveat is that we do not describe H+SC double emissions correctly at NLO, since we use splitting functions instead of full matrix elements.\footnote{Besides 
this caveat, though, note that our use of $1 \to 2$ (as opposed to $1 \to 3$) splitting functions is sufficient at NLL, since $n_\SD$ is a double-logarithmic observable.}
Thus, our approximation should become more accurate as the jet radius $R_0$ becomes smaller, forcing hard emissions in the jet to become collinear.  
We also ignore the effects of logarithms of $\zcut$ that arise from nonglobal radiation~\cite{Dasgupta:2001sh}, and so do not describe emissions in the jet from secondary radiation from outside of the jet.

Despite the extra complications at NLL order, \Eq{eq:NLLevol} is still a linear first-order differential equation, just as in \Sec{subsec:LLevol}. The solution is  
\begin{align}
  p_0^{i \to j(Z)}(\thetacut) &= \delta_{ji} \, \delta(Z-1) \, 
  \exp\left[-I_{i \to \text{any}}(\thetacut,R_0)\right]\,,\\[5pt]
  p_{n \ge 1}^{i \to j(Z)}(\thetacut) &= \sum_{k} \, \int_\thetacut^{R_0}{\df \theta \over \theta} \int_0^{1/2} \df z \, 
  \exp\left[-I_{j(Z) \to \text{any}}(\thetacut,\theta)\right] \nonumber\\[5pt]
  & \hspace{5mm} \times {\as\big(z\,\theta\,{Z \over 1-z} \, p_T\big) \over \pi} \,
  P_{k \to j}(z) \, \Theta_\SD(z,\theta) \, p_{n-1}^{i \to k[Z/(1-z)]}(\theta)~{1 \over 1-z},
\label{eq:NLLsoln}
\end{align}
where
\begin{equation}
  I_{j(Z) \to \text{any}}(\theta_1,\theta_2) = \int_{\theta_1}^{\theta_2} {\df \theta \over \theta} \int_0^{1/2} \df z \,
  {\as(z\,\theta \, Z\,p_T) \over \pi}~P_{j \to \text{any}}(z) \, \Theta_\SD(z,\theta)\,.
\end{equation}
Note that $p_n^{i \to j(Z)}$ vanishes for $Z < 1/2^n$.
The same manipulations that led to \Eq{eq:LLfindPoisson} and the Poisson distribution at LL do not go through at NLL, so we cannot write $p_n^{i \to j(Z)}$ or $p_n^i$ in closed form at this order.
Nonetheless, the integrals in
\Eq{eq:NLLsoln} can be performed numerically by first computing $p_1^{i \to j(Z)}(\thetacut)$, then computing 
$p_2^{i \to j(Z)}(\thetacut)$, and so on until $p_n^j$ is negligible.  In practice, the probability saturates for $n$ of order 10.

\begin{figure}[t]
\centering
\subfloat[]{  \includegraphics[width=0.45\linewidth]{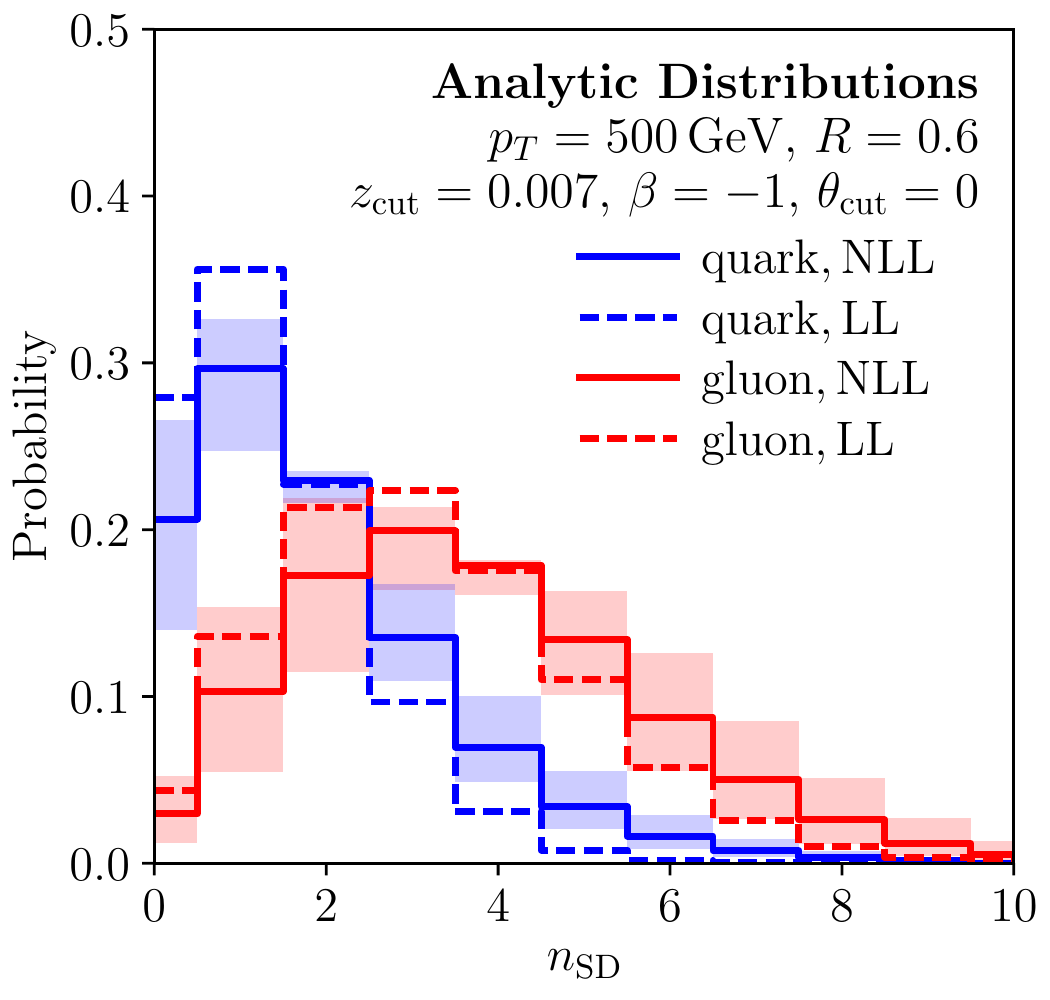}}
\subfloat[]{  \includegraphics[width=0.45\linewidth]{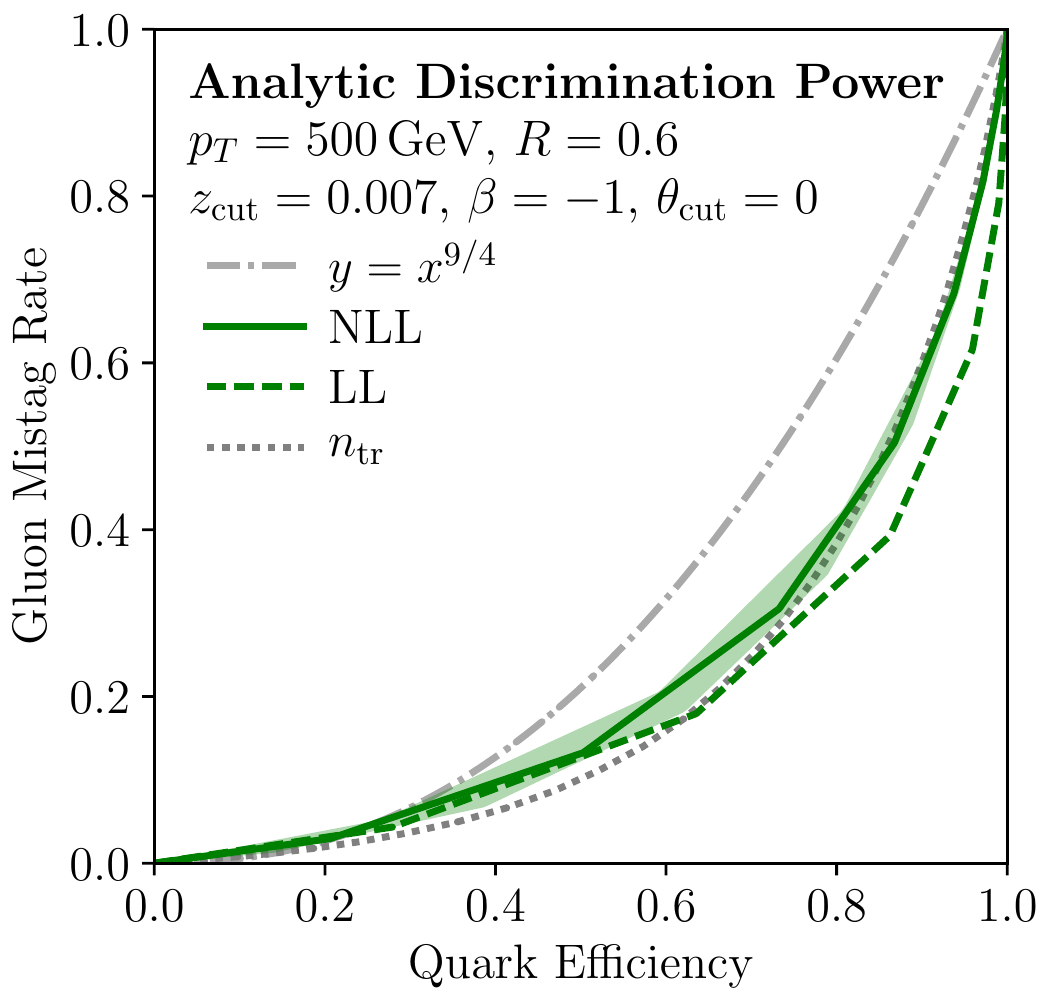}}  \\[-2ex]
\subfloat[]{  \includegraphics[width=0.45\linewidth]{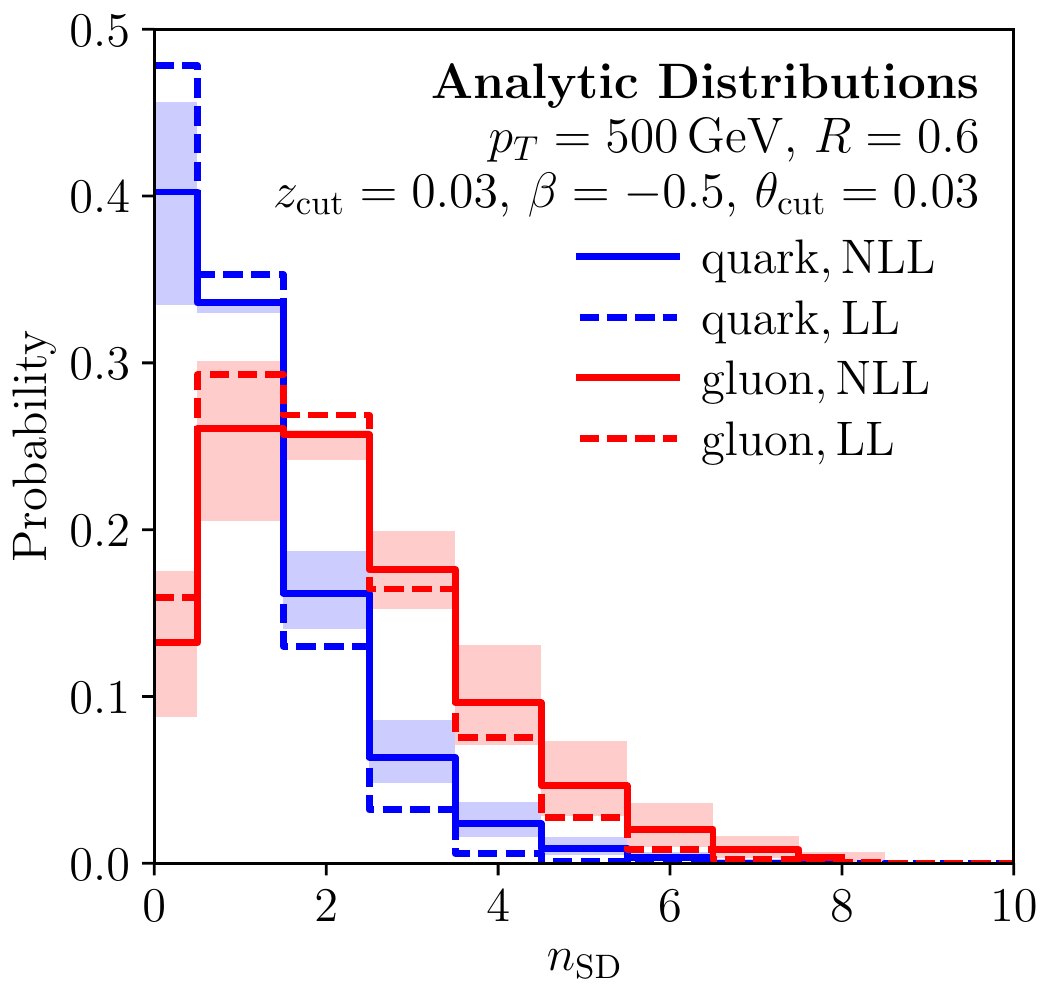}}
\subfloat[]{  \includegraphics[width=0.45\linewidth]{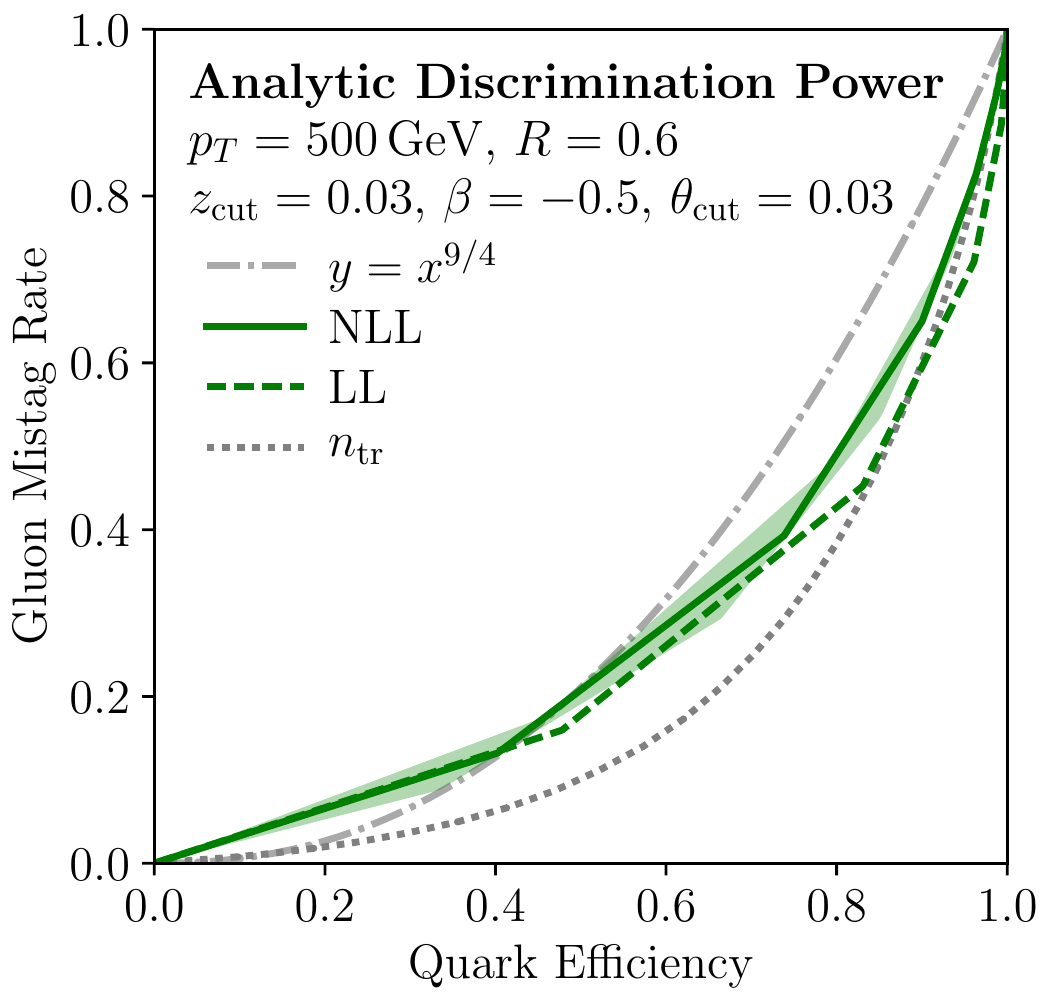}}
  \caption{Calculations at LL and NLL accuracy for (left column) $n_\SD$ distributions and (right column) the corresponding quark/gluon ROC curves. 
    Parameters are chosen according to \Eqss{eq:optparams}{eq:optparams2}{eq:optparams3}
    with $\LambdaNP = 2$ GeV and (top row) $\beta=-1$ and (bottom row) $\beta = -0.5$.  The uncertainties in the NLL calculation come from varying the $\alpha_s$ scale by a factor of 2.
  }
  \label{fig:nll_v_ll}
\end{figure}

The $n_\SD$ distributions and ROC curves at LL and NLL accuracy are displayed in \Fig{fig:nll_v_ll}.  
The uncertainties in the NLL calculation come from varying the $\alpha_s$ scale up and down by a factor of 2.
(Scale variation in the LL calculation does not give a reliable estimate of the uncertainty, since flavor-changing processes are absent at LL; we therefore omit bands around the LL predictions.)
The fact that the uncertainties are abnormally small in one bin is an artifact of this one-dimensional variation procedure, which leaves the scale-varied distributions properly normalized.
Also, the uncertainties in the ROC curve are substantially smaller than the uncertainties in the NLL distributions, since the way we implement the scale variation affects quarks and gluons in a correlated way.  We show both $\beta = -1$ and $\beta = -0.5$ with $\zcut$ and $\thetacut$ chosen to be ``optimal'' according to \Eqss{eq:optparams}{eq:optparams2}{eq:optparams3} with $\LambdaNP = 2$ GeV.  One can see that NLL corrections result in a slight decrease in discrimination power compared to LL, due in part to the flavor changes that occur at this order. 

\subsection{Comparison to Parton Showers}
\label{sec:compare_mc}

It is instructive to compare our NLL calculation of the soft drop multiplicity $n_\SD$ with results
obtained from parton shower generators. In addition to the \textsc{Vincia} setup described in \Sec{subsec:ISD},
we obtained alternative event samples by showering the hard events through \textsc{Pythia} 8.219 \cite{Sjostrand:2006za,Sjostrand:2014zea}, \textsc{Herwig} 7.0.1 \cite{Bahr:2008pv,Bellm:2015jjp}, and \textsc{Sherpa} 2.2.0 \cite{Gleisberg:2008ta}, interfaced to their default hadronization and underlying event models.

\begin{figure}[t]
\centering
\subfloat[]{   \includegraphics[width=0.45\linewidth]{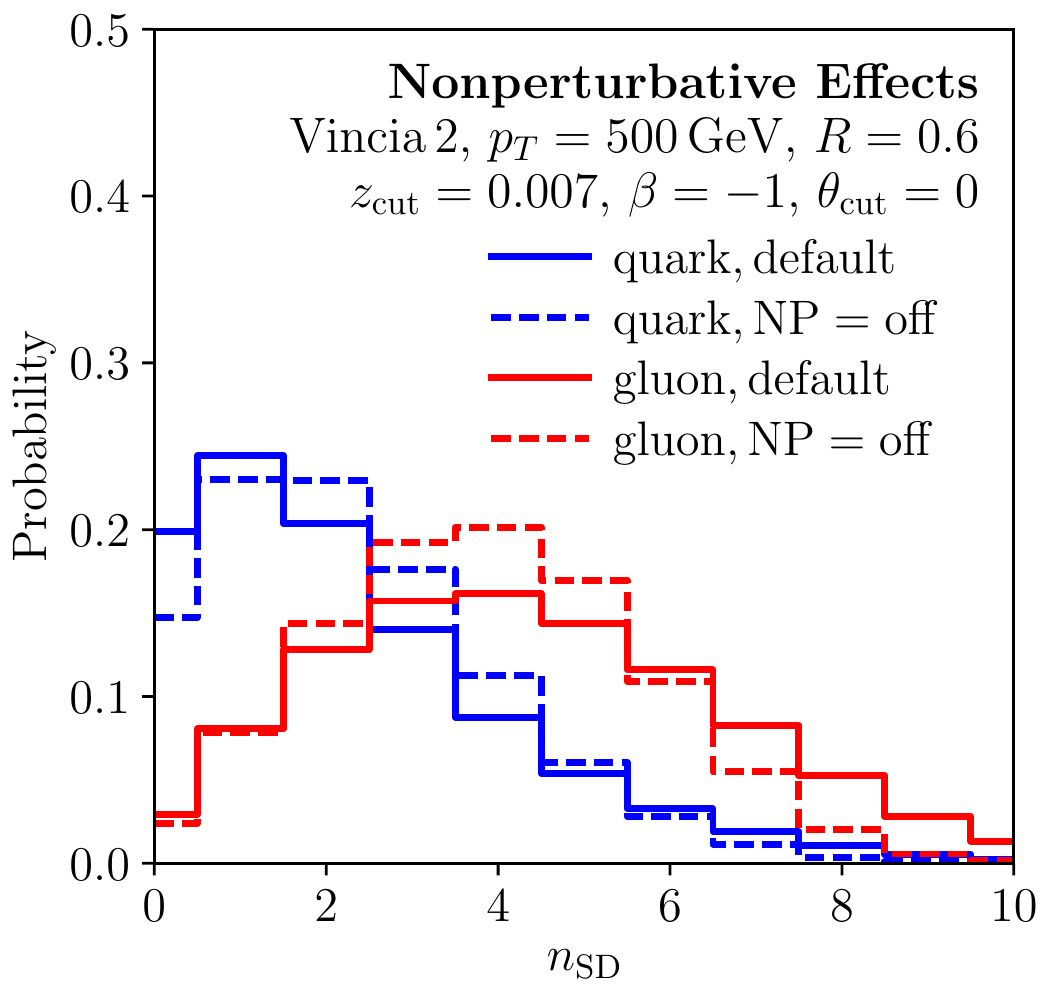}}
\subfloat[]{   \includegraphics[width=0.45\linewidth]{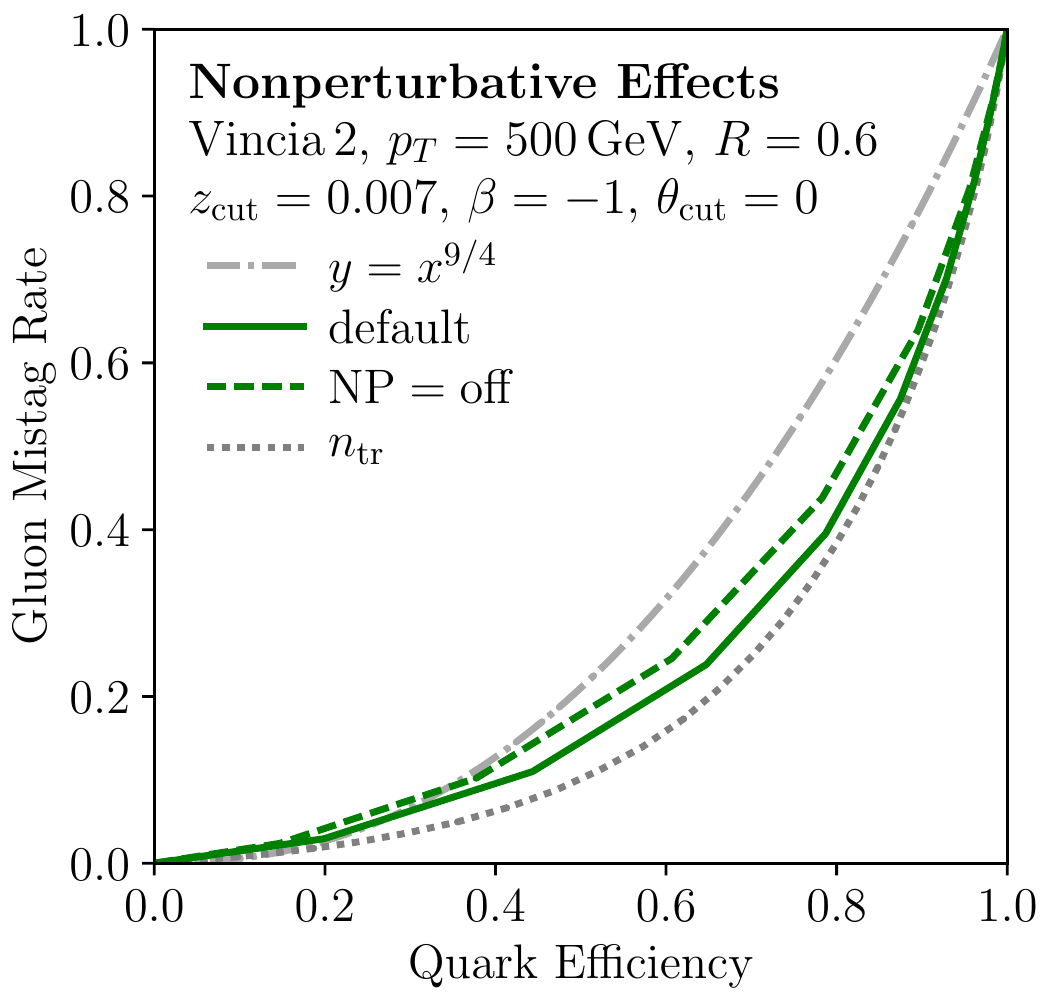}} \\[-2ex]
\subfloat[]{   \includegraphics[width=0.45\linewidth]{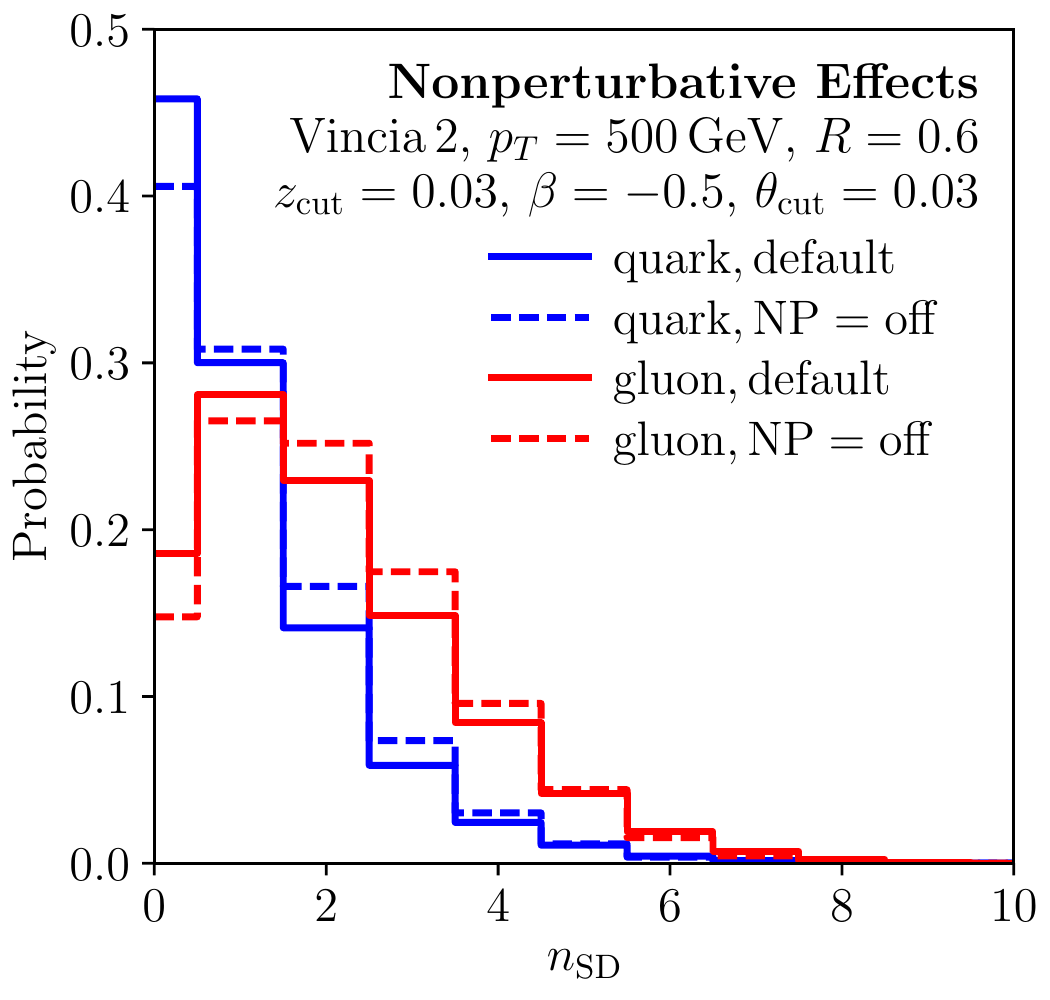}}
\subfloat[]{   \includegraphics[width=0.45\linewidth]{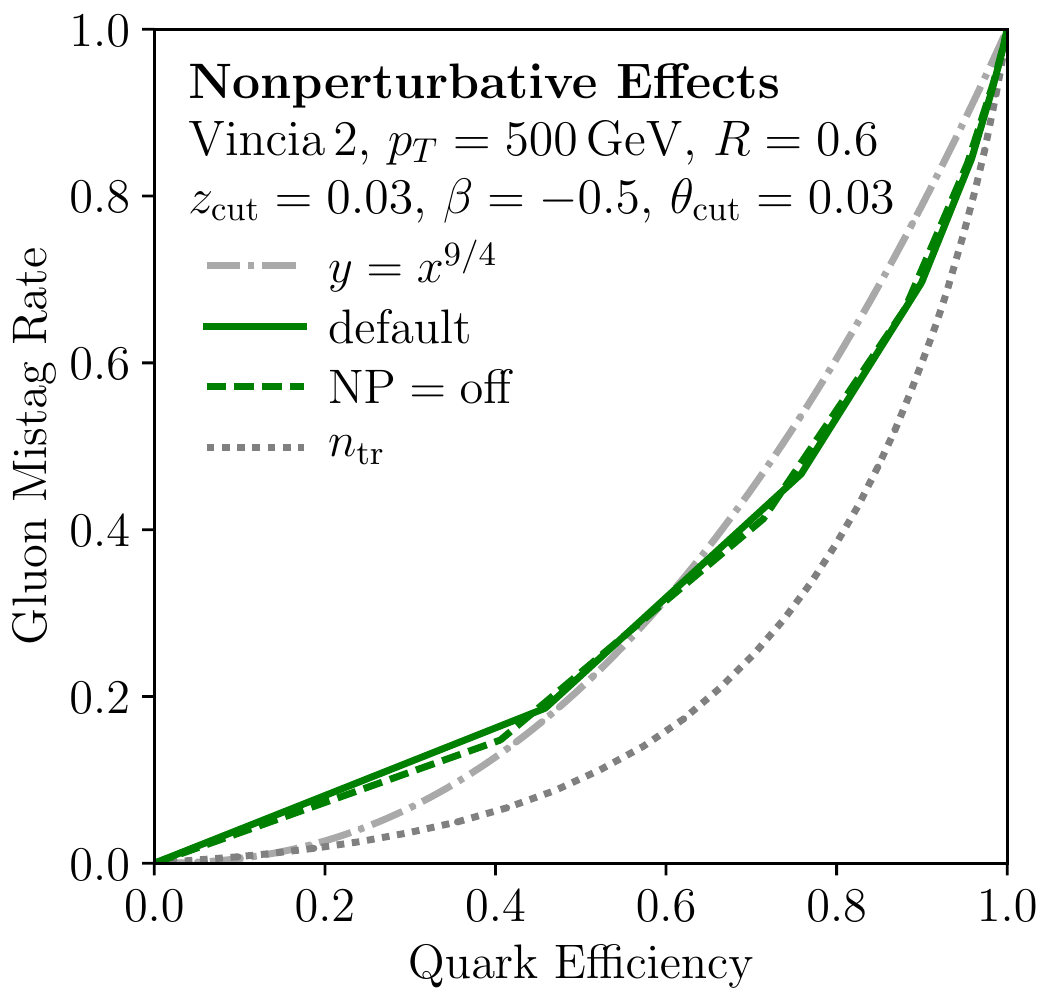}}
  \caption{Impact of nonperturbative effects on (left column) $n_\SD$ distributions and (right column) the corresponding ROC curves.  This study
    employs \textsc{Vincia}, where parameters are chosen according to \Eqss{eq:optparams}{eq:optparams2}{eq:optparams3} with $\LambdaNP = 2$ GeV and (top row) 
    $\beta = -1$ and (bottom row) $\beta = -0.5$.
  }
  \label{fig:had_on_off}
\end{figure}

\begin{figure}[t]
\centering
\subfloat[]{    \includegraphics[width=0.45\linewidth]{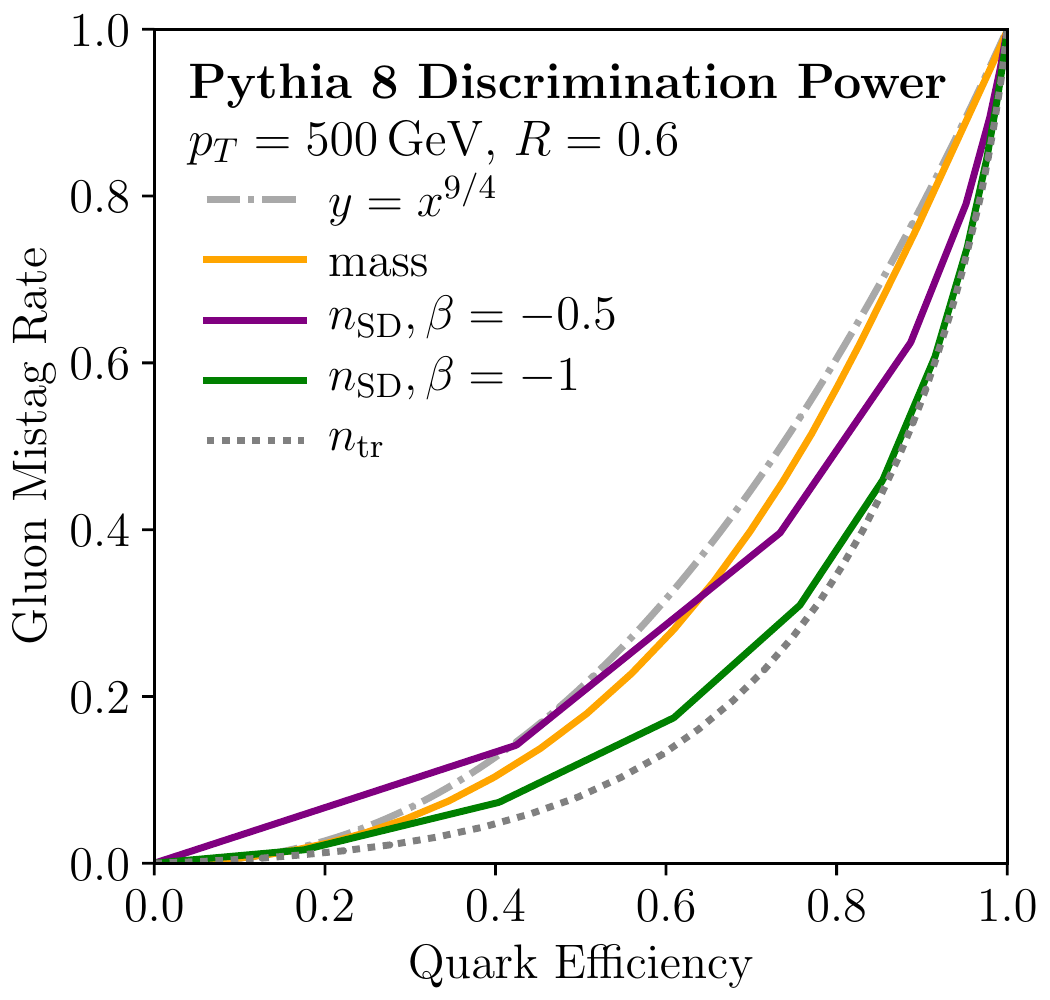}}
\subfloat[]{    \includegraphics[width=0.45\linewidth]{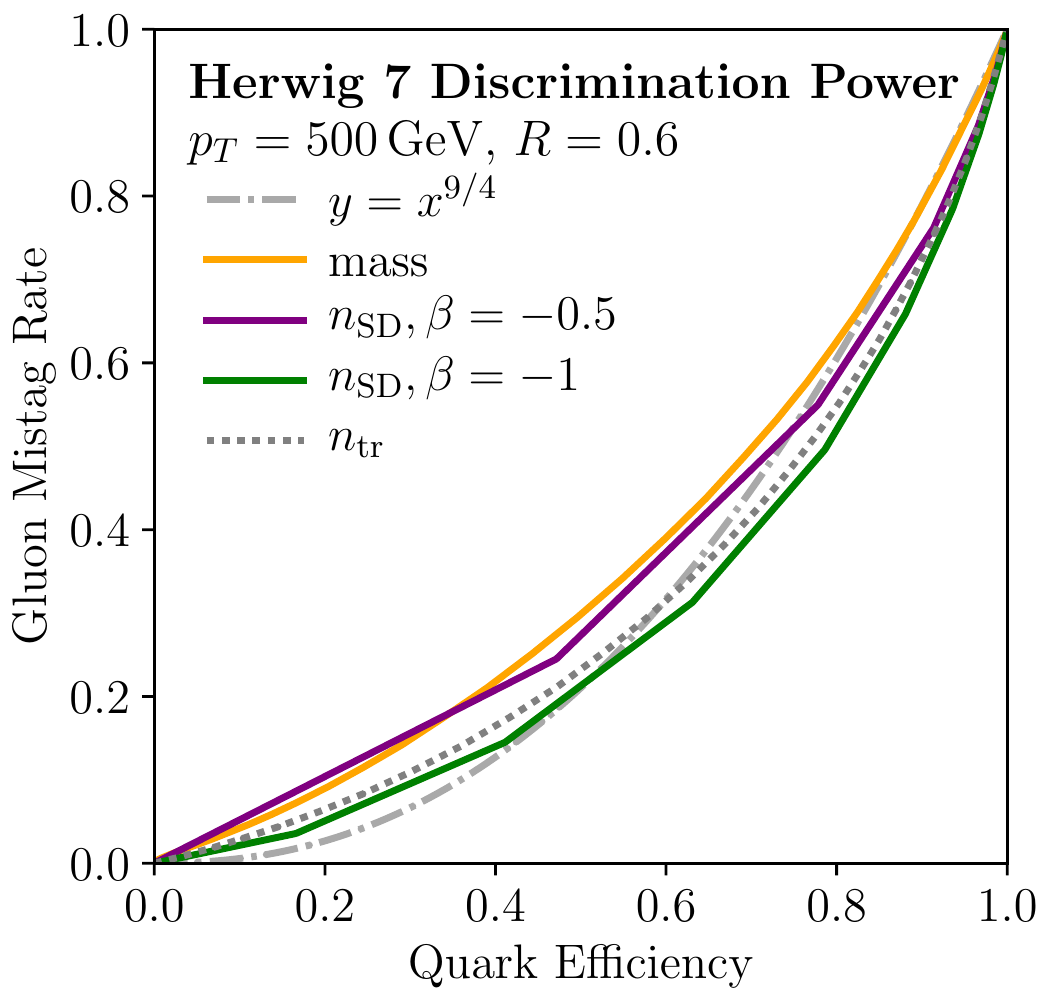}} \\[-2ex]
\subfloat[]{    \includegraphics[width=0.45\linewidth]{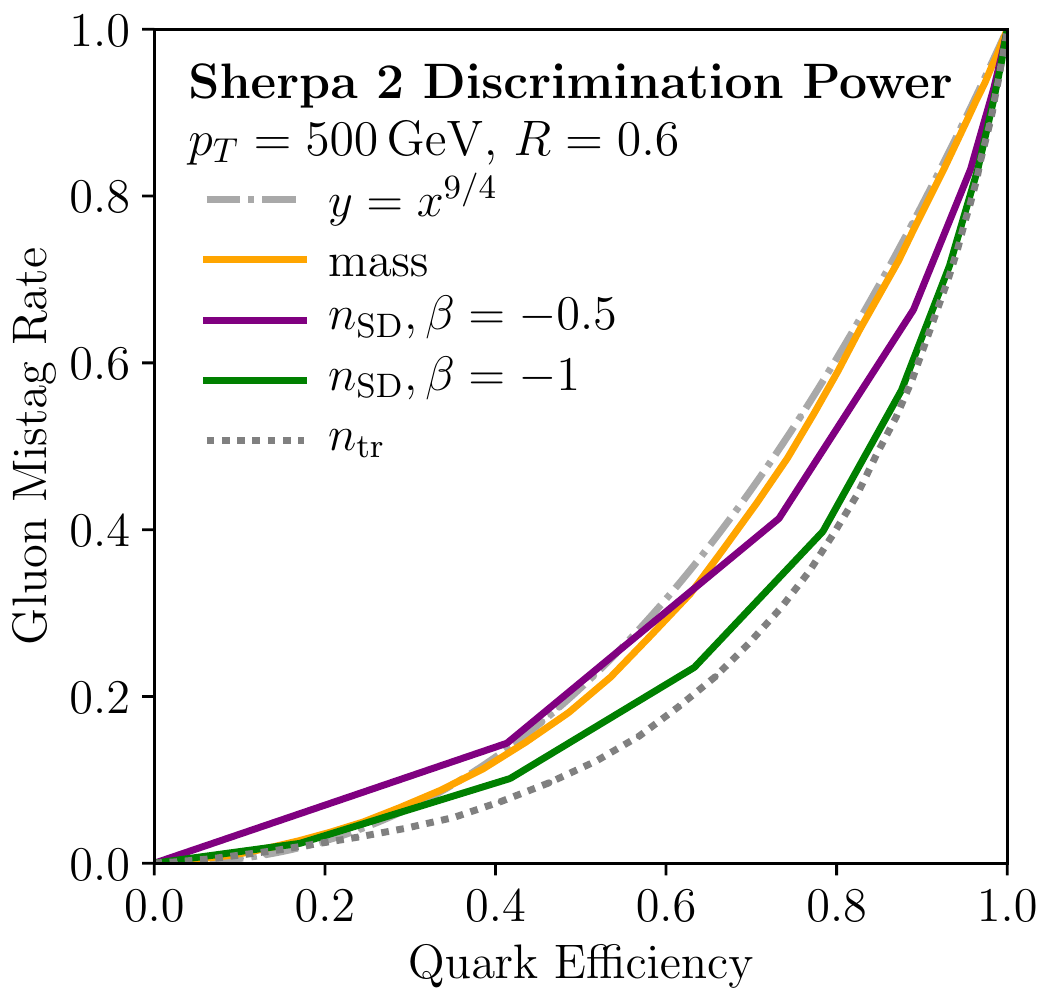}}
\subfloat[]{    \includegraphics[width=0.45\linewidth]{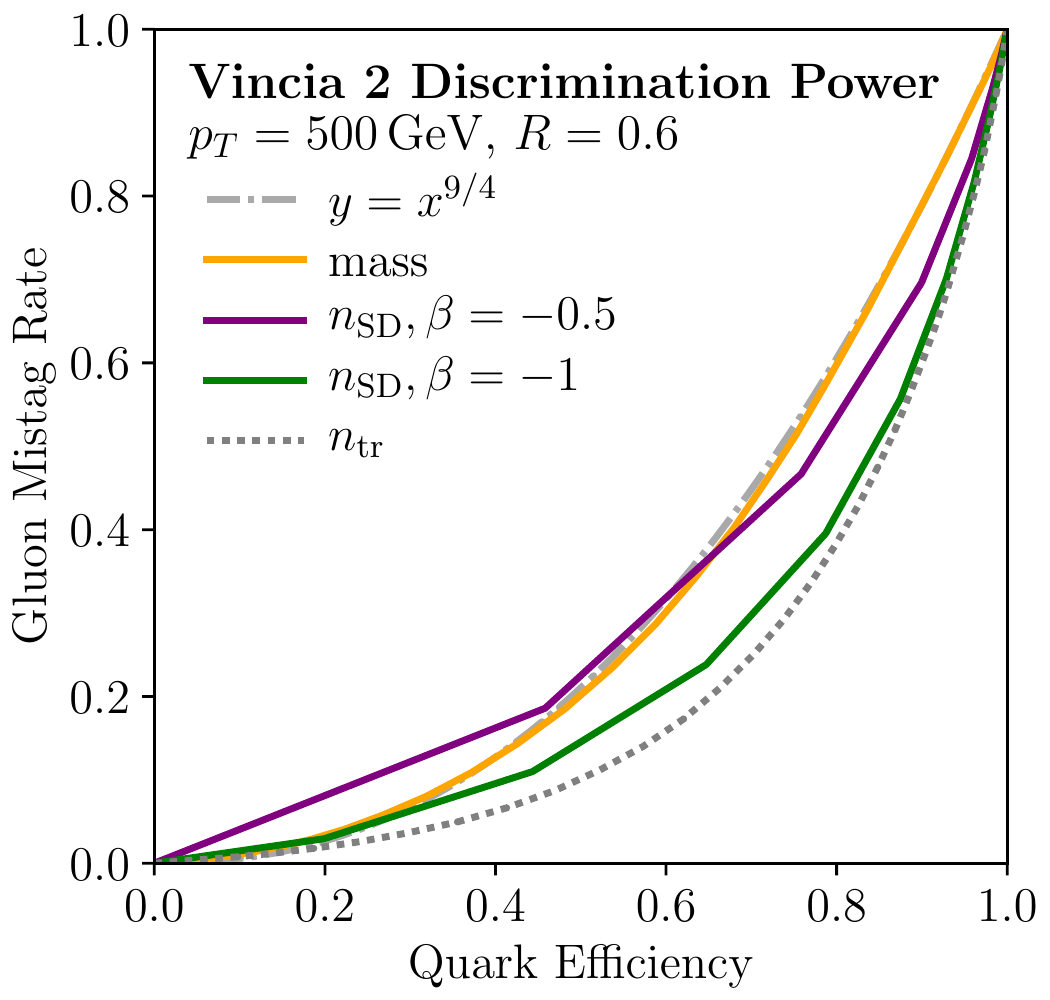}} 
  \caption{Predicted quark/gluon discrimination power from (a) \textsc{Pythia} 8.219, (b) \textsc{Herwig} 7.0.1, (c) \textsc{Sherpa} 2.2.0, and (d) \textsc{Vincia} 2.0.01.  
    While the generators disagree about absolute performance, they agree that $n_\SD$ with $\beta=-1$ outperforms jet mass and approaches the discrimination power of $n_\text{tr}$. 
  }
  \label{fig:nsd_in_mcs}
\end{figure}

\begin{figure}[p]
\captionsetup[subfloat]{farskip=2pt,captionskip=1pt}
\centering
\subfloat[]{   \includegraphics[width=0.42\linewidth]{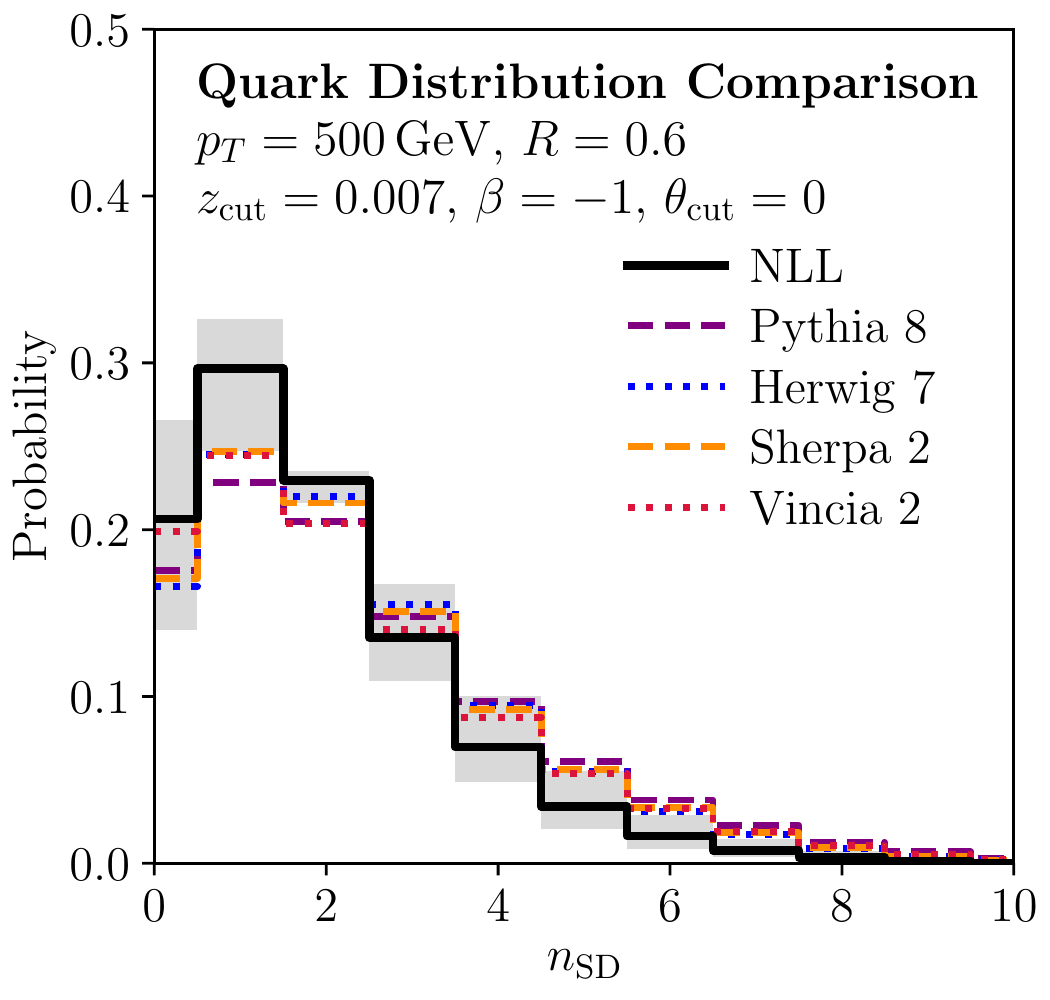}}
\subfloat[]{   \includegraphics[width=0.42\linewidth]{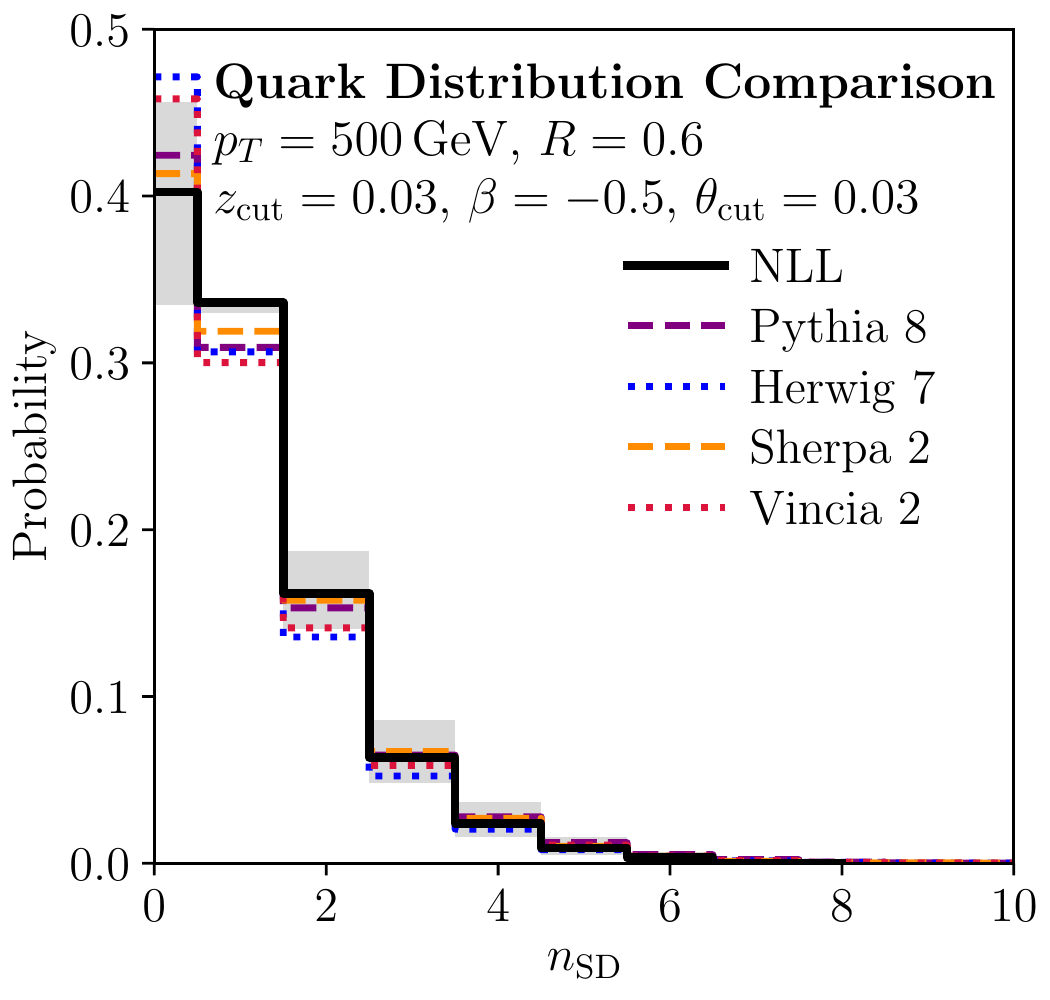}}\\
\subfloat[]{   \includegraphics[width=0.42\linewidth]{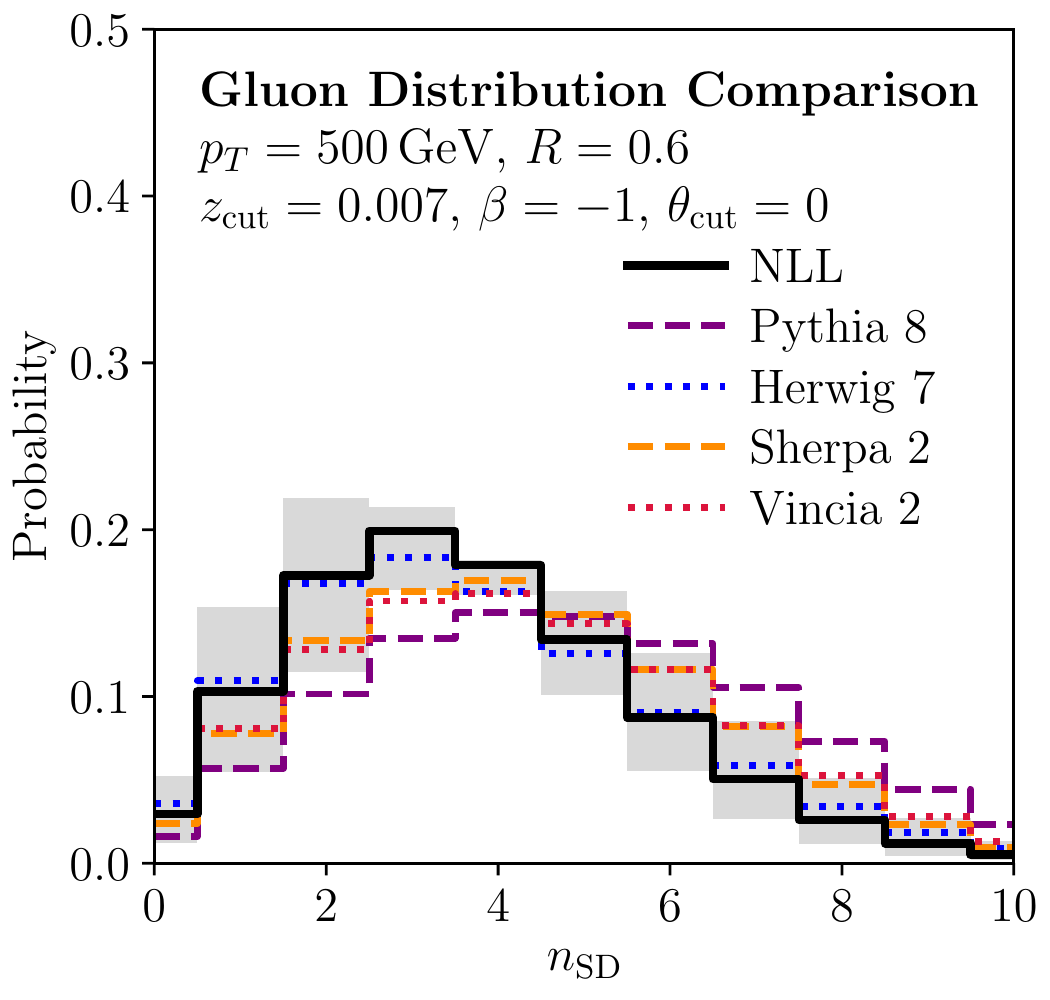}}
\subfloat[]{   \includegraphics[width=0.42\linewidth]{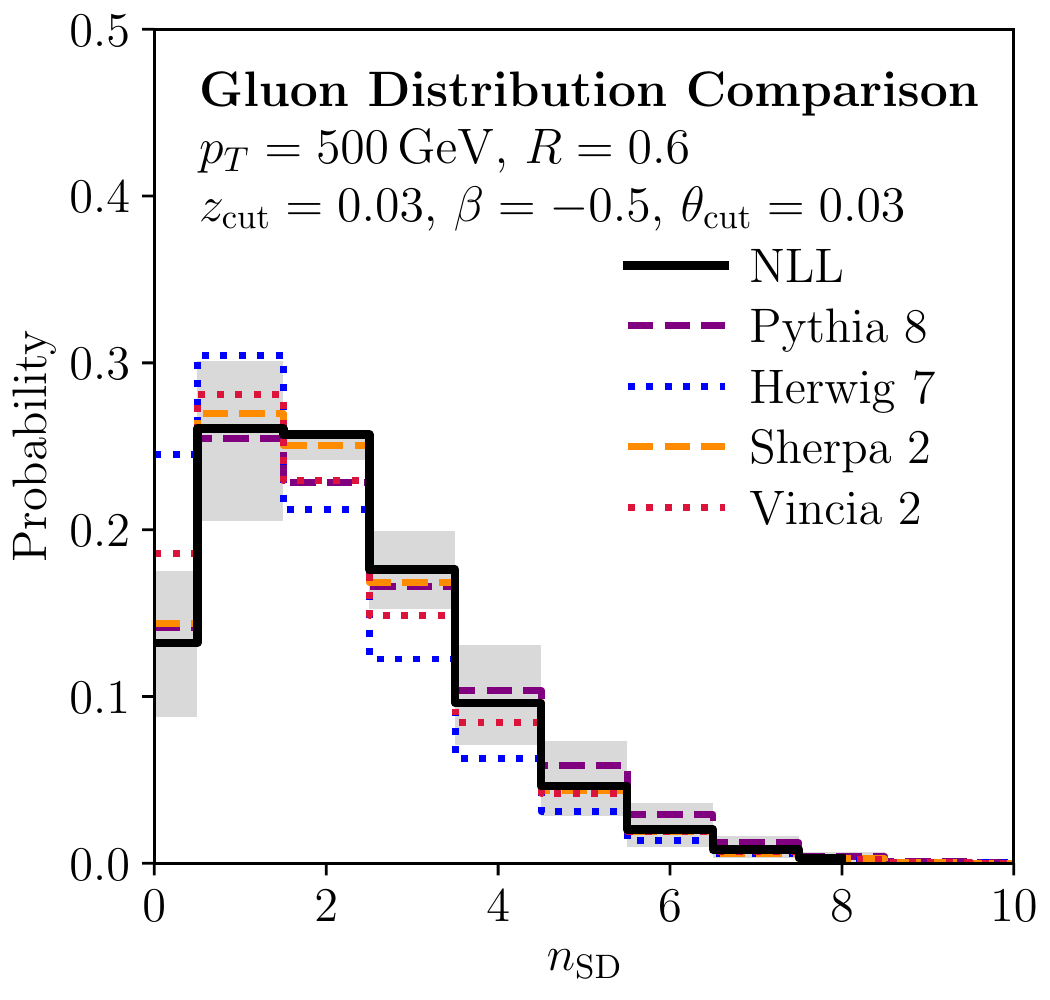}}\\
\subfloat[]{   \includegraphics[width=0.42\linewidth]{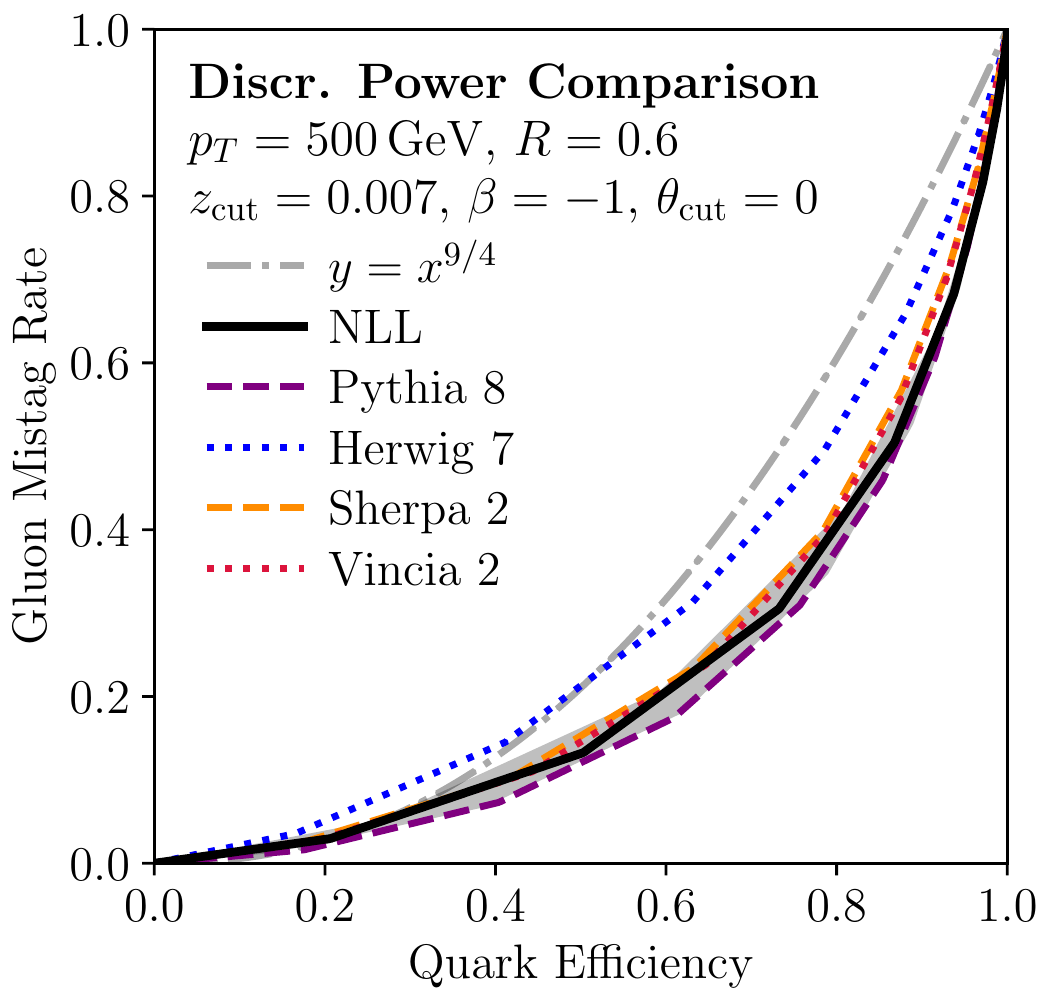}}
\subfloat[]{   \includegraphics[width=0.42\linewidth]{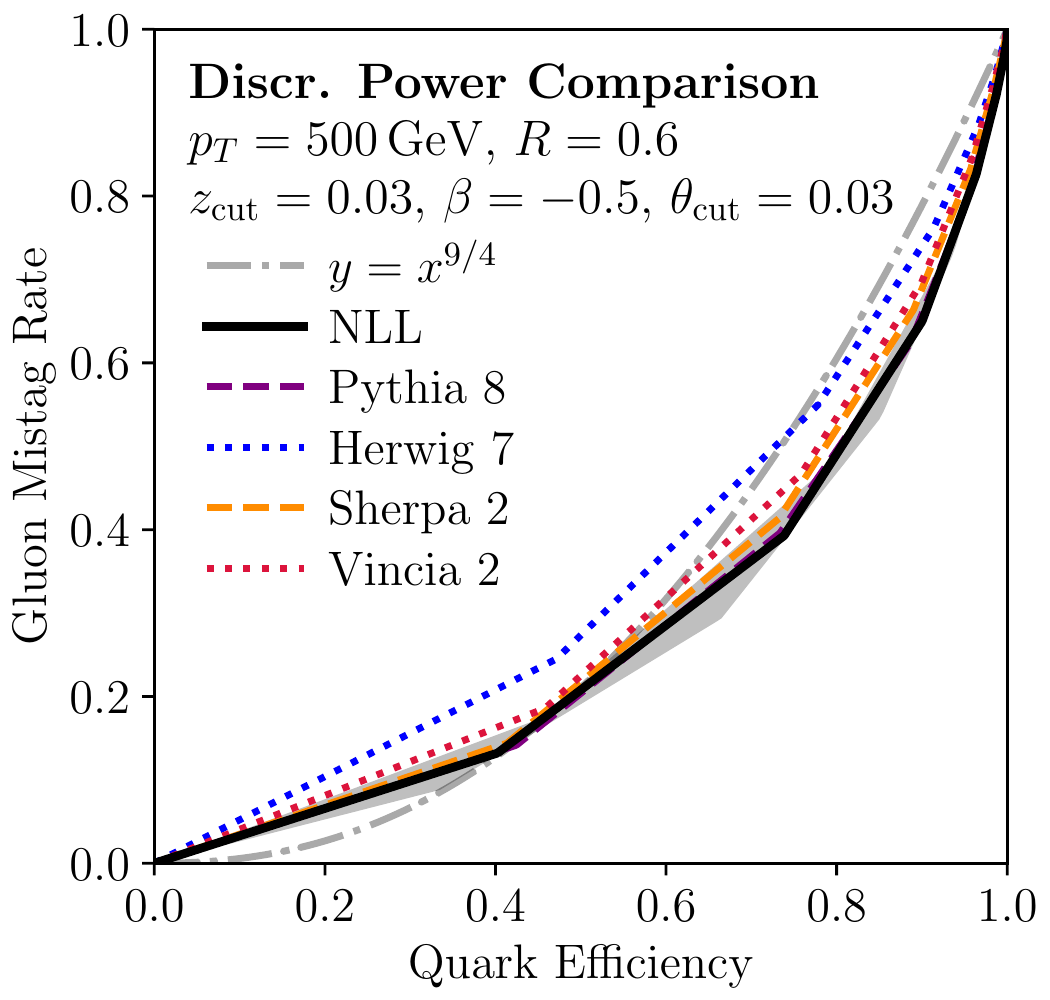}}
  \caption{Analytic NLL distributions compared to parton shower generators for (top row) quark jets, (middle row) gluon jets, along with (bottom row) the corresponding ROC curves.
    Parameters are chosen according to \Eqss{eq:optparams}{eq:optparams2}{eq:optparams3} with $\LambdaNP = 2$ GeV and (left column) $\beta = -1$ and (right column) $\beta = -0.5$.}
  \label{fig:mc_v_nll}
\end{figure}

First, to validate the reliability of our NLL calculation, we want to explore the impact of nonperturbative effects on the parton showers.
In \Sec{sec:mean_discrim} we noted that hadronization effects should generically be minimal provided parameters are 
chosen at or above the values given in \Eqss{eq:optparams}{eq:optparams2}{eq:optparams3}.
To investigate this expectation further, we check the size of nonperturbative corrections in \textsc{Vincia} by turning hadronization and underlying event off and comparing to results obtained using the default settings.
In \Fig{fig:had_on_off}, we show $n_\SD$ with $\beta= -1$ and $\beta=-0.5$, where in each case $\zcut$ and $\thetacut$ are computed using \Eqss{eq:optparams}{eq:optparams2}{eq:optparams3} with $\Lambda_\text{NP} = 2$ GeV. 
As expected, nonperturbative effects are under control, confirming that our perturbative NLL calculations should indeed be reliable in predicting the $n_\SD$ distributions.
Though not shown, the other three parton shower generators also exhibit comparable nonperturbative shifts. 

Next, we show that all parton shower generators predict that soft drop multiplicity is a relatively good quark/gluon discriminant.
In \Fig{fig:nsd_in_mcs}, we compare $n_\SD$ with $\beta = -1$ and $\beta = -0.5$ to jet mass and $n_\text{tr}$ for each generator separately. 
For $\beta = -1$, soft drop multiplicity provides a significant improvement over generic additive observables but does not quite match the performance of track multiplicity.
(See, however, \Fig{fig:sd_lambda_sweep:a} where nonperturbative parameter values push the performance of $n_\SD$ to match $n_\text{tr}$.)
The ordering of the ROC curves is roughly the same between the four generators, though the
absolute discrimination power does differ.

Finally, we can directly compare our NLL predictions to the parton shower generators.
In \Fig{fig:mc_v_nll}, we show the $n_\SD$ distributions and ROC curves for both $\beta = -1$ and $\beta = -0.5$.
When interpreting these curves, one has to remember that the NLL prediction does not include nonperturbative effects.
The quark distributions are roughly similar between the various generators, but there is a larger spread in the gluon distributions, a feature also seen in the study of \Refs{Badger:2016bpw,Gras:2017jty}.
It is interesting to note that both \textsc{Vincia} and \textsc{Sherpa}, as well as our NLL calculation, predict rather strong discrimination power, in better agreement with \textsc{Pythia} than with \textsc{Herwig}.
This highlights the importance of carrying out these analytic calculations to even higher accuracy, in order to better understand the desired behavior for these parton shower generators.


\section{Calculations for Collinear-Unsafe Soft Drop Multiplicity}
\label{sec:irc_unsafe}

Thus far, we have focused on choices of ISD parameters where the quark/gluon discrimination power could be predicted using perturbation theory. In the section, we consider the special case of $\thetacut = 0$ and $\beta = 0$, where the soft drop multiplicity is collinear unsafe but still soft safe, allowing us to calculate its RG evolution. 

\subsection{Review of Generalized Fragmentation Functions}
\label{sec:gff_intro}

To study observables with purely collinear final-state divergences, one can use the formalism of GFFs. Ordinary fragmentation functions are well-known objects in QCD which describe the fragmentation of a quark or gluon into a single hadron. GFFs are nonperturbative objects that describe the fragmentation of a quark or gluon into correlated sets of hadrons. The GFF technique has already been applied successfully to weighted jet charge \cite{Krohn:2012fg,Waalewijn:2012sv}, track functions \cite{Chang:2013rca,Chang:2013iba}, and generalized angularities \cite{Larkoski:2014pca}, and a forthcoming paper explores the broader space of observables described by GFFs \cite{Elder:2017bkd}.  

Each collinear-unsafe observable $x$ has an associated set of GFFs, $\GFF_i(x, \mu)$, where $i$ labels each quark flavor, anti-quark flavor, and gluon.  They are normalized to have unit integral,
\begin{equation}
\int_{-\infty}^\infty \text{d}x \, \GFF_i(x, \mu) = 1,
\end{equation}
and at leading order, they have the interpretation of the probability of parton $i$ to yield the observable value $x$.  In higher-order partonic calculations, the GFFs absorb collinear divergences and pick up dependence on the RG scale $\mu$.  While the GFFs themselves cannot be calculated using perturbation theory, their RG evolution is calculable.  Ordinary fragmentation functions exhibit linear DGLAP evolution \cite{Altarelli:1977zs,Gribov:1972ri,Dokshitzer:1977sg}, whereas GFFs in general have non-linear evolution equations which can even involve mixing between different sets of GFFs.  

As shown in \Ref{Elder:2017bkd}, though, for observables defined on a pairwise clustering tree, the evolution equations for the GFFs greatly simplify.  These observables are called fractal jet observables, since their RG evolution is reminiscent of the fractal structure of the parton shower.  For $\thetacut = 0$ and $\beta = 0$, soft drop multiplicity (and its weighted variant) is an example of a fractal jet observable, allowing us to use the GFF formalism.

It is important to emphasize that the GFF formalism only works for purely collinear divergences.  For $\thetacut = 0$ but $\beta > 0$, there are mixed soft-collinear divergences in the simultaneous $z \to 0$ and $\theta \to 0$ limits.  These correlated diverges would require additional regulators, similar in spirit to rapidity regularization \cite{Chiu:2012ir} (see also \cite{Bertolini:2015pka}).  The use of fragmentation functions to study the $\beta = 0$ limit was previously considered in \Ref{Larkoski:2015lea} to study the soft-dropped $z_g$ distribution (which is the same as $z_1$ for ISD).

Following \Ref{Elder:2017bkd}, consider a fractal observable $x$ defined recursively on an IRC-safe binary clustering tree as follows.  
Each final-state hadron is assigned a starting weight $w_a$, which serves as the initial seed for the observable, and the observable $x$ is built recursively according to 
\begin{equation}
\label{eq:recursion}
x = \hat{x}(z, x_1, x_2),
\end{equation}
where $z \in [0,1]$ is the momentum fraction of the $2 \to 1$ merging, and $x_1$ and $x_2$ are the values of the observable (or the starting weight $w_a$) on the daughter nodes.  
Note that $\hat{x}$ is independent of the opening angle $\theta$ of the merging, and the only angular dependence comes through the choice of clustering tree.  The leading-order RG evolution for the GFFs associated with $x$ is
\begin{equation}
\label{eq:rg_evolution}
\mu \frac{\mathrm{d} }{\mathrm{d}  \mu} \GFF_i(x, \mu) = \frac12 \sum_{jk} \int \mathrm{d} z \, \mathrm{d}  x_1 \, \mathrm{d}  x_2 \, \frac{\as(\mu)}{\pi} P_{i\to jk}(z) \, \GFF_j(x_1, \mu)\,  \GFF_k(x_2, \mu) \, \delta\left[x - \hat{x}(z,x_1, x_2) \right],
\end{equation}
where $P_{i \to jk}(z)$ is the splitting function.  
At this order, the evolution equation (but not the observable itself) is independent of the choice of clustering tree.  
Note that the evolution equation is also independent of the starting weights $w_a$, which are effectively encoded in the low-scale initial conditions for $\GFF_i$.
Even though the clustering tree is IRC safe, $x$ is generally collinear unsafe, since \Eq{eq:recursion} allows an exactly collinear splitting to change the observable.

The canonical RG scale for a generic GFF is
\begin{equation}
\label{eq:mu_canonical}
\mu = E_{\rm jet} R_0,
\end{equation}
and if we can extract the functional form of $\GFF_i(x, \mu)$ at a low scale, we can use \Eq{eq:rg_evolution} to predict their form at a higher scale.  The RG equations have the same recursive structure as a parton shower, and we can use the numerical techniques of \Ref{Elder:2017bkd} to evolve the GFFs in $\mu$.  As we will see, our observable of interest actually has a linear evolution equation, which greatly simplifies the numerical treatment.

\subsection{Linear Evolution for Soft Drop Multiplicity}

For $\thetacut = 0$ and $\beta = 0$, soft drop multiplicity is an example of a fractal observable.  More generally, any ISD observable of the form 
\begin{equation}
x = \sum_n f(z_n)
\end{equation}
is a fractal observable. Using C/A for the binary clustering tree with starting weights $w_a = 0$, the recursion relation for this general observable is 
\begin{equation}\label{eq:sd_mult_cases}
\hat{x}(z,x_1, x_2) =
\begin{cases}
x_2 & \quad 0 \leq z < \zcut, \\
x_2+f(z) & \quad \zcut \leq z \leq 1/2, \\
x_1+f(1-z) & \quad 1/2 \leq z \leq 1-\zcut, \\
x_1 & \quad 1-\zcut < z \leq 1.
\end{cases}
\end{equation}
The four cases check which subjet is harder and whether the softer subjet passes soft drop. If the softer subjet fails soft drop (i.e.\ $\min(z,1-z) < \zcut$), then the observable value is unchanged.  If the softer subjet passes soft drop, then the $f(z)$ (or $f(1-z)$) value of the splitting enters linearly into the observable.

The recursion relation in \Eq{eq:sd_mult_cases} takes a particularly simple form, since each of the four cases involves either $x_1$ or $x_2$, but not both. This allows us to rewrite the RG evolution from \Eq{eq:rg_evolution} in the form
\begin{equation}
\label{eq:linear_evolution_again}
\mu \frac{\mathrm{d} }{\mathrm{d}  \mu} \GFF_i(x, \mu) = \sum_{jk} \frac{\as(\mu)}{\pi} \left( \int_0^{\zcut} \mathrm{d}z \, P_{i \to jk}(z) \GFF_k(x, \mu) + \int_{\zcut}^{1/2} \mathrm{d}z\, P_{i \to jk}(z) \GFF_k(x - f(z), \mu) \right),
\end{equation}
where we have simplified using the identity $P_{i \to jk}(z) = P_{i \to kj}(1-z)$. This evolution equation is linear, and hence is numerically no more difficult to solve than the ordinary DGLAP equations.
This form holds both for the ordinary soft drop multiplicity as well as for the weighted variants in \App{sec:weighted_nSD}, just with a different choice of $f(z)$.

\subsection{Evolution for Pure Yang-Mills}

Before showing numerical results, it is instructive to consider the case of $n_f = 0$, where there is only a gluon GFF and the evolution can be studied analytically.  Of course, this limit cannot teach us anything about quark/gluon discrimination directly, but we will see that the gluon GFF asymptotes to an exact Poisson distribution at sufficiently large $\mu$, such that it behaves like an idealized counting observable.

For pure Yang-Mills, we can drop flavor labels, and write the gluon GFF as $\GFF \equiv \GFF_g$ and the relevant splitting function as $P(z) \equiv P_{g \to gg}(z)$. Specializing to soft drop multiplicity (i.e. $f(z) = 1$), the evolution equation in \Eq{eq:linear_evolution_again} becomes
\begin{align}
\label{eq:linear_evolution_sd}
\mu \frac{\mathrm{d} }{\mathrm{d}  \mu} \GFF(x, \mu) &= \frac{\as(\mu)}{\pi} \left( \int_0^{\zcut} \mathrm{d}z \, P(z) \GFF(x, \mu) + \int_{\zcut}^{1/2} \mathrm{d}z\, P(z) \GFF(x - 1, \mu) \right) \\
\label{eq:linear_sd_mult_evolution}
&= P_{\rm ave} \, \frac{\as(\mu)}{2 \pi} \Bigl(\GFF (x-1, \mu) - \GFF (x, \mu)\Bigr),
\end{align}
where we have defined
\begin{equation}
P_{\rm ave} = \int_{z_{\rm{cut}}}^{1-z_{\rm{cut}}} \text{d}z \, P(z).
\end{equation}

The interpretation of \Eq{eq:linear_sd_mult_evolution} is that gluon emissions that pass soft drop are added at a rate of $P_{\rm ave} \, \as(\mu) / 2 \pi$ in $\log \mu$ evolution. Specifically, in evolving from $\mu_i$ to $\mu_f$, the expected number of additional emissions is 
\begin{equation}
\lambda(\mu_i, \mu_f) = \frac{P_{\rm ave}}{2 \pi} \int_{\log \mu_i}^{\log \mu_f} \text{d}(\log \mu)\, \as(\mu),
\end{equation}
so the GFF at $\mu_f$ is
\be
\label{eq:finalG}
\GFF(x, \mu_f) = \GFF(x, \mu_i) \otimes \text{Pois}(\lambda(\mu_i, \mu_f))[x],
\ee
where the convolution is in $x$.\footnote{The reader who finds this derivation too slick can explicitly check that \Eq{eq:finalG} solves \Eq{eq:linear_sd_mult_evolution}.  It is helpful to note that $\frac{\text{d}}{\text{d} \lambda} \text{Pois}(\lambda)[x] = \text{Pois}(\lambda)[x-1] - \text{Pois}(\lambda)[x]$.}

As $\mu_f$ increases, more emissions are added, so the initial GFF distributions at $\mu_i$ becomes less and less important. Substituting in the one-loop running of the strong coupling constant in pure Yang-Mills, 
\begin{equation}\label{eq:one_loop_alpha}
\as(\mu) = \frac{1}{\beta_0 \log(\mu^2/\Lambda_{\rm QCD}^2)}, \qquad \beta_0 = \frac{11}{3} C_A,
\end{equation}
the number of expected emissions is
\begin{equation}
\label{eq:rg_result}
\lambda(\mu_i, \mu_f) = \frac{P_{\rm ave}}{4 \pi \beta_0} \log \left(\frac{\log \frac{\mu_f}{\Lambda_{\rm QCD}}} {\log \frac{\mu_i}{\Lambda_{\rm QCD}}} \right).
\end{equation}
Since this quantity continues to grow at high $\mu_f$, the IR boundary condition $\GFF(x, \mu_i)$ is irrelevant in the $\mu_f \to \infty$ limit, yielding the asymptotic form
\begin{equation}
\GFF(x, \mu \gg \Lambda_{\rm QCD}) \approx \text{Pois}(\lambda(\mu))[x], \quad \lambda(\mu) = \frac{P_{\rm ave}}{4 \pi \beta_0} \log \log \frac{\mu}{\Lambda_{\rm QCD}}.
\end{equation}
Thus, we find a Poisson distribution whose mean scales as $\log \log \mu$, such that the soft drop multiplicity acts like an idealized counting observable.

\subsection{Comparison to Parton Showers}

\begin{figure}[t]
\centering
\subfloat[]{    \includegraphics[width=0.45\linewidth]{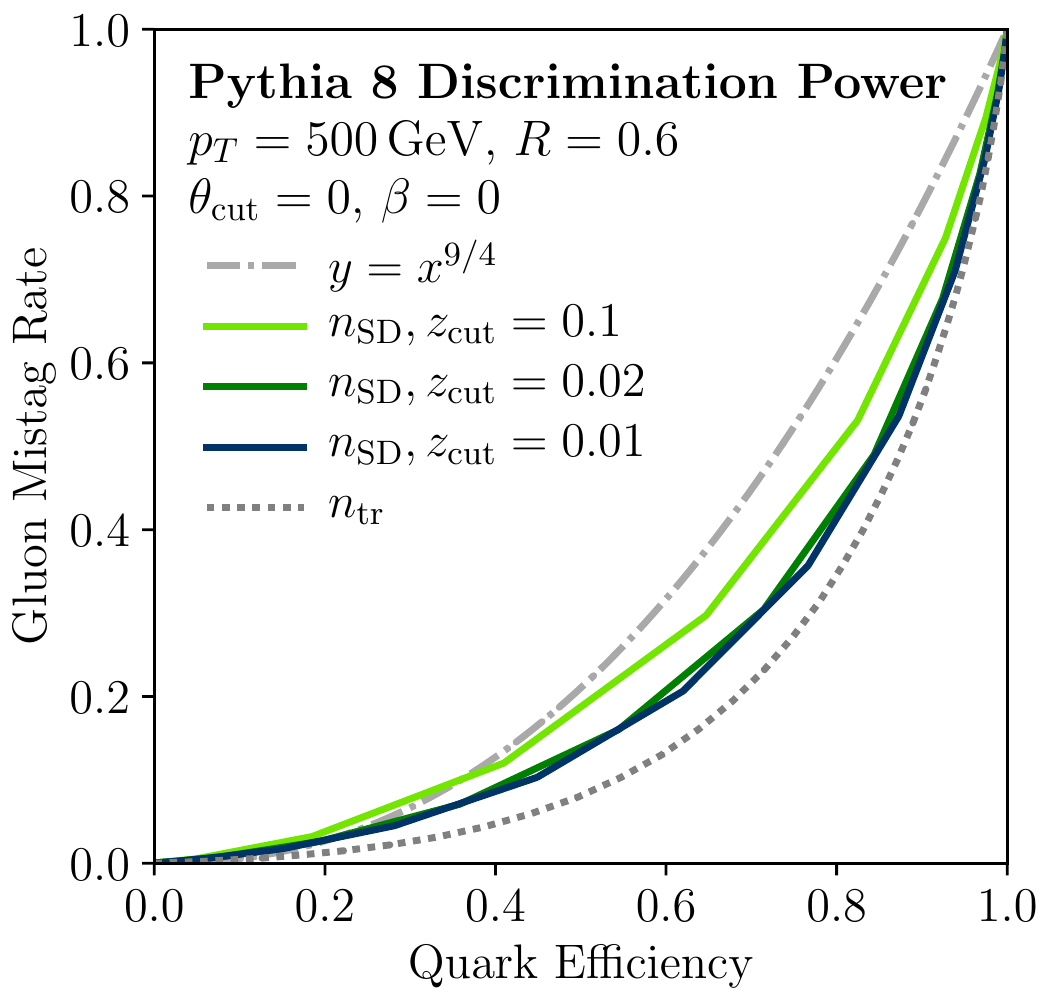}}
\subfloat[]{    \includegraphics[width=0.45\linewidth]{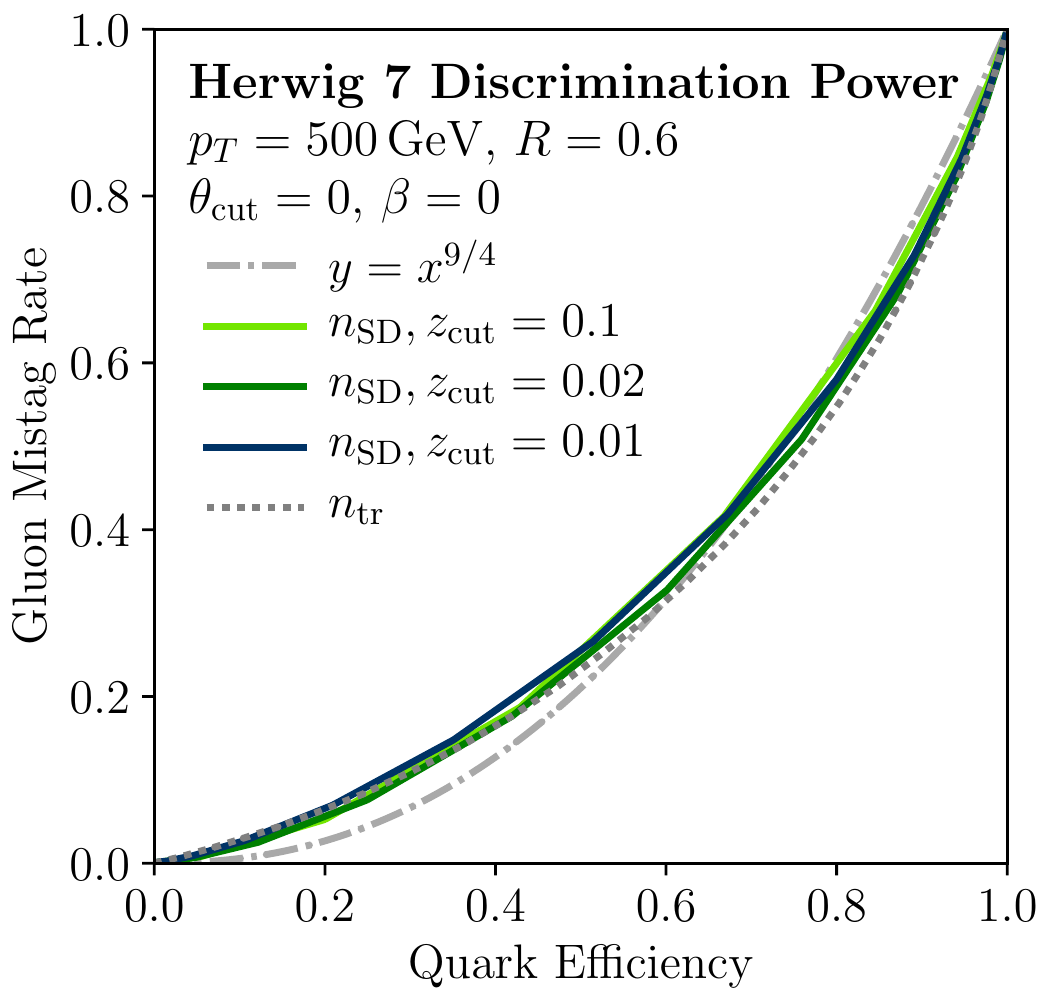}} \\[-2ex]
\subfloat[]{    \includegraphics[width=0.45\linewidth]{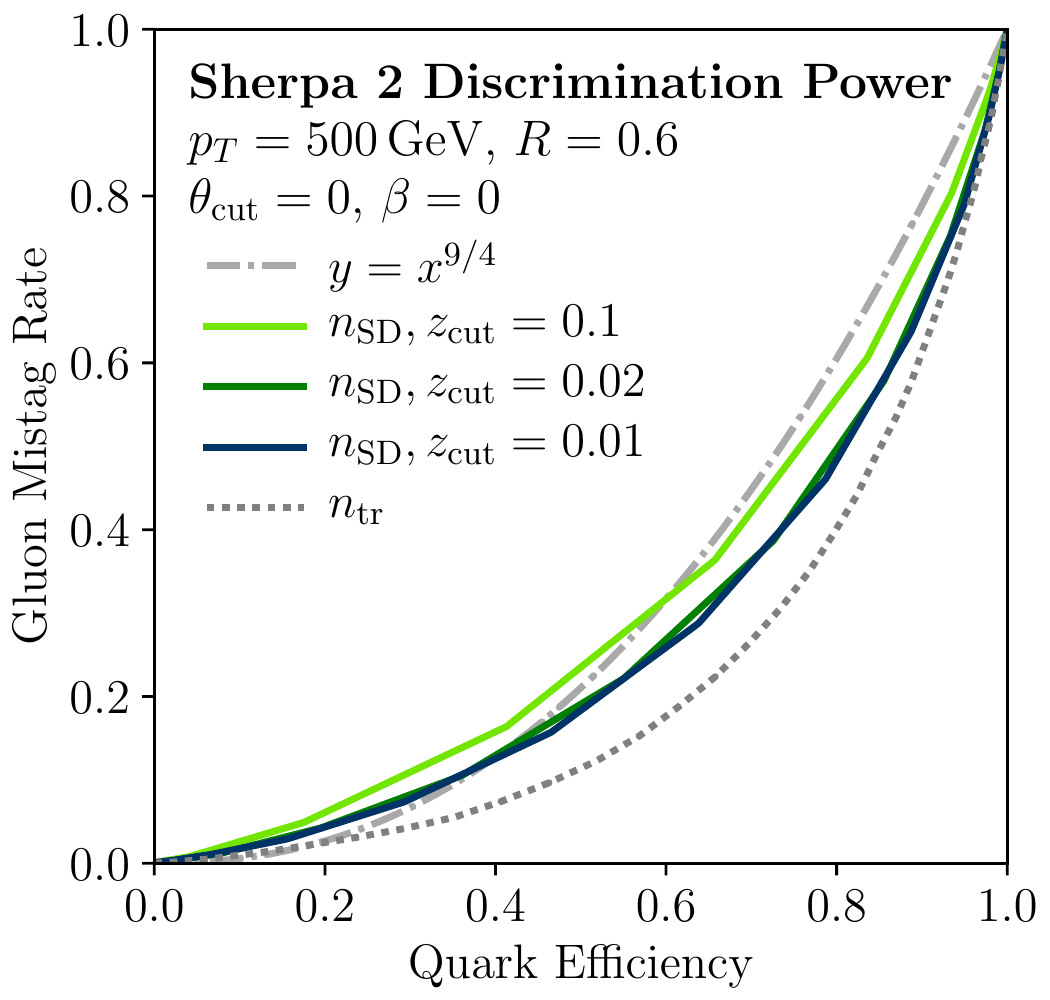}}
\subfloat[]{    \includegraphics[width=0.45\linewidth]{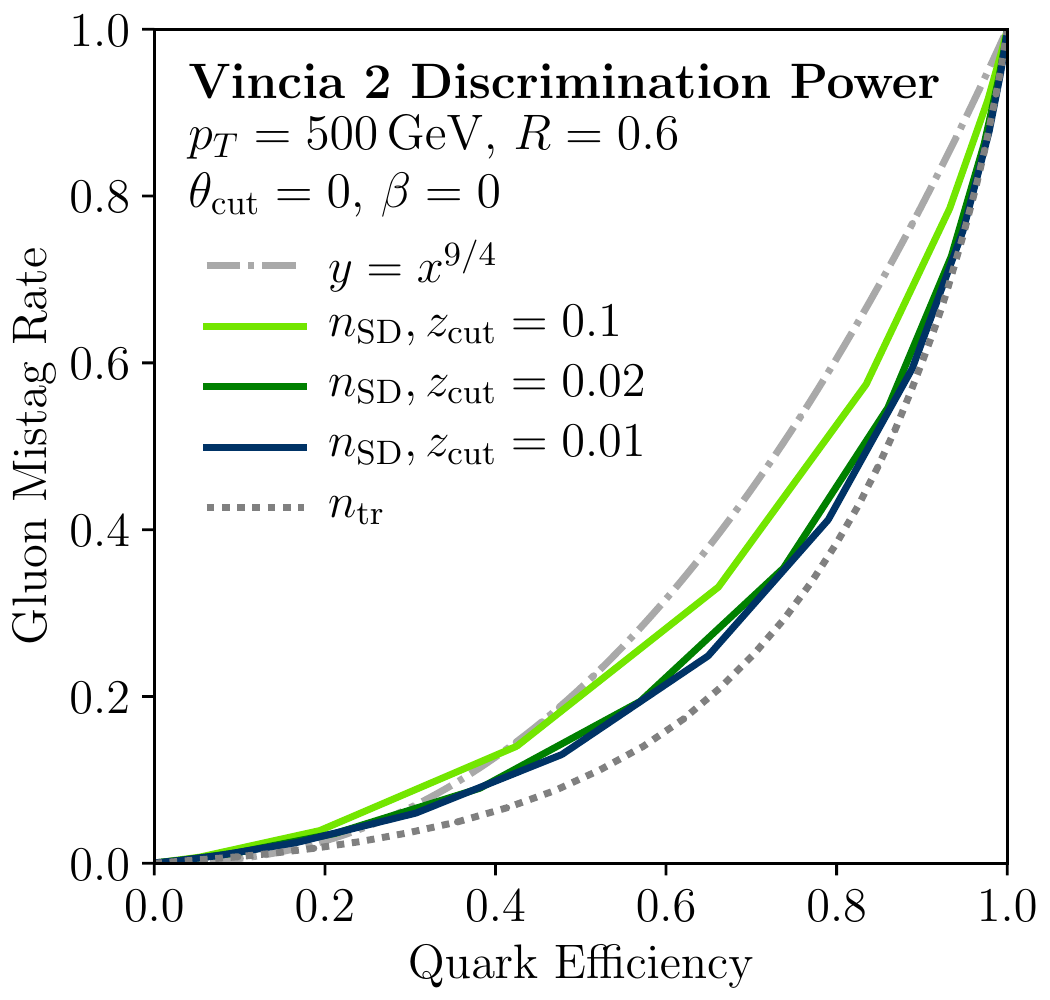}} 
  \caption{Same as \Fig{fig:nsd_in_mcs}, but for the collinear-unsafe soft drop multiplicity with $\thetacut = 0$ and $\beta = 0$.
  }
  \label{fig:nsd_in_mcs_colunsafe}
\end{figure}

We now compare the results of the GFF approach to parton shower predictions. First, in \Fig{fig:nsd_in_mcs_colunsafe}, we show the predicted discrimination power for the collinear-unsafe $n_{\SD}$ from the same four parton showers studied in \Sec{sec:compare_mc}. We see that for low $\zcut$ values, the discrimination power of the collinear-unsafe soft drop multiplicity approaches that of our benchmark IRC-safe soft drop multiplicity, previously shown in \Fig{fig:nsd_in_mcs}. (It does not, however, reach the power of the nonperturbative soft drop multiplicities shown in \Fig{fig:sd_lambda_sweep:a}.)  Making $z_{\rm cut}$ any smaller does not significantly improve discrimination power, so we use $z_{\rm cut} = 0.02$ as our baseline parameter choice.

To make a prediction using the GFF approach, we need to extract the nonperturbative distributions at a low scale and then evolve them to a higher scale.  In a full analysis, the low scale distributions would be extracted from data, but here we can use the parton shower generators.
For this, we switch to $e^+ e^-$ collisions, generating pure quark and gluon samples through the processes $e^+e^- \to \gamma/Z^* \to q \bar{q}$ and $e^+e^- \to H^* \to g g$ in \textsc{Vincia} 2.0.01.
Setting $R_0 = 0.6$ as our baseline, we generate jets with energies in a 10\% window of $E_{\text{jet}} = \unit[400]{GeV}$, corresponding to $\mu = E_{\text{jet}} R_0 = \unit[240]{GeV}$.
We then extract $n_{\rm SD}$ from the generated events, which at leading order, is a direct measure of the corresponding GFFs.\footnote{At higher orders, one has to perform a matching calculation; see further discussion in \Ref{Elder:2017bkd}.}

\begin{figure}[t]
\centering
\subfloat[]{ \includegraphics[width=0.45\textwidth]{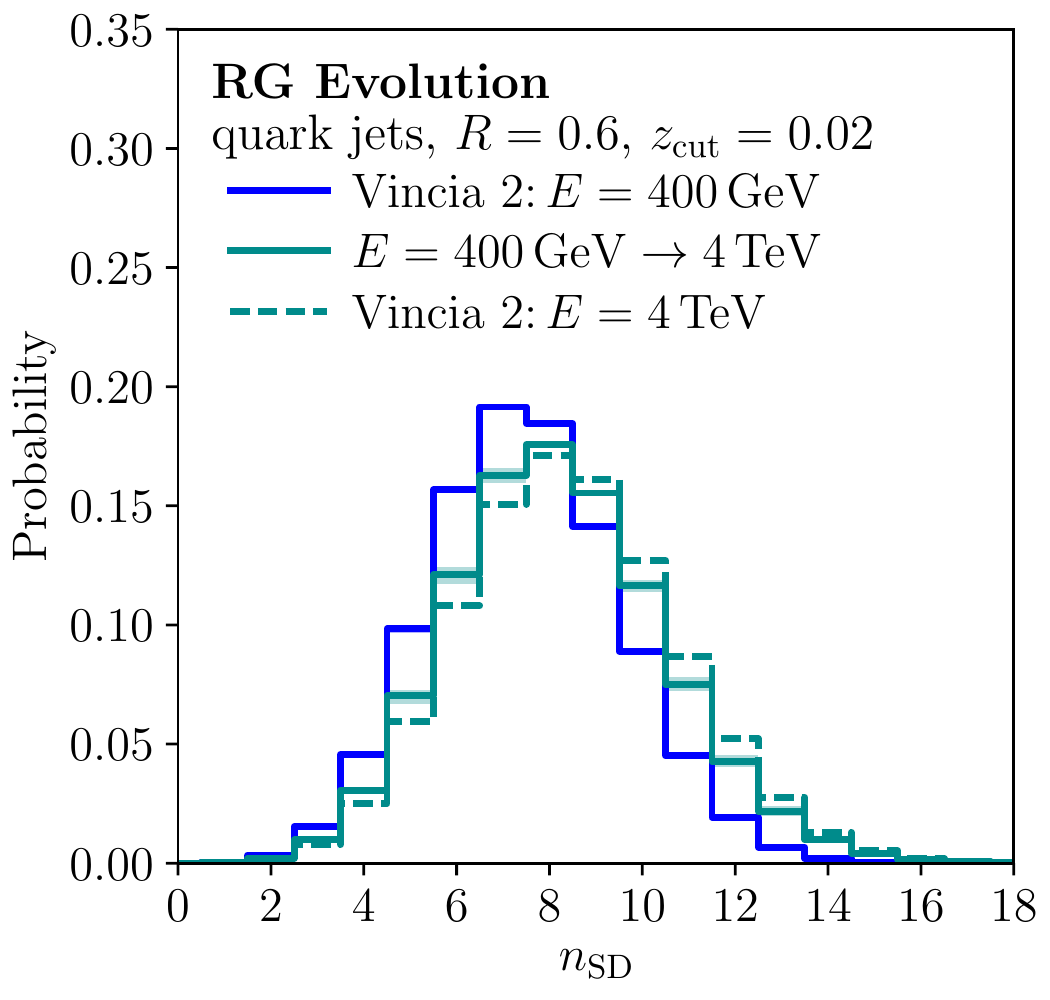}} 
\subfloat[]{ \includegraphics[width=0.45\textwidth]{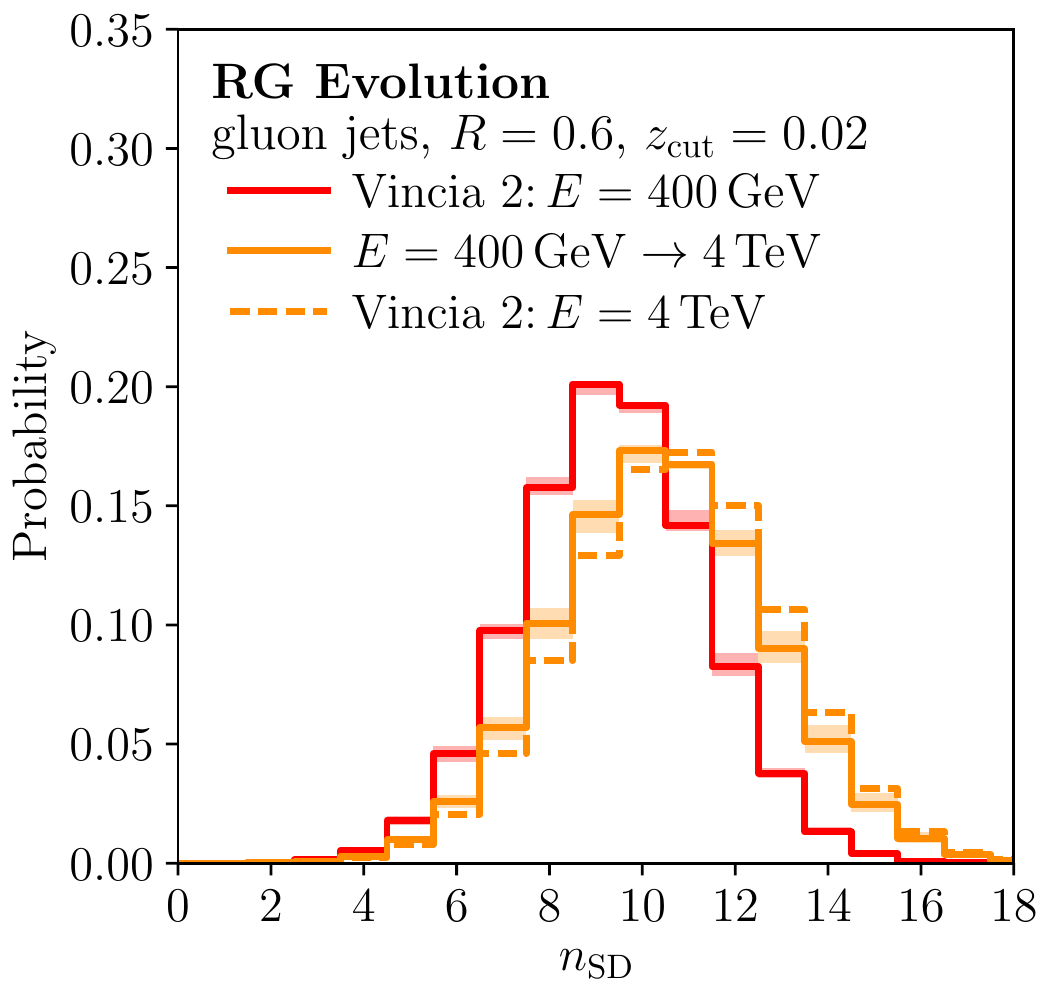}} \\[-2ex]
\subfloat[]{ \includegraphics[width=0.45\textwidth]{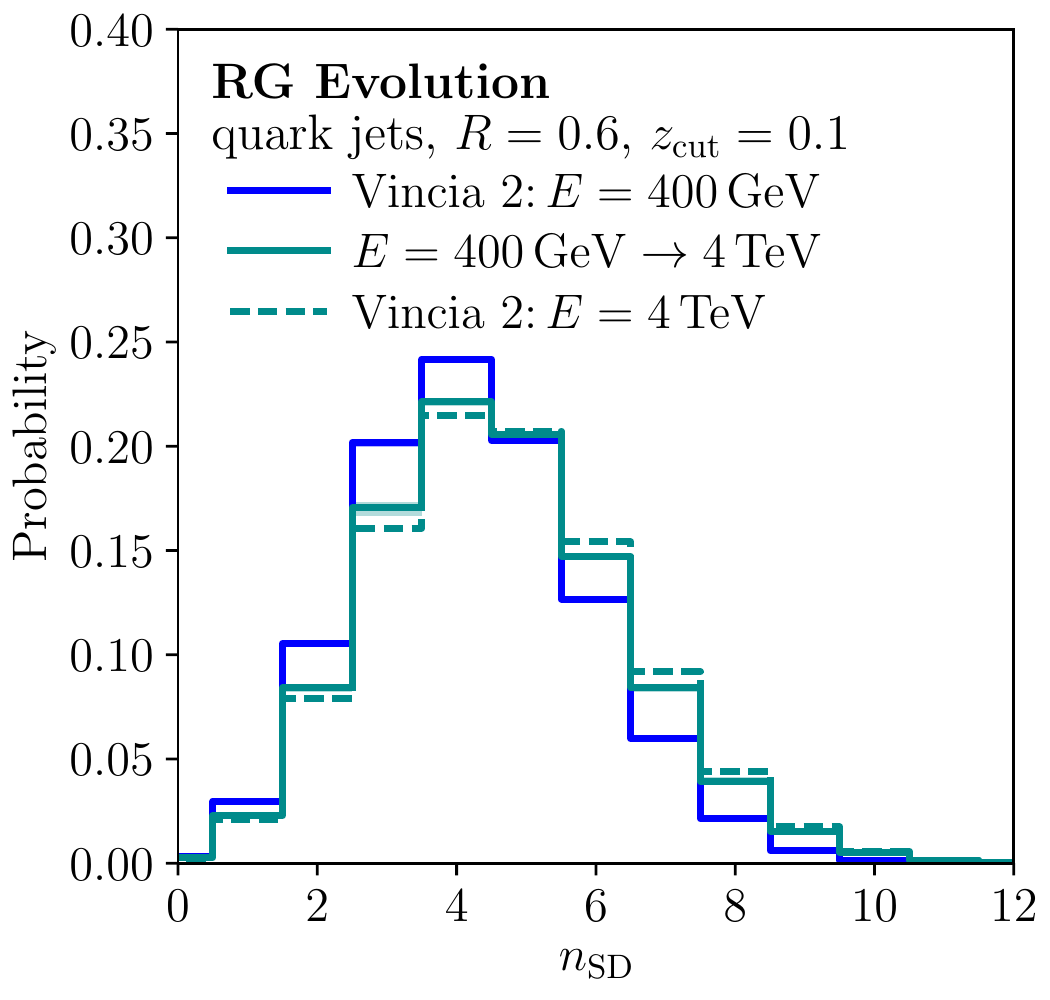}}
\subfloat[]{ \includegraphics[width=0.45\textwidth]{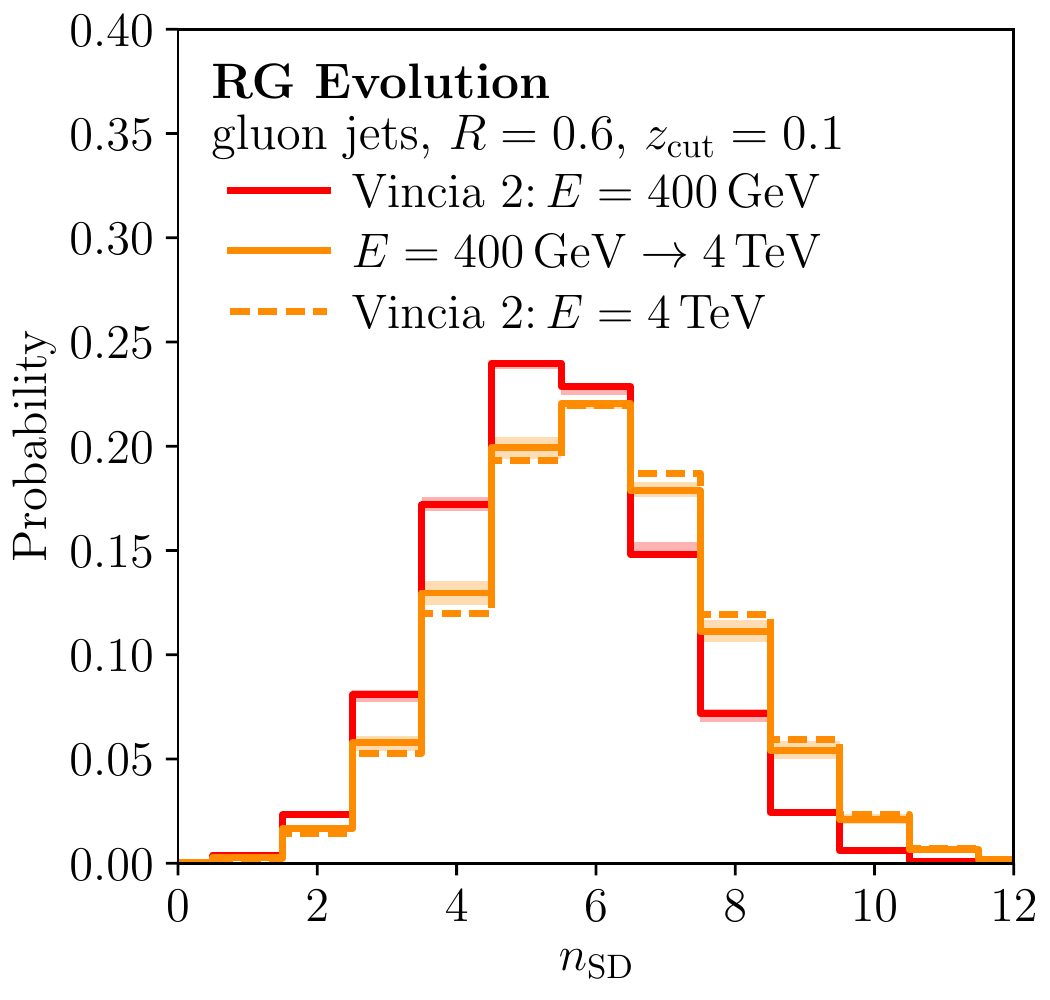}}
  \caption{RG evolution of the collinear-unsafe soft drop multiplicity for (left column) the quark singlet GFF and (right column) the gluon GFF.  Shown are the results for (top row) $\zcut = 0.02$ and (bottom row) $\zcut = 0.1$, taking distributions extracted from \textsc{Vincia} at a low scale and evolving them to a higher scale.  The uncertainties in the evolved distributions come from varying the jet radius used for GFF extraction and the $\mu$ scale for the RG evolution.}
  \label{fig:ev_dist}
\end{figure}

Using \Eq{eq:linear_evolution_again}, we evolve the GFFs to $\unit[4]{TeV}$ using the energy scale in \Eq{eq:mu_canonical} and the two-loop running of $\as$.\footnote{Since 
we only consider the leading-order evolution of the GFFs, strictly speaking, only leading-order evolution of $\as$ is needed at this order.  
Switching to one-loop running has a negligible effect on the results of this section.}  
This evolution includes all 10 active quark and antiquark flavors, as $n_f = 5$ in this energy range.\footnote{For simplicity, we ignore effects due to the $g\to t\bar{t}$ splitting, which would require a matching calculation to the top quark electroweak decay.} There are various sources of theoretical uncertainties in the evolved result, and we highlight two of them in this study.  
The first contribution is due to the fact that the energy scale \Eq{eq:mu_canonical} only depends on the product $E_{\text{jet}} R_0$, though the initial distributions could be extracted with any $R_0$.  
To estimate this uncertainty, which serves as a consistency check of the choice of $\mu$ scale, we also extract GFFs with $R_0 = 0.3$ and $R_0 = 0.9$, keeping $\mu$ fixed.  
The second contribution is from uncertainty in the absolute value of the energy scale itself.  To address this, we perform evolution with both half and double the energy scale of \Eq{eq:mu_canonical}.  We plot the envelope of these $9$ results in a shaded uncertainty band.  Of course, this is only a subset of the possible GFF uncertainties, but a full study is beyond the scope of this work.

\begin{figure}[t]
\centering
\subfloat[]{\includegraphics[width=.45\textwidth]{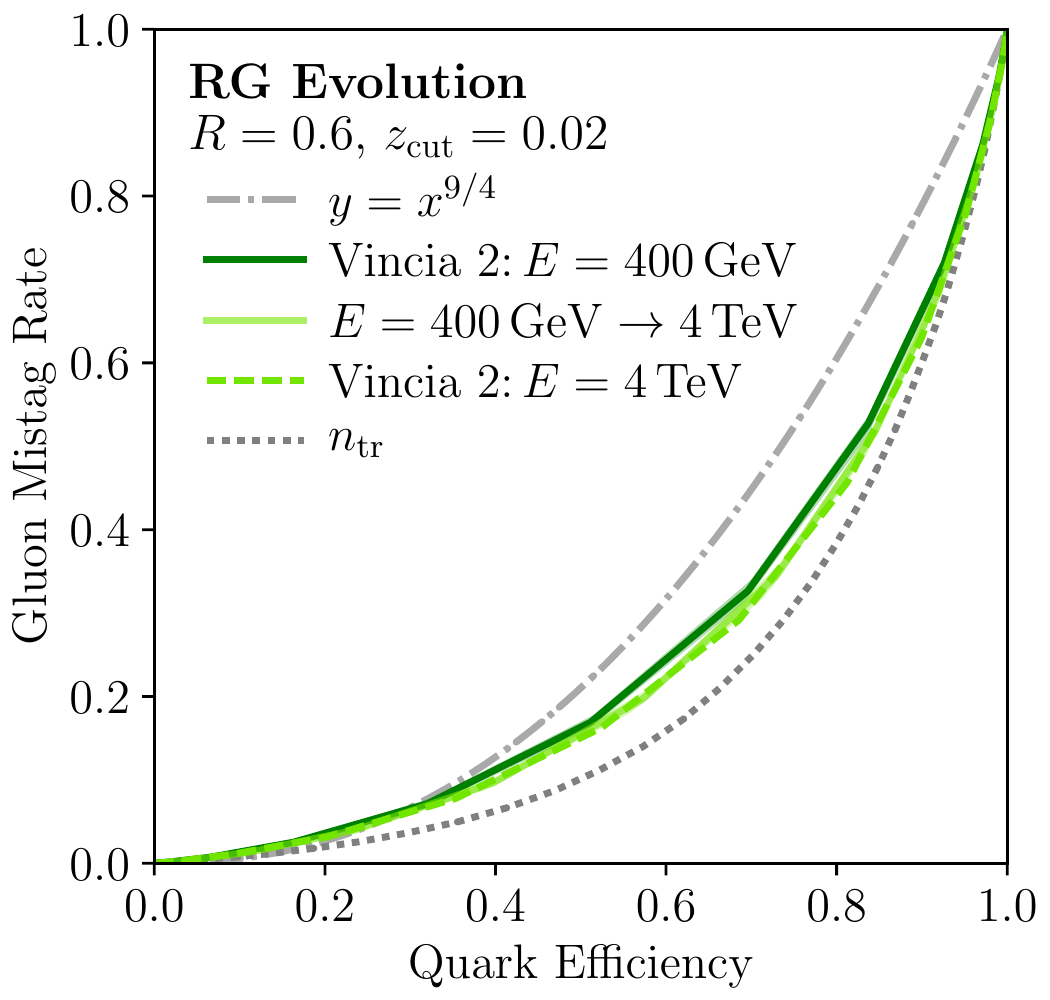}}
\subfloat[]{\includegraphics[width=.45\textwidth]{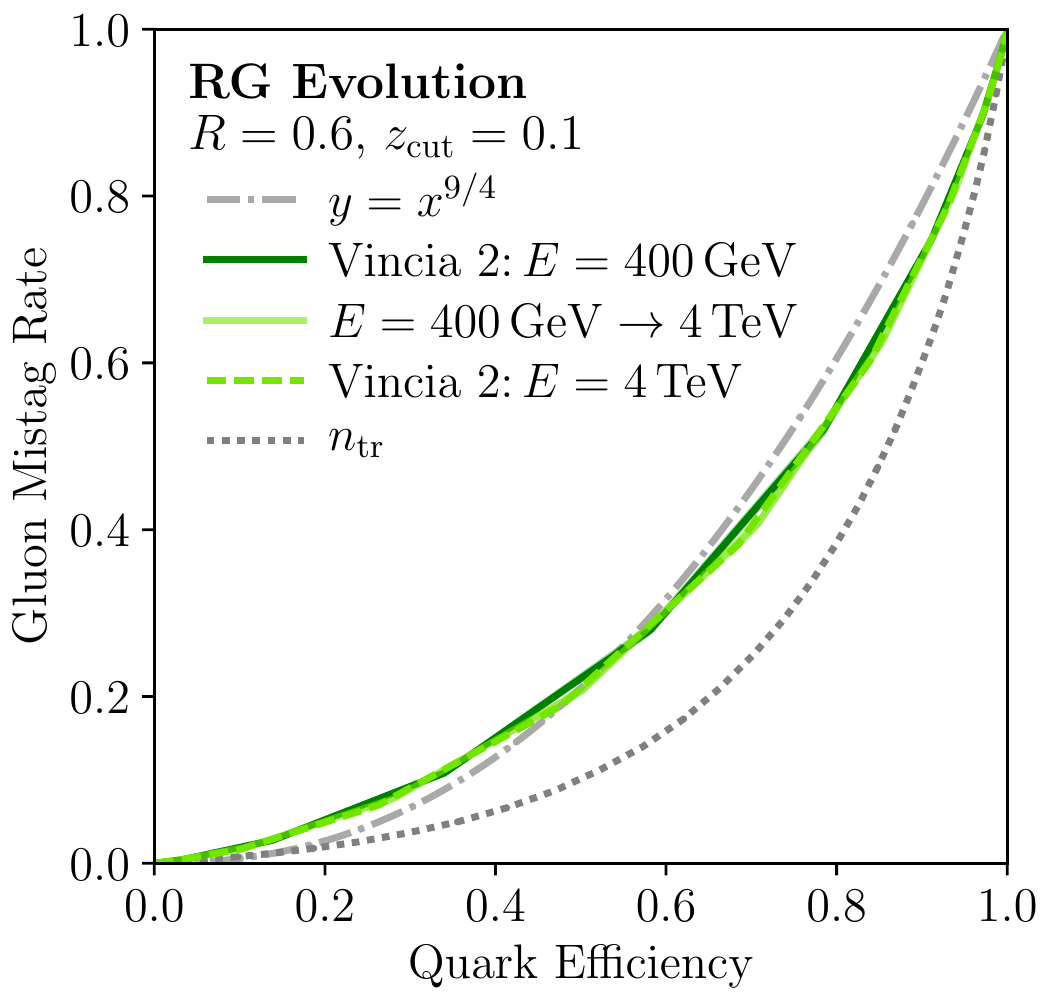}}
\caption{\label{fig:ev_rocs} RG evolution of ROC curve (quark singlet vs. gluon) for the collinear-unsafe soft drop multiplicity with (a) $\zcut = 0.02$ and (b) $\zcut = 0.1$. In both cases, there is very little evolution in the discrimination power with energy scale.}
\end{figure}

The results for $\zcut = 0.02$ and $\zcut = 0.1$ are shown in \Fig{fig:ev_dist}, comparing the RG-evolved results to \textsc{Vincia} distributions extracted at the high scale.  To show a single curve for quark jets, we plot the quark-singlet distribution
\begin{equation}
\label{eq:quark_singlet_def}
\mathcal{Q}(x, \mu) = \frac{1}{2n_f} \sum_{i \in \{u, \bar{u}, \ldots, b, \bar{b}\}} \mathcal{F}_i(x, \mu)
\end{equation}
as defined in \Ref{Elder:2017bkd} (where it is instead denoted by $\mathcal{S}$). We find reasonable agreement between the RG evolution and \textsc{Vincia} for both $\zcut$ values, with a larger range of evolution for the case of $\zcut = 0.02$.  The uncertainties in the RG evolution do not fully cover the high-scale \textsc{Vincia} distribution, though it is worth emphasizing that we are only using the LO evolution equations.

In \Fig{fig:ev_rocs}, we show the RG evolution of the quark/gluon ROC curves.  Despite the fact that the $n_\SD$ distributions themselves exhibit significant RG evolution, the corresponding ROC curves do not change significantly with the energy scale $\mu$.  This is a key prediction of the GFF approach, and one that we can better understand by studying the moments of the GFF distributions.

\subsection{Moment Space Evolution}

\newcommand{\gmom}{\overline{\mathcal{G}}}
\newcommand{\qmom}{\overline{\mathcal{Q}}}
\newcommand{\fmom}{\overline{\GFF}}

To understand the slow evolution of the quark/gluon discrimination power, consider the evolution in moment space.  Following \Ref{Elder:2017bkd}, the $n^\text{th}$ moment of a GFF is defined as
\begin{equation}
\fmom_i(n, \mu) = \int \df x\, x^n \GFF_i(x, \mu).
\end{equation}
In moment space, we denote the gluon GFF by $\gmom(n,\mu)$, and the quark-singlet GFF (as defined in \Eq{eq:quark_singlet_def}) by $\qmom(n,\mu)$.  To derive the moment space evolution equations, we integrate both sides of \Eq{eq:linear_evolution_sd} against $x^n$, shift the final integral by $x \to x+1$, and then simplify the $n^\text{th}$ moments with the splitting function identities
\begin{equation}
\int_0^1 \mathrm{d}z \, \left[ P_{g \to gg}(z) + 2 n_f P_{g \to q \bar{q}}(z) \right] = 0, \quad \int_0^1 \mathrm{d}z \, P_{q \to qg}(z) = 0.
\end{equation}
After these manipulations, the moment evolution equation for the $n^\text{th}$ gluon or quark-singlet GFF can be written solely in terms of the difference $\gmom(n) - \qmom(n)$, along with lower moments $\gmom(k)$, $\qmom(k)$ for $k < n$. 

For $n = 1$, the evolution equation for the means is
\begin{equation}
\label{eq:added_reference}
\mu \frac{\mathrm{d} }{\mathrm{d} \mu} \begin{pmatrix} \gmom(1) \\ \qmom(1) \end{pmatrix} = \frac{\alpha_s}{\pi} \left[  \big(\gmom(1) - \qmom(1) \big)  \begin{pmatrix} \bar{P}_{g\to gg}^{0, 1/2} \\ \bar{P}_{q\to qg}^{0, 1/2} \end{pmatrix} + \begin{pmatrix} \bar{P}_{g\to gg}^{\zcut, 1/2} + 2 n_f \bar{P}_{g\to \bar{q}q}^{\zcut, 1/2} \\ \bar{P}_{q\to qg}^{\zcut, 1/2} + \bar{P}_{q\to gq}^{\zcut, 1/2} \end{pmatrix}\right]
\end{equation}
where we are suppressing the $\mu$ arguments and using the abbreviated notation 
\begin{equation}
\bar{P}_{i\to jk}^{z_1,z_2} = \int_{z_1}^{z_2} \mathrm{d}z \, P_{i \to jk}(z).
\end{equation}
The appearance of the difference of the moments on the right-hand side has a dramatic effect on the high-energy limit of the evolution.
Specifically, the difference in the means evolves as 
\begin{equation}
\label{eq:mean_ev}
\mu \frac{\mathrm{d} }{\mathrm{d} \mu} \big(\gmom(1) - \qmom(1)\big) = \frac{\alpha_s}{\pi} \left[c_1 - c_2 \big(\gmom(1) - \qmom(1)\big)\right],
\end{equation}
where $c_1$ and $c_2$ are positive constants defined by integrals of the splitting functions.  Thus, at high energies, the difference in the means asymptotes to a constant, 
\begin{equation}
\gmom(1) - \qmom(1) \Rightarrow \frac{c_1}{c_2} = \frac{\bar{P}_{g\to gg}^{\zcut, 1/2} + 2 n_f \bar{P}_{g\to \bar{q}q}^{\zcut, 1/2} - \bar{P}_{q\to qg}^{\zcut, 1/2} - \bar{P}_{q\to gq}^{\zcut, 1/2}}{\bar{P}_{q\to qg}^{0, 1/2} - \bar{P}_{g\to gg}^{0, 1/2}}.
\end{equation}

This asymptotic behavior is strikingly different from the LL analysis of IRC-safe multiplicity in \Sec{sec:ll_scale}.  In the IRC-safe case, the LL prediction is that the gluon and quark means should have a constant ratio determined by $C_A/C_F$.  Here, in the collinear-unsafe case, the gluon and quark means asymptote to having a constant difference.
Physically, this occurs because the RG evolution takes flavor mixing effects into account, so that at sufficiently high energies, the $n_\SD$ distributions for quark and gluon jets become essentially the same. While we have only presented the calculation for the quark-singlet mean, it is straightforward to show that the means for each individual quark flavor behave in the same way, with differences between different quark flavors evolving to zero.  

Moving to higher moments, a useful simplification occurs for the variances,
\begin{equation}
\sigma^2_i = \fmom_i(2) - \fmom_i(1)^2.
\end{equation}
In this case, the evolution of the variances only depends on the difference of the variances and the difference of the means,
\begin{equation}
\label{eq:variance_ev}
\mu \frac{\mathrm{d} }{\mathrm{d} \mu} \begin{pmatrix} \sigma^2_{\gmom} \\ \sigma^2_{\qmom} \end{pmatrix} = \frac{\alpha_s}{\pi} \left[ \begin{pmatrix} \bar{P}_{g\to gg}^{0, 1/2} \\ \bar{P}_{q\to qg}^{0, 1/2} \end{pmatrix} \Big(\sigma^2_{\gmom} - \sigma^2_{\qmom} - \big(\gmom(1) - \qmom(1)\big)^2\Big) + \begin{pmatrix} \bar{P}_{g\to gg}^{\zcut, 1/2} + 2 n_f \bar{P}_{g\to \bar{q}q}^{\zcut, 1/2} \\ \bar{P}_{q\to qg}^{\zcut, 1/2} + \bar{P}_{q\to gq}^{\zcut, 1/2} \end{pmatrix}\right].
\end{equation}
At sufficiently high energies, $\gmom(1) - \qmom(1)$ approaches a constant, so the evolution equation for the variances is of the same form as the evolution equation for the means. We find that, like the means, the difference of variances asymptotes to a constant,
\begin{equation}
\label{eq:variance_limit}
\sigma^2_{\gmom} - \sigma^2_{\qmom} \Rightarrow \mathrm{const}.
\end{equation}
Substituting our asymptotic results back into \Eq{eq:added_reference} and \Eq{eq:variance_ev}, we see that both the mean and variance simply grow linearly in $\alpha_s(\mu) \df (\log \mu)$ at high energies, so that they become proportional in the UV limit. Therefore, even with flavor-mixing effects, the soft drop multiplicity maintains a Poisson-like distribution, with $\sigma^2 = O(\mu)$.

\begin{figure}[t]
\centering
\subfloat[]{
\label{fig:ev_moments_1}
\includegraphics[width=.45\textwidth]{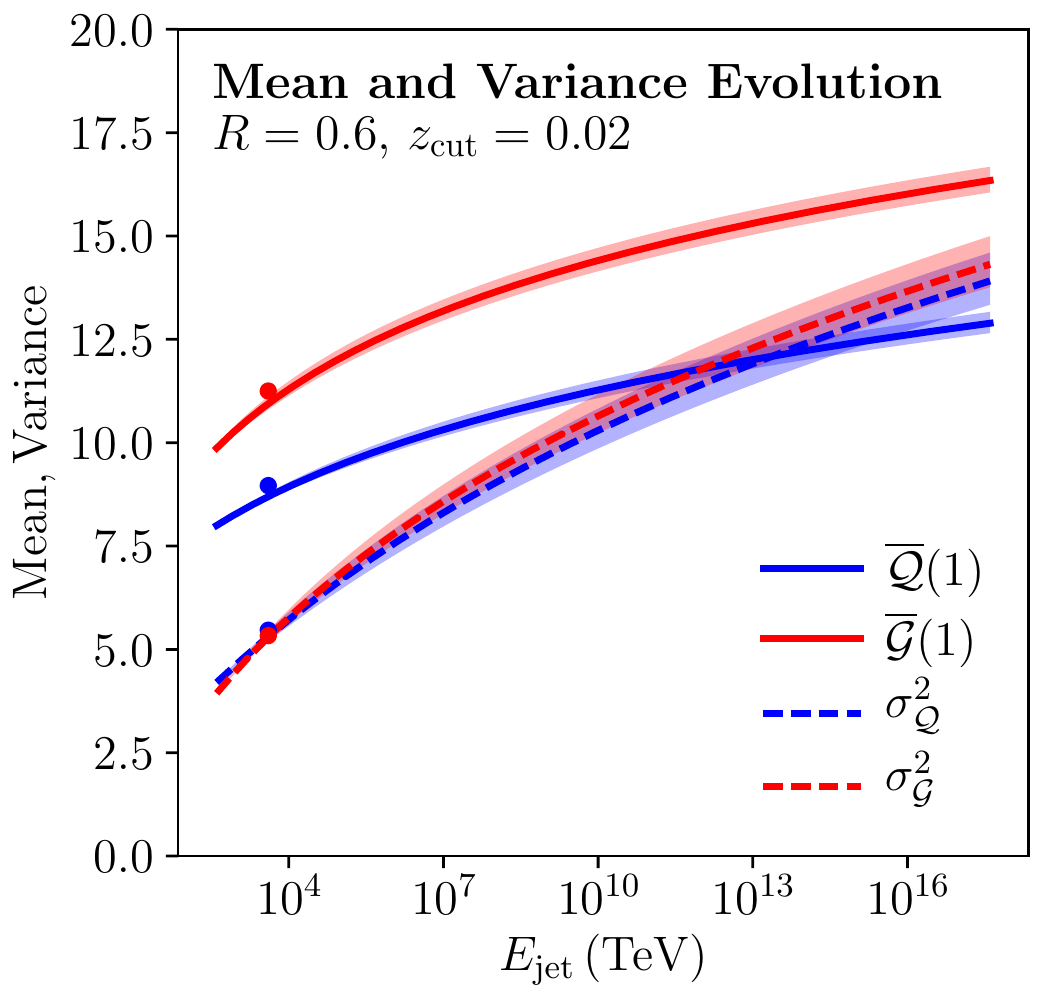}}
\subfloat[]{
\label{fig:ev_moments_2}
\includegraphics[width=.45\textwidth]{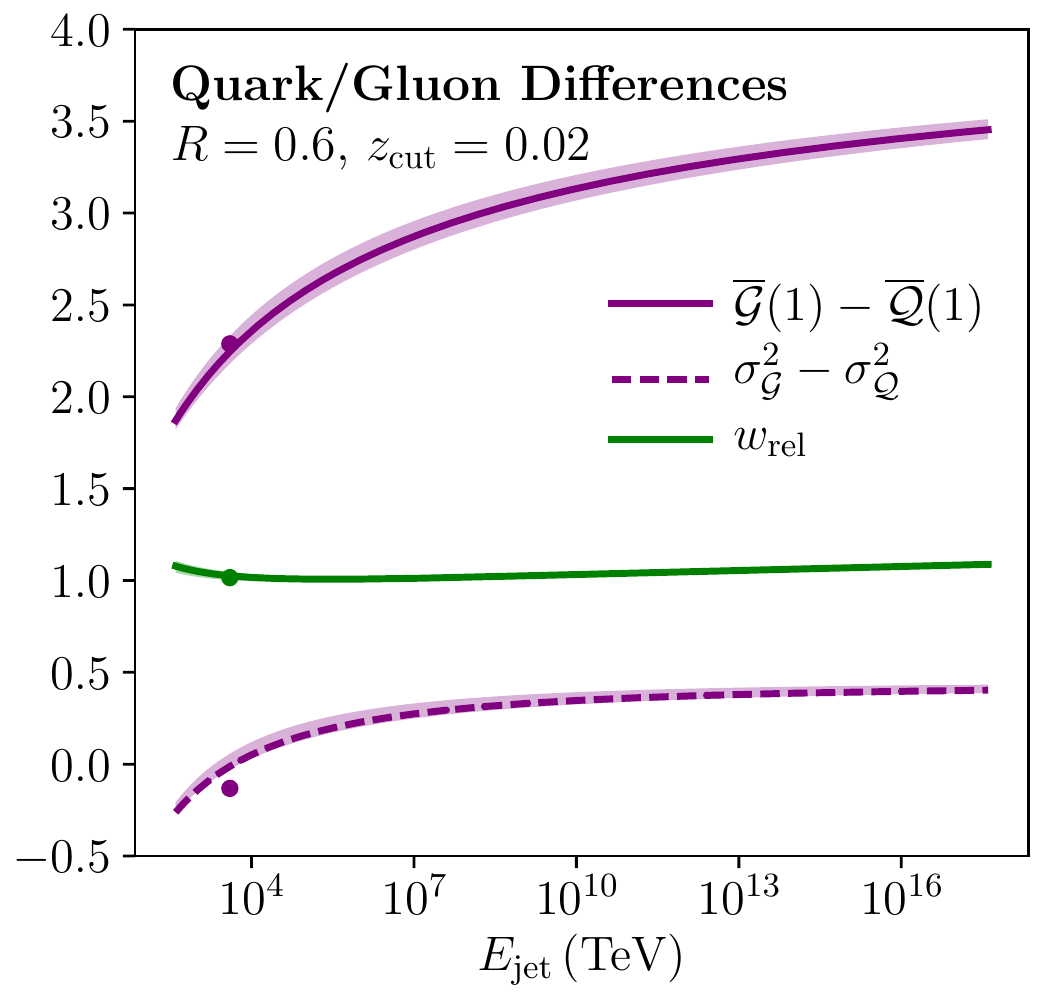}}
\caption{\label{fig:ev_moments} (a) RG evolution of means and variances of the quark-singlet and gluon GFFs for the soft drop multiplicity with $\zcut = 0.02$.  (b) RG evolution of the mean/variance differences, which asymptotically approach constants.  Also shown is the relative width $w_{\mathrm{rel}}$ defined in \Eq{eq:relativewidthunsafe}, which increases slowly.  For comparison, quantities extracted from $\textsc{Vincia}$ at $E_{\mathrm{jet}} = \unit[4]{TeV}$ are shown as dots.}
\end{figure}

We can roughly estimate the discrimination power of the soft drop multiplicity using a relative width, similar to that of \Eq{eq:relative_width_first}. Since Casimir scaling no longer holds, the distance between the quark-singlet and gluon distributions is no longer characterized by the means, but rather the difference in means. Moreover, in the UV limit, the standard deviations of the quark singlet and gluon distributions approach each other.   Thus, the quantity
\begin{equation}
\label{eq:relativewidthunsafe}
w_{\text{rel}} \equiv \frac{\sqrt{\sigma_{\gmom}^2}}{\gmom(1) - \qmom(1)}
\end{equation}
characterizes the extent to which the distributions overlap, and hence measures the discrimination power of the soft drop multiplicity.   We see that, as a result of flavor-mixing effects, the relative width is now expected to increase somewhat as more emissions are counted, roughly as the square root of the mean.

To verify these results, we numerically evolve the GFFs according to \Eq{eq:linear_evolution_again}, starting from an initial condition extracted from $\textsc{Vincia}$ 2.0.01 at $E_{\mathrm{jet}} = \unit[400]{GeV}$ and $R = 0.6$.  As in \Fig{fig:ev_dist}, we show a theoretical uncertainty band constructed from the envelope of 9 results.   In \Fig{fig:ev_moments_1}, we show the evolution of the mean and variance of the soft drop multiplicity for quark singlets and gluons. As expected from the above analysis, the mean and variance curves become parallel at sufficiently large values of $\mu$.  This is confirmed in \Fig{fig:ev_moments_2}, which shows that the differences do indeed asymptote to constant values.

Crucially, the relative width in \Fig{fig:ev_moments_2} remains approximately constant over a large energy range, as the increase in the standard deviation is canceled by the increase in the mean difference as it approaches its asymptotic value.  This explains the slow evolution of discrimination power seen in \Fig{fig:ev_rocs}.  In this way, even though these collinear-unsafe distributions cannot be predicted directly from first principles, the GFF approach gives us a valuable analytic handle on their RG evolution.


\section{Conclusions}
\label{sec:conclusion}

Quark/gluon discrimination has a long history, with many proposed discriminants \cite{Nilles:1980ys,Jones:1988ay,Fodor:1989ir,Jones:1990rz,Lonnblad:1990bi,Pumplin:1991kc,Gallicchio:2011xq,Chatrchyan:2012sn,Krohn:2012fg,Larkoski:2014pca,FerreiradeLima:2016gcz,Komiske:2016rsd,Davighi:2017hok} though relatively few analytic calculations \cite{Larkoski:2013eya,Larkoski:2014pca,Bhattacherjee:2015psa}.
Because $C_A/C_F$ is an order 1 number, distinguishing quark- from gluon-initiated jets is an intrinsically hard problem.
Moreover, to gain a quantitative understanding of quark/gluon separation power, one has to account for physics effects beyond the LL approximation, including the impact of nonperturbative physics.
These physics effects are modeled to differing degrees in parton shower generators, but ultimately one wants quark/gluon studies to be based on systematically-improvable analytic calculations.

In this paper, we introduced an IRC-safe counting observable which approaches the quark/gluon discrimination performance of IRC-unsafe track multiplicity.
Through a LL analysis, we demystified the power of multiplicity, showing that Poisson distributions typically yield better quark/gluon separation than Sudakov distributions, even though they are both controlled by the same $C_A$ and $C_F$ Casimir factors.
Specifically, we introduced soft drop multiplicity, which depends on multiple soft gluon emissions even at LL accuracy, allowing it to outperform observables like jet mass whose value is dominated by a single gluon emission.
Remarkably, there is a choice of ISD parameters where soft drop multiplicity is controlled by perturbative physics, such that its behavior can be reliably studied from first principles.  

To gain a more quantitative understanding of $n_{\SD}$, we introduced NLL evolution equations, which allowed us to make interesting comparisons to parton shower generators.
We also studied a collinear-unsafe (but infrared-safe) version of $n_{\SD}$, whose RG evolution could be studied using the formalism of GFFs.
In both cases, analytic understanding was aided by the recursive structure of the observable.
This motivates further studies into jet measurements performed on (groomed) clustering trees, which can depart significantly from the more commonly studied additive observables.

Ultimately, any single observable will never match the performance of multivariate jet tagging methods.
This has been emphasized recently in the context of deep neural networks which exploit subtle correlations to maximize separation power \cite{Cogan:2014oua,Almeida:2015jua,deOliveira:2015xxd,Baldi:2016fql,Conway:2016caq,Guest:2016iqz,Barnard:2016qma,Komiske:2016rsd,deOliveira:2017pjk,Kasieczka:2017nvn,Louppe:2017ipp}.
Still, we are encouraged by observables like soft drop multiplicity which offer a balance between discrimination power and analytic tractability.
Going beyond LL order where $n_{\SD}$ can saturate the discrimination power (see \Sec{sec:mean_discrim}), it would be interesting to study correlations between $n_{\SD}$ and other IRC-safe observables like jet mass to see if there is additional information in their combination.
Because the physics basis for $n_{\SD}$ is so transparent, we suspect it will be a useful benchmark for both parton shower tuning and experimental jet analyses.
Because the analytic structure of $n_{\SD}$ is so unique, we hope it inspires new precision calculations in QCD.

\acknowledgments

We thank Ben Elder, Massimiliano Procura, and Wouter Waalewijn for insights on GFF evolution.
We also benefitted from discussions with Fr\'{e}d\'{e}ric Dryer and Ben Nachman.
C.F. is supported by the U.S. Department of Energy under grant DE-SC0013607.
The work of K.Z. and J.T. is supported by the DOE under grant contract numbers DE-SC-00012567 and DE-SC-00015476.


\appendix

\section{Weighted Soft Drop Multiplicity}
\label{sec:weighted_nSD}

At the end of \Sec{sec:ll_scale}, we used LL reasoning to argue that soft drop multiplicity $n_\SD$ extracts all of the 
quark/gluon discriminatory information from the $(z_n,\theta_n)$ variables recorded by ISD. 
In this appendix, we study a variant of $n_\SD$, the weighted soft drop multiplicity, defined in \Eq{eq:nSD_again} and repeated for convenience:
\begin{equation}
\label{eq:nSD_again_again}
n_\SD^{(\kappa)} = \sum_{n} z_n^\kappa\,.
\end{equation}
While quark/gluon performance is not improved by weighting, the purpose of this appendix is to demonstrate that the techniques of this paper are applicable to a variety of observables. 

\subsection{Discrimination Power}
\label{subsec:weightedcase}

\begin{figure}[t]
\centering
 \includegraphics[width=.45\textwidth]{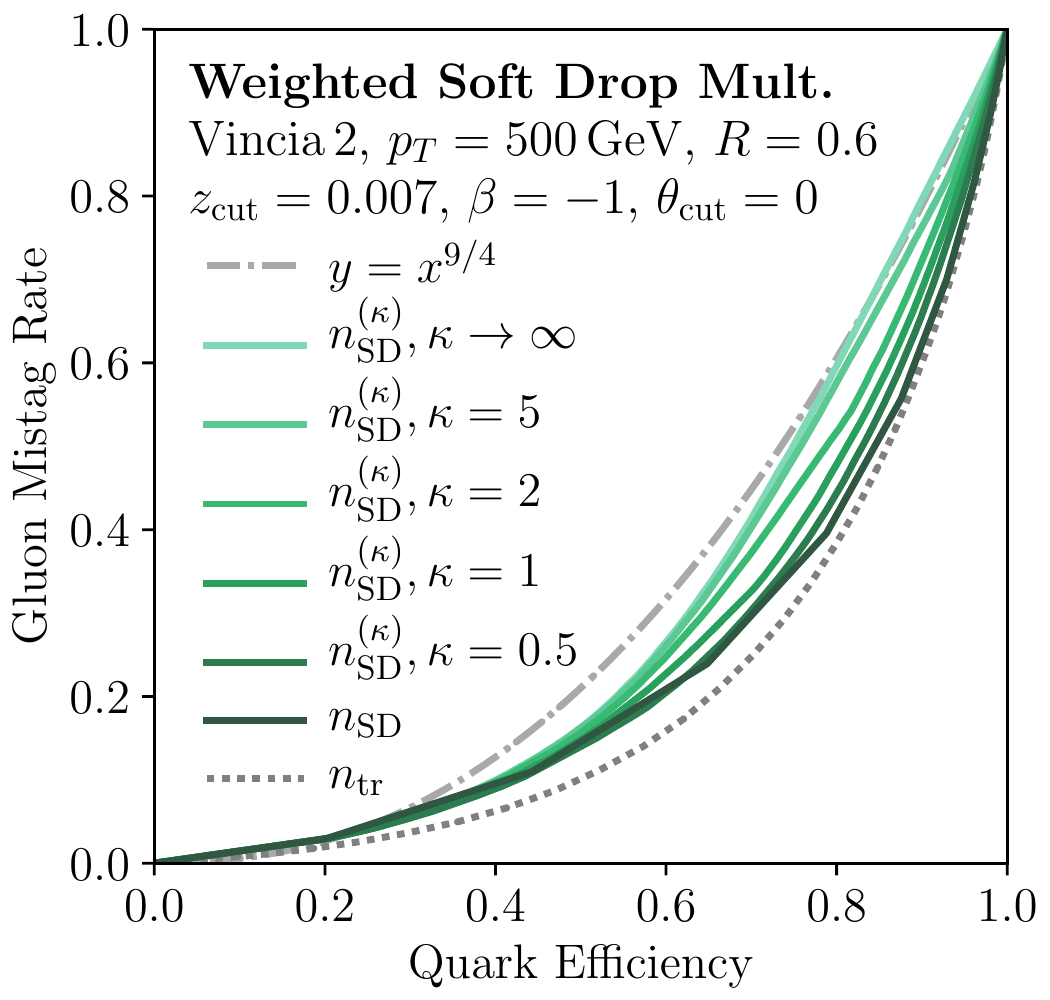}
\caption{
Quark/gluon discrimination power of weighted soft drop multiplicity as a function of $\kappa$, at the benchmark parameters from \Eq{eq:benchmark}.  We also show the limit $\kappa \to \infty$, which is equivalent to $\max(z_n)$. 
}
\label{fig:sd_kappa_sweep}
\end{figure}

For small values of $\kappa$, the weighted soft drop multiplicity is still sensitive to all emissions in the region $A_{\rm emit}$. 
On the other hand, as $\kappa \to \infty$, only the largest $z_n$ value contributes significantly to the observable. 
As a result, the weighted multiplicity interpolates between counting and additive behavior, in the limits $\kappa \to 0$ and $\kappa \to \infty$, respectively. 
The $\kappa$ dependence of the discrimination power, extracted from \textsc{Vincia}, is shown in \Fig{fig:sd_kappa_sweep}. 
One can see that the quark/gluon performance decreases monotonically as $\kappa$ increases. 

The LL distribution of the weighted soft drop multiplicity is analytically complicated.
Indeed, any analytic expression for it must contain a sum of distributions, one for each value of the number $n$ of counted emissions. 
For example, when $\beta \leq 0$, each emission contributes at least $z_\text{cut}^\kappa$, so at most $n$ emissions can contribute to $n_\SD^{(\kappa)}$ if its value is below $n \, z_\text{cut}^\kappa$. 
A full analysis along these lines is carried out in \App{sec:weight_emissions} below. 

To qualitatively understand the trend in \Fig{fig:sd_kappa_sweep}, consider the limit in which ISD records many emissions.
Strictly speaking, this analysis is not quantitatively applicable in the perturbative regime, where $n \lesssim 10$ emissions are counted.
Nor is this reasoning applicable in the collinear-unsafe regime studied in \App{sec:weight_emissions_RG}, where solely perturbative reasoning is insufficient.
Nonetheless, the many-emission limit serves to build intuition.

In the double-logarithmic approximation, where emissions are soft and collinear and $\as$ is a fixed coupling, the weighted multiplicity distribution can be found from summing independent identically distributed numbers.
By the central limit theorem, this converges to a normal distribution in the limit of many recorded emissions. 
In this limit, it suffices to compute the mean and variance of $n_\SD^{(\kappa)}$ to estimate its discrimination power. 
These are determined at lowest order from the average values of $z^\kappa$ and $z^{2\kappa}$ in the allowed emission region as
\begin{equation}
\label{eq:mean_var_weightedcase}
\langle n_\SD^{(\kappa)} \rangle_i = \rho_i A_{\rm emit} \langle z^\kappa \rangle\,, \qquad \text{Var}\left(n_\SD^{(\kappa)}\right)_i = \rho_i A_{\rm emit} \langle z^{2\kappa} \rangle\,,
\end{equation}
where
\begin{align}
\langle z^\kappa \rangle &= \frac{1}{A_{\rm emit}} \int_{\thetacut}^{R_0}  \frac{\df \theta}{ \theta}\int_{\zcut}^{1/2}\frac{\df z}{z} \, z^\kappa \, \Theta\left[z - z_{\rm cut} \left(\frac{\theta}{R_0}\right)^\beta \right]\,.
\end{align}
With a fixed coupling, the mean value of $z^\kappa$ for $\beta > 0$ is
\begin{equation}\label{eq:bpkappa}
A_\text{emit}\langle z^\kappa\rangle^{\beta > 0} = \frac{1}{2^\kappa \kappa}\log\frac{R_0}{\thetacut} - \frac{\zcut^\kappa}{\beta \kappa^2}\left(
1-\left(\frac{\thetacut}{R_0}\right)^{\beta \kappa}\right)\,.
\end{equation}
For $\beta < 0$, the mean value is
\begin{align}\label{eq:bnkappa}
A_\text{emit}\langle z^\kappa\rangle^{\beta < 0} &=\Theta\left[\thetacut-(2\zcut)^{\frac{1}{|\beta|}}R_0\right] \left(\frac{1}{2^\kappa \kappa}\log\frac{R_0}{\thetacut}-\frac{\zcut^\kappa}{\beta \kappa^2}\left[1-\left(\frac{\thetacut}{R_0}\right)^{\beta \kappa}\right]\right)\\
&
\hspace{1cm}
+\Theta\left[(2\zcut)^{\frac{1}{|\beta|}}R_0-\thetacut\right] \left(\frac{1}{2^\kappa \beta \kappa}\log(2\zcut)-\frac{\zcut^\kappa}{\beta\kappa^2}\Big[1-(2\zcut)^{-\kappa}\Big]\right)
\,.\nonumber
\end{align}
Because of the $\rho_i$ prefactor in \Eq{eq:mean_var_weightedcase}, we see that the mean and variance once again satisfy Casimir scaling as in \Eq{eq:mean_var_scaling}. 
Moreover, both the variance and mean scale with the counted area $A_{\rm emit}$, establishing that the weighted soft drop multiplicity is Poisson-like distributed as defined in \Sec{sec:ll_scale}.

The discrimination power is determined by the relative width
\begin{equation}
\label{eq:kappa_wrel}
w_{\text{rel}} \equiv \frac{\sqrt{\text{Var}\left(n_\SD^{(\kappa)}\right)_i}}{\left\langle n_\SD^{(\kappa)} \right\rangle_i} = \frac{1}{\sqrt{\rho_i A_{\rm emit}}} \frac{\sqrt{\langle z^{2\kappa} \rangle}}{\langle z^\kappa \rangle}.
\end{equation}
We can get a sense for the behavior of $w_{\text{rel}}$ by considering two extreme limits. 
For $\kappa\to 0$ and any choice of $\beta$, the mean value $\langle z^\kappa\rangle$ (and hence $w_{\text{rel}}$) approaches a constant, independent of $\kappa$. 
For $\kappa\to \infty$, the mean value scales with $\kappa$ like
\begin{equation}
A_\text{emit}\langle z^\kappa\rangle^{\kappa\to \infty} \sim \frac{1}{2^\kappa \kappa}\,,
\end{equation}
with $\zcut < 1/2$, such that the relative width scales as
\begin{equation}
w_\text{rel}^{\kappa\to \infty}\sim\sqrt{\kappa}\,.
\end{equation}
Since the relative width increases with increasing $\kappa$, this reasoning predicts that the discrimination power decreases as $\kappa$ increases. 
This implies the best discrimination power is attained for $\kappa = 0$ (i.e.~ordinary soft drop multiplicity) and decreases for higher $\kappa$. 
Physically, the discrimination power of $n_\SD^{(\kappa)}$ comes from sensitivity to multiple emissions, and for higher $\kappa$, sensitivity to softer emissions is decreased. 
In the extreme limit of $\kappa\to \infty$, the weighted soft drop multiplicity reduces to the energy fraction of the hardest emission, $n_\text{SD}^{\kappa\to\infty}=\max(z_n)$.

This qualitatively explains the trend seen in \Fig{fig:sd_kappa_sweep}, i.e.~that the discrimination power monotonically decreasing as $\kappa$ increases. 
In the limit $\kappa \to \infty$, the discrimination power reaches the universal result predicted by Casimir scaling (slightly off due to small nonperturbative corrections), 
as the observable $\max(z_n)$ is determined by a Sudakov form factor.

\subsection{Analytic Calculation}
\label{sec:weight_emissions}

Using evolution equations similar to those employed in \Sec{sec:irc_safe}, we can compute the distribution of IRC-safe weighted soft drop multiplicities. 
We will demonstrate this here at LL for simplicity; by taking into account flavor changes and energy losses, one could obtain NLL evolution equations as in \Sec{sec:NLL}. Since $n_\SD^{(\kappa)}$ is a continuous observable, however, significantly more computation time would be required 
to compute its NLL distribution, in comparison to the discrete unweighted case.

Let $p^i(n_\SD,\thetacut) \, \df n_\SD$ denote the differential probability that, given a flavor $i$ jet, its weighted soft drop multiplicity is measured to be $n_\SD$. Here, we leave the $\zcut$, $\beta$, and $\kappa$ dependence implicit.
Though the weighted soft drop multiplicity does not directly count emissions, it is still useful to keep track of the number of contributing emissions, using \begin{equation}
  {p^i(n_\SD,\thetacut)} = \sum_{n=0}^\infty {p^i_n(n_\SD,\thetacut)}\,,
\end{equation}
where $n$ labels the number of counted emissions as before. 
If we change the resolution angle from $\thetacut$ to $\thetacut - \delta\thetacut$, then 
\begin{align}
  p^i_n(n_\SD,\thetacut-\delta\thetacut) &= p^i_n(n_\SD,\thetacut) \, \left( 1- {\delta \thetacut \over \thetacut} \, \int_0^{1/2} \df z \, 
  {\as(z\,\theta\,p_T) \over \pi} \, P_{i \to i}(z) \, \Theta_\SD(z,\theta) \right) \nonumber \\
  &
  \hspace{-1cm}
  + {\delta\thetacut \over \thetacut} \, \int_0^{1/2}
  \df z~ {\as(z\,\theta \,p_T) \over \pi} \, P_{i \to i}(z) \, \Theta_\SD(z,\theta) \, 
  p^i_{n-1}(n_\SD-z^\kappa,\thetacut)\,.
\label{eq:kappaEvol}
\end{align}
This leads to a linear differential equation. 
Instead of the Poisson distribution found in \Sec{subsec:LLevol}, the solution in this case is differential in $n_\SD = \sum_i z_i^\kappa$:
\begin{align}
  p^i_n(n_\SD,\thetacut) &\\
&
\hspace{-2cm}
  = \frac{e^{-I_{i \to i}(\thetacut,R_0)}}{n!} \left( \, \prod_{i=1}^n \int_\thetacut^{R_0} 
  {d\theta_i \over \theta_i} \int_0^{1/2} \df z_i \, {\as(z_i \, \theta_i\,p_T) \over \pi} \, P_{i \to i}(z_i) \, 
  \Theta_\SD(z_i,\theta_i) \right) \delta\Big( n_\SD - \sum_{i=1}^n z_i^\kappa\Big)\,.\nonumber
\label{eq:kappaLL}
\end{align}

\begin{figure}[t]
\centering
\subfloat[]{\includegraphics[width=0.45\textwidth]{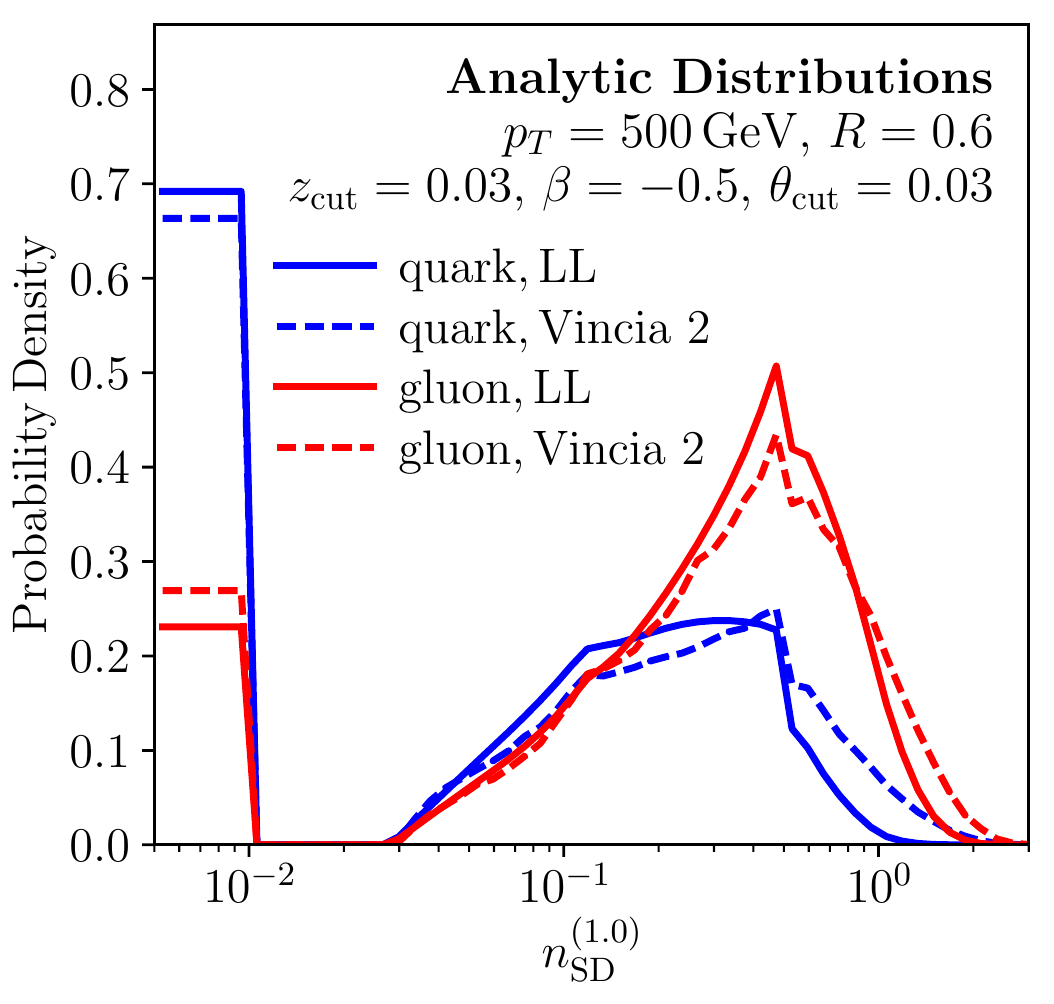}}
\subfloat[]{\includegraphics[width=0.45\textwidth]{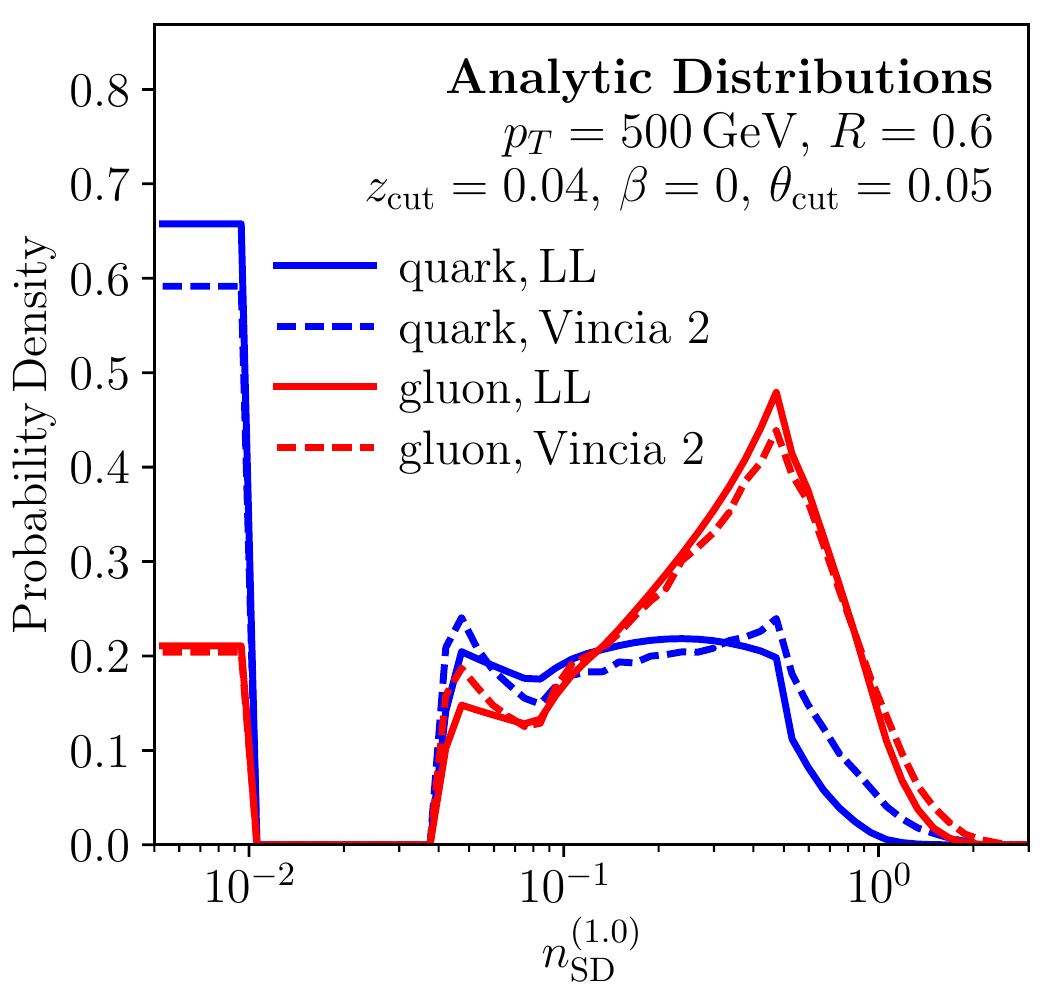}}
\caption{LL calculation of weighted soft drop multiplicity distributions with $\kappa =1$, compared to \textsc{Vincia}.
The plots have two different sets of ISD parameters which were chosen to display the sharp features characteristic of $n_\SD^{(\kappa)}$ in the perturbative regime.
The curves shown are the probability distribution functions of $\log n_{\mathrm{SD}}^{(1.0)}$, so that they integrate to one in logarithmic space. The leftmost bin is an underflow bin, showing the probability that no emissions were counted by ISD, such that $n_\SD^{(1.0)} = 0$. 
}
\label{fig:kappa_LL}
\end{figure}

In the perturbative regime, the behavior of $n_\SD^{(\kappa)}$ is most clearly seen on a logarithmic scale.
Two example LL distributions are displayed in \Fig{fig:kappa_LL} and compared to results from \textsc{Vincia}. 
In these examples, soft drop parameters were chosen to demonstrate that the sharp features of the $n_\SD^{(\kappa)}$ distributions are indeed captured by the LL evolution equations. 
These sharp features result from the edges of the $p_n^i(n_\SD,\thetacut)$ distributions for different values of $n$. 
For example, with $\beta \leq 0$, the $p_n^i(n_\SD,\thetacut)$ distribution only has support on the interval $[n\,z_\text{cut}^\kappa, {n \over 2^\kappa}]$. 

\begin{figure}[t]
\centering
\subfloat[]{\includegraphics[width=.45\textwidth]{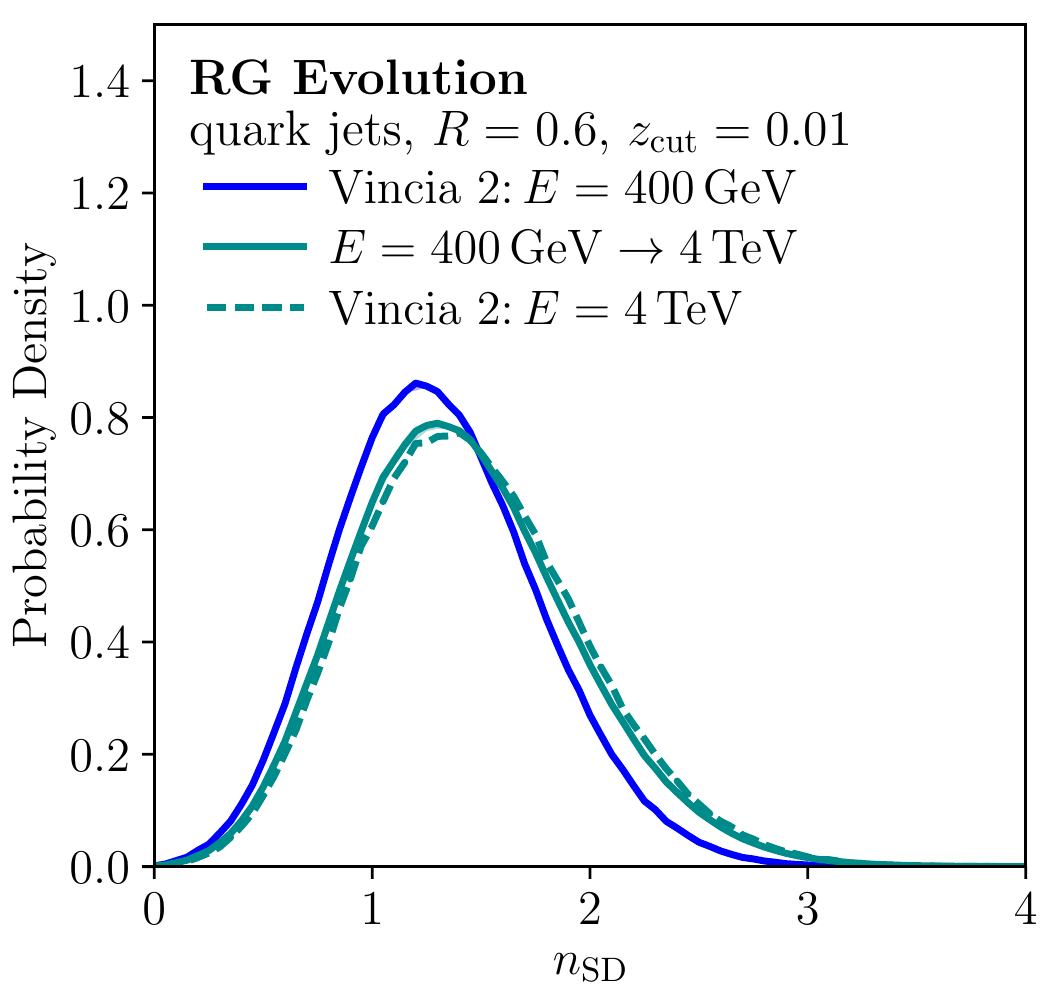}}
\subfloat[]{\includegraphics[width=.45\textwidth]{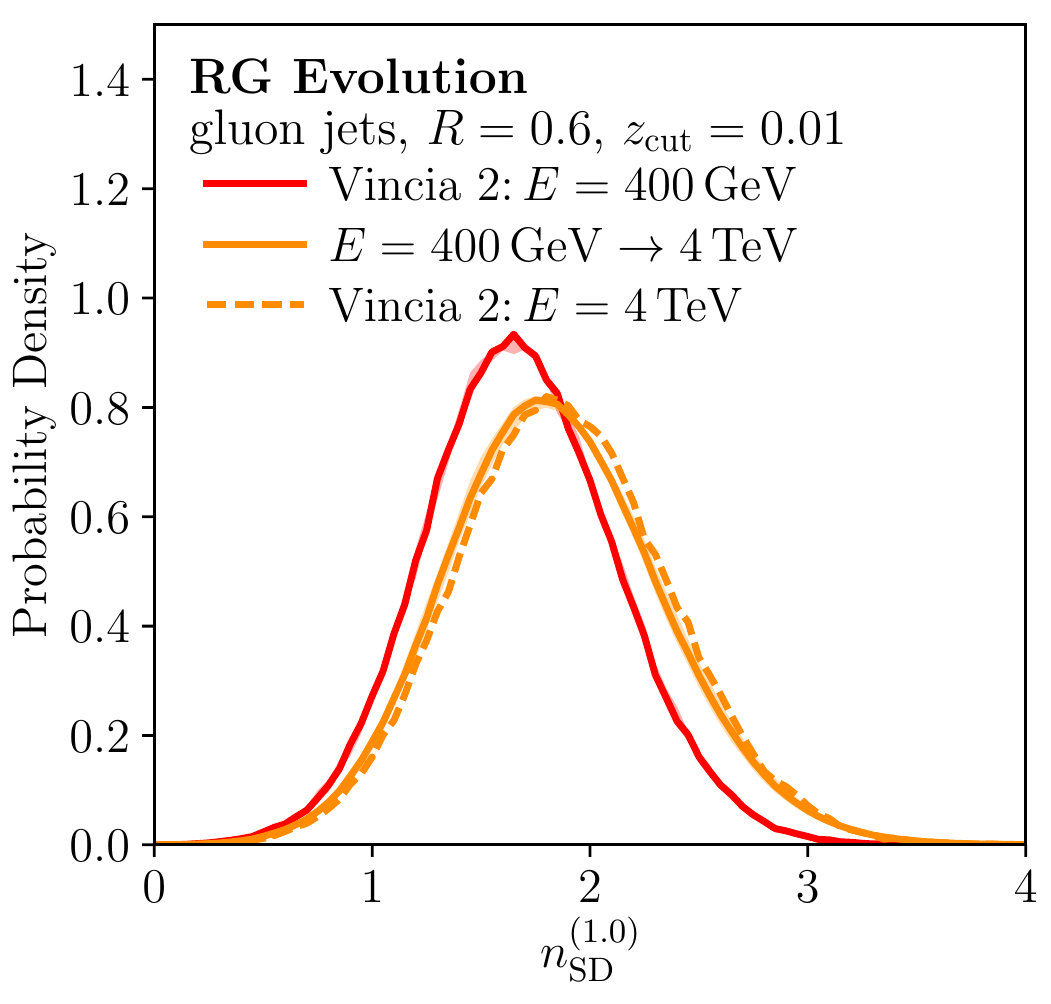}}
\caption{\label{fig:ev_dist_1} RG evolution of collinear-unsafe weighted soft drop multiplicity with $\zcut = 0.01$ and $\kappa = 1$, for the (a) quark-singlet and (b) gluon cases.}
\end{figure}

\subsection{Collinear-Unsafe Evolution}
\label{sec:weight_emissions_RG}

In the case of a collinear-unsafe weighted soft drop multiplicity with $\beta = 0$ and $\theta_{\rm cut} = 0$, we can apply the methods of \Sec{sec:irc_unsafe}.  Specifically, after extracting the GFF at some RG scale $\mu$, we can use \Eq{eq:linear_evolution_again} with the particular choice $f(z) = z^\kappa$ to predict the upwards evolution.  In \Fig{fig:ev_dist_1}, we compare the result of the RG evolution for $\zcut = 0.01$ and $\kappa = 1$ to \textsc{Vincia}, finding overall good agreement.  By eye, one can see that these $\kappa = 1$ distributions do not yield as good separation power as the $\kappa = 0$ distributions shown in \Fig{fig:ev_dist}, though the degree of RG evolution is similar for both the weighted and unweighted cases.

\bibliography{bibliography}
\bibliographystyle{JHEP}

\end{document}